\documentclass[aps,prl,reprint,superscriptaddress,floatfix]{revtex4-2}
\usepackage{graphicx,bm,epsfig,textcomp,color,dcolumn,setspace,array,calrsfs,mathrsfs,amsmath,amssymb,mathptmx,gensymb,booktabs,url,bibentry,natbib,subfigure,braket,lipsum,xcolor,relsize,diagbox}
\usepackage[T1]{fontenc}
\usepackage[mathscr]{eucal}
\usepackage[bookmarks=true,colorlinks=true,allcolors=blue,plainpages=false,pdfpagelabels,final,breaklinks=true]{hyperref}
\usepackage{cleveref}
\usepackage{tikz}
\usepackage[compat=1.1.0]{tikz-feynman}
\tikzfeynmanset{warn luatex=false}

\DeclareMathAlphabet{\mathcal}{OMS}{cmsy}{m}{n}
\makeatletter
\newcommand{\supplementarytableofcontents}{%
  \@starttoc{stoc}% 
}
\newcommand{\writesupplementarycontents}{%
    \let\latex@addcontentsline\addcontentsline
    \renewcommand{\addcontentsline}[3]{%
        \def\toc@extension{toc}%
        \def\target@extension{##1}%
        \ifx\target@extension\toc@extension
            \latex@addcontentsline{stoc}{##2}{##3}%
        \else
            \latex@addcontentsline{##1}{##2}{##3}%
        \fi
    }%
}
\def\fps@figure{!t}
\def\fps@table{!t}
\makeatother  

\newcommand{\Tr}{\mathrm{Tr}}
\newcommand{\tr}{\mathrm{tr}}

\begin{document}

\title{Fractional Quantum Geometry in Topological Mott Regime}
\author{Junyu Tang}
\affiliation{International Center for Quantum Materials, School of Physics, Peking University, 
Beijing 100871, China}
\author{Hongquan Lv}   
\affiliation{International Center for Quantum Materials, School of Physics, Peking University, 
Beijing 100871, China} 

\author{Gang v.~Chen}  
\email{chenxray@pku.edu.cn}
\affiliation{International Center for Quantum Materials, School of Physics, Peking University, 
Beijing 100871, China}
\affiliation{Beijing Key Laboratory of Quantum Devices, Peking University, Beijing 100871, China}
\affiliation{Collaborative Innovation Center of Quantum Matter, 100871, Beijing, China}

\begin{abstract}
Inspired by recent interests in quantum geometry in electronic bands, we consider a two-dimensional topological Mott insulator, 
 {\sl i.e.} chiral spin liquid, whose spinon bands
 develop nontrivial band topological and quantum geometry. 
 The virtual polarization of the gapped charge sector
transfers the electromagnetic drive to the charge-neutral spinons 
through the emergent $U(1)$ gauge field, 
producing a tensor Ioffe-Larkin response and 
enabling two complementary optical protocols for measurements.  
In the off-shell (low-frequency) regime, a self-calibrated ratio of the physical longitudinal and Hall
conductivities yields the spinon Chern number without the microscopic knowledge of the charge response. 
In the on-shell (resonant) regime, combining the inverse Ioffe-Larkin response with the Kramers-Kronig 
relation enables reconstruction of the quantum-metric and Berry-curvature spectral densities. Our approach directly applies to the triangular lattice Hofstadter-Hubbard model in Moir\'e systems, 
can be well adapted to other triangular lattice antiferromagnets such as Nb$_3$Br$_8$ with appropriate modifications. We establish a unified framework for quantitatively probing the quantum geometry of fractionalized quasiparticles in topological Mott regime. 
\end{abstract}

% composite fermion quantum geometry ??? 

\maketitle

\noindent\textit{Introduction.---}The quantum geometric tensor 
plays a pivotal role in quantum materials,
governing transport, optical, and collective phenomena beyond what is encoded
in the band dispersion~\cite{XiaoChangNiu2010,Ahn2022}.
Its antisymmetric part, the Berry curvature, determines Hall-type responses,
while its symmetric part, the quantum metric, controls interband spectral
weight, Wannier localization, and geometric contributions to flat-band
superfluidity~\cite{ProvostVallee1980,SouzaWilkensMartin2000,Marzari2012,NeupertChamonMudry2013,PeottaTorma2015}. Electromagnetic probes can access these quantities for
charged quasiparticles~\cite{OzawaGoldman2018,Tran2017}, 
but not directly for the fractionalized, electrically neutral excitations of a quantum spin
liquid in the Mott regime~\cite{Anderson1973,SavaryBalents2017,ZhouKanodaNg2017,Broholm2020}.
 
This obstruction is especially interesting in the
topological Mott regime, where the neutral excitations may 
develop the nontrivial band structure topology
even though the charge transport 
is frozen~\cite{KalmeyerLaughlin1987,KalmeyerLaughlin1989,WenWilczekZee1989,Szasz2020,Cookmeyer2021}. 
One representative example in 2D is the chiral spin liquid (CSL)
where the spinon band has a nontrivial Chern number. 
Recently, several materials, including the moir\'e heterostructure~\cite{Huang2024CSL,Kuhlenkamp2024}, 
the triangular-lattice organics~\cite{Shimizu_PRL_2003,Szasz2020}, and the cluster magnets Nb$_3$Br$_8$~\cite{Yao2026SurfaceCSL} have been proposed to realize the CSL. 
Unlike an ordinary Chern insulator, a CSL has no electronic 
dc Hall conductance from which the spinon Chern number can be read 
off directly. 
Previous studies have investigated the sub-Mott-gap optical and magneto-optical signatures 
of spinon Fermi surfaces and fractionalization~\cite{NgLee2007,Potter2013,Colbert2014,MaNg2015,MaNg2016}, as well as electromagnetic signatures of chiral order~\cite{Banerjee2023}. More recently, Raman circular dichroism has been shown to be sensitive to the quantum geometry and handedness of fractionalized excitations~\cite{Koller2025}. These works primarily identified qualitative spectroscopic signatures of fractionalization and chirality. Here, we instead develop a quantitative optical protocol that explicitly extracts the spinon Chern number and reconstructs the hidden quantum geometry of gapped spinon bands without a Fermi surface.

%%%%

\begin{figure}[b]
    \centering
    \includegraphics[width=7.5cm]{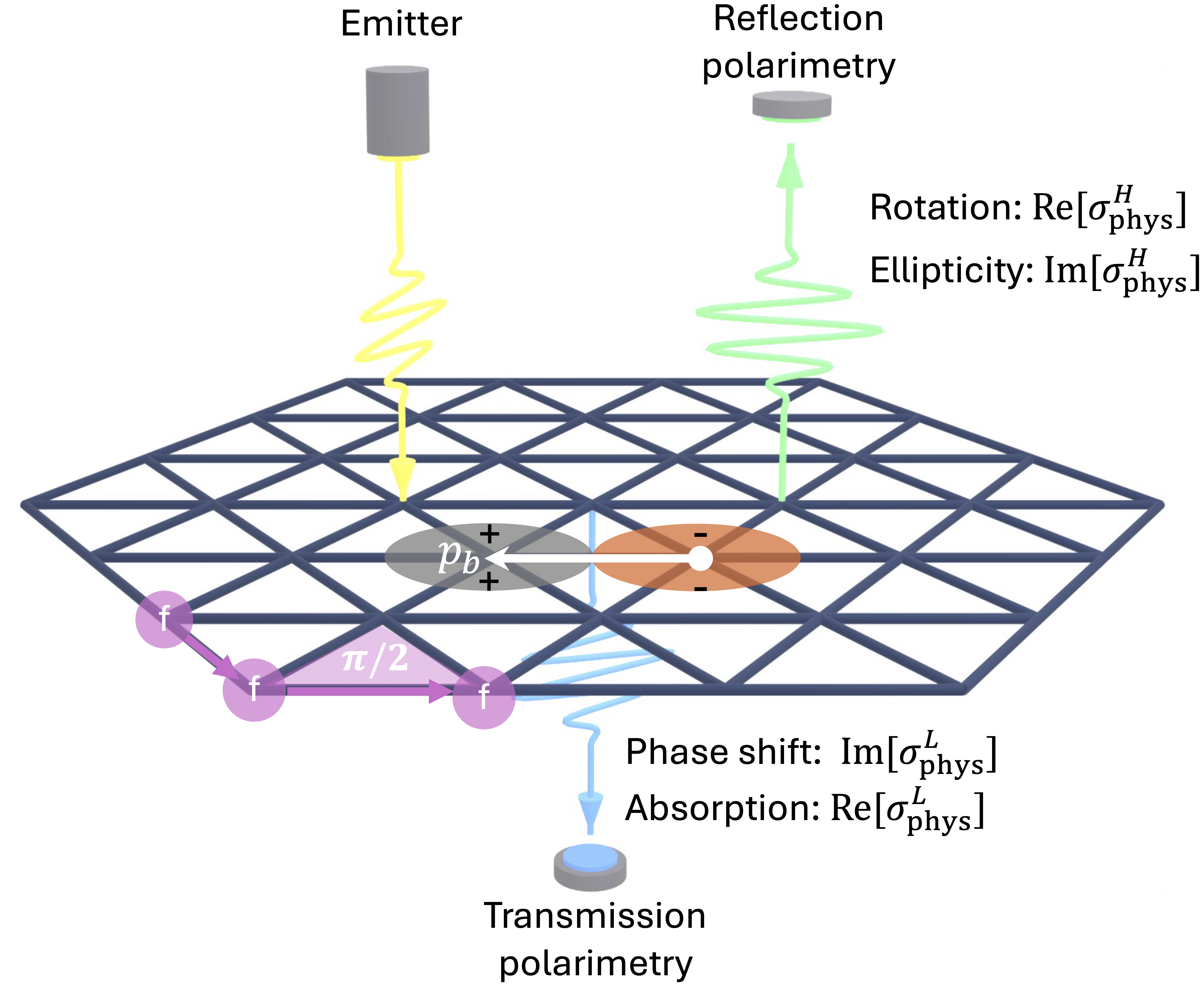}
    \caption{Optical probe of spinon quantum geometry on a triangular lattice CSL. Linearly polarized incident light (yellow) couples to the charge sector and, 
    through imbalanced doublon-holon fluctuations, induces a virtual polarization $p_b$ (white arrow), 
    which mediates the coupling to the internal gauge field seen by the spinons (purple). 
    Transmission and reflection polarimetry resolve the longitudinal and Hall conductivities, respectively.}
    \label{fig:model} 
\end{figure}

We start with a slave-rotor formulation 
that simultaneously captures the charge and spin fluctuations together in a CSL. 
In this framework, a sub-Mott-gap electromagnetic field can virtually polarize the gapped charge sector, thereby driving neutral spinons through the induced internal gauge field~\cite{IoffeLarkin1989,NgLee2007}, as illustrated in Fig.~\ref{fig:model}. This mediation produces a frequency-dependent Mott filter that dresses the spinon response. Consequently, the physical optical response is not a direct measure of the spinon quantum geometry, 
but rather a tensor Ioffe-Larkin composition of the chargon and spinon responses. 
In this Letter, we show that the off-shell, low-frequency longitudinal and Hall responses 
together determine the spinon Chern number without independent knowledge of the charge polarizability, 
whereas the on-shell resonant responses reconstruct the energy-resolved 
quantum-metric and Berry-curvature spectral densities.

\noindent\textit{Parton optical response.---}We decompose the electron operator 
as ${c_{i\sigma}=b_i f_{i\sigma}}$, 
where 
$b_i=e^{i\theta_i}$ is the charged rotor (chargon) and $f_{i\sigma}$ is a
neutral spinon~\cite{FlorensGeorges2004,LeeNagaosaWen2006}. 
As a specific model, we consider the half-filled triangular-lattice 
Hofstadter--Hubbard (TLHH) model with electron hopping $t$, repulsion $U$, 
and orbital flux ${\Phi_\triangle=\pm\pi/2}$ per triangle plaquette from the external magnetic field. 
In the uniform CSL solution, 
the spinons entirely inherit this flux while the chargons experience zero static flux, 
thus becoming trivially gapped~\cite{Yang_PRL_2026,ZhangLiuSong2026,Divic2025}. 
Writing the static electromagnetic and internal backgrounds as $\bar{\bm A}$ and $\bar{\bm a}$, respectively, 
this saddle point admits the gauge choice $\bar{\bm a}=\bar{\bm A}$. 
We use the unbarred fields $\bm A$ and $\bm a$ to denote the optical probe 
and the internal gauge fluctuation about these backgrounds. Therefore, the
spinons couple to $\bm a$, whereas the chargons couple to ${\bm A-\bm a}$. 
Since $\bar{\bm a}=\bar{\bm A}$, the chargon sector indeed experiences zero average flux. 
Throughout this Letter, we use the natural units $e=\hbar=k_B=1$ for simplicity.

%%%%%

Integrating out the parton fields yields the effective quadratic action, 
which in the uniform optical limit has the form~\cite{SM},
\begin{equation}
    S^{\rm eff}[A,a]
    =\frac12(A-a)_i\Pi_b^{ij}(A-a)_j+\frac12a_i\Pi_f^{ij}a_j.
    \label{eq:quadratic_action_letter}
\end{equation}
Here, $\Pi_b$ and $\Pi_f$ are the chargon and spinon response kernels 
that depend on the driving frequency $\omega$, with $\{i,j,k\}\in \{x,y,z\}$. 
Both spatial and frequency sums are assumed implicitly. 
Integrating out the gauge field $\bm a$ yields the tensor Ioffe--Larkin rule 
for the physical response kernel~\cite{IoffeLarkin1989,NagaosaLee1990,SchofieldWheatley1993},
\begin{equation}
    \Pi_{\rm phys}
    =\Pi_b-\Pi_b(\Pi_b+\Pi_f)^{-1}\Pi_b.
    \label{eq:Pi_phys}
\end{equation}
where we have omitted the spatial indices, and the inverse relation reads $\Pi_{\rm phys}^{-1}=\Pi_b^{-1}+\Pi_f^{-1}$.
%The physical retarded response is obtained by analytic continuation $i\omega_n\to \omega+i0^+$ from the Matsubara frequency $\omega_n$ to the real frequency $\omega$. 
The physical optical conductivity is given 
by $\sigma_{\rm phys}(\omega)=i\Pi_{\rm phys}(\omega)/\omega$.

%%% 

In the CSL Mott regime, the chargon is gapped and thus is not condensed. 
The chargon has a real hopping and a vanishing static relative flux. 
Its antiunitary symmetry forbids a Hall response at all frequencies, 
while the unbroken $C_3$ rotation symmetry makes the longitudinal response isotropic~\cite{SM}. 
Thus, we have the decomposition along the longitudinal and Hall channels 
for the spinon and chargon kernels as
\begin{equation}
    \Pi_b^{ij}(\omega)=p_b(\omega)\delta^{ij},\quad
    \Pi_f^{ij}(\omega)=\Pi_f^L(\omega)\delta^{ij}+\Pi_f^H(\omega)\epsilon^{ij},
    \label{eq:parton_tensor_structure_letter}
\end{equation}
with $\epsilon^{ij}$ the Levi-Civita symbol. Below the chargon pair energy threshold
$\omega_b^{\rm th}=2\Delta_b$, the real doublon-holon pair creation is forbidden;
virtual doublon-holon fluctuations instead produce an electric polarization 
response. The clean chargon kernel is therefore real, and we denote this
scalar kernel by $p_b$ to emphasize its polarization character.
Its low-frequency expansion is~\cite{SM} 
\begin{equation}
    p_b(\omega)=-\chi_b \omega^2+O(\omega^4).
    \label{eq:chargon_low_frequency_letter}
\end{equation}
where $\chi_b=-\lim_{\omega\to 0}p_b/\omega^2$ is the chargon polarizability and the minus sign is consistent with a positive electric polarizability convention. The full expression for the Gaussian chargon response is derived in the SM~\cite{SM}. For an isotropic tensor $X^{ij}=X^L\delta^{ij}+X^H\epsilon^{ij}$, define
the circular eigenvalues $X^\pm=X^L\pm iX^H$. Eq.~\eqref{eq:Pi_phys}
then becomes 
\begin{equation}
    \Pi_{\rm phys}^\pm
    =\frac{p_b\Pi_f^\pm}{p_b+\Pi_f^\pm}
    \equiv\mathcal F^\pm \Pi_f^\pm,\quad
    \mathcal F^\pm=\frac{p_b}{p_b+\Pi_f^\pm}.
    \label{eq:IL_circular_general}
\end{equation}
The complex, frequency-dependent Mott filter $\mathcal F^\pm$ expresses the mediation of the optical drive by virtual chargon polarization. 
Consequently, the physical response does not directly measure the pure spinon kernel; instead, it encodes both the chargon and spinon responses in a nontrivial way. The TLHH model provides a perfect example of a gapped CSL where the chargon sector that directly couples to the probe electromagnetic field is trivial, while the spinon sector that couples to the internal gauge field is topologically nontrivial. In the following, we first show how the spinon quantum geometry is encoded in the spinon kernel, and then demonstrate how to extract it from the physical response.

%%%%%%%

The TLHH with the uniform flux $\Phi_\triangle=\pi/2$ or $-\pi/2$ per triangle plaquette 
has two magnetic sublattices once the gauge is fixed. For each spin flavor its Bloch Hamiltonian 
can be written as $h_f(\bm k)=\bm d(\bm k)\cdot\bm\tau$, where $\bm\tau$ acts on the two sublattices. 
For $\Phi_\triangle=\pi/2$, a convenient gauge gives~\cite{SM}
\begin{equation} 
    \bm d(\bm k)
    =-2t_f\bigl(\cos k_2,-\cos(k_2-k_1),\cos k_1\bigr),
    \label{eq:triangular_spinon_letter}
\end{equation}
with $k_1=ak_x$, $k_2=a(k_x+\sqrt3k_y)/2$, and lattice spacing $a$.
The band energies are $E_\pm=\pm d$, where $d=|\bm d|$.
The lower band is filled for each of the $N_\sigma=2$ spin flavors and has
$C_s=+1$ in our convention. Complex conjugation reverses the flux and
$C_s$, giving the opposite chirality of the
$U(1)_{\pm2}$ CSL~\cite{KalmeyerLaughlin1987,WenWilczekZee1989}. 
%Note that the Zeeman term is assumed to be much smaller than the spinon gap and has been ignored since it does not affect the spinon quantum geometry. 
The hopping amplitudes for spinon and chargon are determined by mean-field solutions 
$t_f=t|\langle b_i^\dagger b_j\rangle|$ and $t_b=t|\sum_\sigma\langle f_{i\sigma}^\dagger f_{j\sigma}\rangle|$, together with the Hilbert space constraint fixed $\lambda$. 
The details of self-consistent solutions are given in the SM~\cite{SM}. The spinon $\Delta_s$ and chargon gaps $\Delta_b$ 
are known once $t_f$ and $t_b$ are determined at a specific value of $U/t$. 
The spinon interband threshold is then $\Delta_s=2\min_{\bm k}d(\bm k) =2\sqrt3t_f$, 
and the interband continuum spans $\Delta_s\leq\omega\leq2\Delta_s$~\footnote{Here $\Delta_s$ is a particle--hole threshold, whereas $\Delta_b$ corresponds to the single-chargon gap}.

%%%

Including both spin flavours, the spinon model has four bands. 
Since the total spin is conserved and the Hamiltonian separates 
into two identical two-band blocks, each with one filled lower band. 
The eigenstates factorize as 
$\ket{u_{\pm,\sigma}(\bm k)}=\ket{u_\pm(\bm k)}\otimes\ket{\sigma}$,
with the momentum-independent spin states. 
The two flavours contribute the identical geometric responses, 
accounted for by the factor $N_\sigma=2$ below. 
We therefore define the quantum geometric tensor of 
the occupied lower band $\ket{u_-}$ within a single spin flavour as
\begin{equation}
    Q_{ij}
    =\langle\partial_i u_-|(1-|u_-\rangle\langle u_-|)
    |\partial_j u_-\rangle
    =g_{ij}-\frac{i}{2}\Omega_{ij}.
    \label{eq:QGT_letter}
\end{equation}
whose real and imaginary parts correspond to the quantum metric 
$g_{ij}$ and Berry curvature $\Omega_{xy}$, 
respectively~\cite{ProvostVallee1980,XiaoChangNiu2010,Ahn2022}. 
After the imaginary-time path integral and Matsubara summation~\cite{SM}, 
the zero-temperature retarded kernels are  
related to the quantum geometric tensor as 
\begin{align}
\Pi_{f}^{L}(\omega)
    &=
    \mathcal P\int [d\bm k]
    \frac{4d(\bm k)\omega^2}{
        \omega^2-[2d(\bm k)]^2
    }
    g_{xx}(\bm k)\notag\\
    &-i \pi\omega^2
    \int [d\bm k]
    g_{xx}(\bm k)
    \delta[
        \omega-2d(\bm k)
    ] ,
    \label{eq:Pi_L}\\
    \Pi_{f}^{H}(\omega)=
    &-\frac{\pi\omega^2}{2}
    \int [d\bm k]
    \Omega_{xy}(\bm k)
    \delta[\omega-2d(\bm k)],\nonumber\\
    &-i \mathcal P
    \int [d\bm k] 
    \frac{4\omega d^2(\bm k)\Omega_{xy}(\bm k)}{\omega^2-[2d(\bm k)]^2}
    \label{eq:Pi_H},
\end{align}
where $\int[d\bm k]\equiv
N_\sigma\int_{\rm MBZ}d^2k/(2\pi)^2$ runs over the magnetic Brillouin
zone and $\mathcal{P}$ denotes the principal value. 
These expressions connect the spinon longitudinal and Hall kernel 
(in response to internal gauge fields) with the spinon quantum metric 
and Berry curvature, providing the basis for the two optical extraction protocols below.

\noindent\textit{Off-shell measurement.---}Our first goal is to extract the spinon Chern number 
from the measurable
subgap conductivities. The starting point is the dc spinon Hall conductivity. 
Taking the low-frequency limit of Eq.~\eqref{eq:Pi_H} gives
\begin{align}
    \sigma_{f,0}^H
    &\equiv\operatorname{Re}\sigma_f^H(\omega\to0)
    =-\lim_{\omega\to0}\frac{\operatorname{Im}\Pi_f^H(\omega)}{\omega}
    \notag\\
    &=-\int[d\bm k]\,\Omega_{xy}(\bm k)
    =-\frac{N_\sigma C_s}{2\pi}.
    \label{eq:sigma_f_H_0}
\end{align}
where $C_s$ is the Chern number per spin flavour. Unfortunately, 
this quantized response is to the internal gauge field and cannot be measured directly. 
To determine
how it enters the physical optical response, we decompose the Ioffe-Larkin
rule in Eq.~\eqref{eq:IL_circular_general} into the longitudinal and Hall channels with
\begin{align}
    \Pi_{\rm phys}^L
    &=p_b-\frac{p_b^2(p_b+\Pi_f^L)}
    {(p_b+\Pi_f^L)^2+(\Pi_f^H)^2},
    \label{eq:Pi_phys_L_exact}\\
    \Pi_{\rm phys}^H
    &=\frac{p_b^2\Pi_f^H}
    {(p_b+\Pi_f^L)^2+(\Pi_f^H)^2}.
    \label{eq:Pi_phys_H_exact}
\end{align}
The physical Hall response thus inherits the spinon Hall kernel, but its
magnitude is modulated by the charge sector. We here use the accompanying
longitudinal response to eliminate this unknown contribution in the
low-frequency limit. For $0<\omega<\Delta_s\ll\omega_b^{\rm th}=2\Delta_b$,
both sectors are nonabsorptive. $p_b$ and $\Pi_f^L$ are real, and
$\Pi_f^H$ is purely imaginary. Eq.~\eqref{eq:Pi_phys_H_exact} can be simplified as 
\begin{equation}
    \operatorname{Re}\sigma_{\rm phys}^H(\omega)
    =\mathcal F^H(\omega)\operatorname{Re}\sigma_f^H(\omega),
    \label{eq:Sigma_H_0}
\end{equation}
where $\mathcal F^H(\omega)$ is a real Hall Mott filter and has the form
\begin{equation}
    \mathcal F^H(\omega)
    =\frac{p_b^2(\omega)}
    {[p_b(\omega)+\Pi_f^L(\omega)]^2+[\Pi_f^H(\omega)]^2}.
    \label{eq:mott_filter_exact}
\end{equation}
To identify the conductivity combination that cancels the charge contribution,
we expand the kernels at low frequency,
\begin{subequations}
\label{eq:low_frequency_kernels_letter}
\begin{align}
    p_b&=-\chi_b\omega^2+O(\omega^4),\\
    \Pi_f^L&=-\chi_f^L\omega^2+O(\omega^4),\\
    \Pi_f^H&=-i\omega\sigma_{f,0}^H+O(\omega^3),
\end{align}
\end{subequations}
where $\chi_f^L=\int[d\bm k]\,g_{xx}(\bm k)/d(\bm k)>0$ and
$\chi_b>0$ comes from the definition in Eq.~\eqref{eq:chargon_low_frequency_letter}.
For $C_s\ne0$, the Hall term controls the leading order of the
Ioffe--Larkin denominator. Thus, based on Eq.~\eqref{eq:mott_filter_exact}, we have $\mathcal F^H=-\chi_b^2\omega^2/(\sigma_{f,0}^H)^2+O(\omega^4)$. On the other hand, substituting the expansions from Eq.~\eqref{eq:low_frequency_kernels_letter} into Eq.~\eqref{eq:Pi_phys_L_exact} gives
$\Pi_{\rm phys}^L=-\chi_b\omega^2+O(\omega^4)$.
Collecting all the frequency-dependent terms, we obtain
\begin{align}
    \operatorname{Re}\left[\sigma_{\rm phys}^H\right]
    &=\frac{2\pi\chi_b^2}{N_\sigma C_s}\omega^2+O(\omega^4),
    \label{eq:low_frequency_H}\\
    \operatorname{Im}\left[\sigma_{\rm phys}^L\right]
    &=-\chi_b\omega+O(\omega^3).
    \label{eq:low_frequency_L}
\end{align}
From the above expressions, we see that the charge polarizability $\chi_b$ appears in both the longitudinal and Hall responses. 
This leads to a self-calibrated estimator for the spinon Chern number $C_s$ based solely on 
the measurable physical conductivities, without requiring any microscopic knowledge of the chargon response. Specifically, we define the finite-frequency Chern-number estimator $C_s(\omega)$ as
\begin{equation}
    \begin{aligned}
    C_s(\omega)\equiv\frac{2\pi}{N_\sigma}
    \frac{[\operatorname{Im}\sigma_{\rm phys}^L(\omega)]^2}
    {\operatorname{Re}\sigma_{\rm phys}^H(\omega)}.
    \end{aligned}
    \label{eq:cs}
\end{equation}
From Eqs.~\eqref{eq:low_frequency_H} and \eqref{eq:low_frequency_L}, we know that $C_s(\omega)$ approaches the quantized spinon Chern number $C_s=\lim_{\omega\to0}C_s(\omega)$ in the $\omega\to0$ limit~\cite{cond_dim}, providing a direct optical probe of the topological invariant of the neutral spinon bands.

%%%%%%%%%

\begin{figure}[!t]
    \centering
    \includegraphics[width=0.9\linewidth]{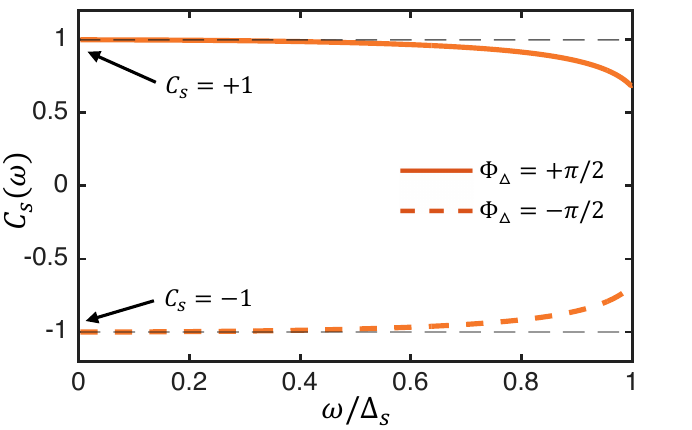}
    \caption{Finite-frequency Chern-number estimator $C_s(\omega)$ from
    Eq.~\eqref{eq:cs} for the triangular-lattice CSL at
    $U/t=15$. The solid (dashed) orange curves correspond to
    $\Phi_\triangle=+\pi/2$ ($-\pi/2$), and approach
    $C_s=+1$ ($-1$) as $\omega\to0$. The horizontal gray dashed lines mark the
    quantized values.}
    \label{fig:conductivity}
\end{figure}

To verify our proposal, we numerically evaluate Eq.~\eqref{eq:cs} based on our TLHH model 
with the self-consistent ansatz. Fig.~\ref{fig:conductivity} shows that $C_s(\omega)$ 
indeed approaches $+1$
and $-1$ in the low-frequency limit for the two opposite fluxes. 
Reversing the flux sign reverses the 
Hall response and the estimator, while leaving the longitudinal response
unchanged. The numerical result also shows that an accurate estimate does not require
an asymptotically small probe frequency: even at $\omega/\Delta_s=0.5$, we have $C_s(\omega)\simeq\pm0.98$, 
which is already very close to the ideal quantized value. Therefore, the estimator remains close
to the topological invariant over an appreciable subgap frequency window.
This is useful experimentally because, in the insulating regime, ${\rm Re}[\sigma^H_{\rm phys}]$ and ${\rm Im}[\sigma^L_{\rm phys}]$
vanish quickly as $\omega^2$ and $\omega$ at low frequencies, respectively. 
Thus, measurement at finite subgap frequencies supports larger measurable signals. 
Exact quantization refers to extrapolation to the $\omega=0$ limit. 
Finally, we note that in the presence of a finite spinon broadening $\eta_s$, the estimator deviates from the quantized value for $\omega\lesssim\eta_s$ and diverges as $\omega\to0$, although the underlying band Chern number remains unchanged. Reliable extraction therefore requires the window $\eta_s\ll\omega\ll\Delta_s$, as explained by the low-frequency analysis~\cite{SM}.

%%%%%%%%%

\noindent\textit{On-shell measurement.---}We next turn to the energy-resolved 
quantum geometry probed by the spinon
interband transitions, for which the on-shell condition is
${\omega=2d(\bm k)}$. Define the geometric spectral densities as
$O(\omega)=\int[d\bm k]\,O(\bm k)\delta[\omega-2d(\bm k)]$,
where $O=g_{xx}$ or $\Omega_{xy}$. Eqs.~\eqref{eq:Pi_L} and
\eqref{eq:Pi_H} relate these spectra directly to the spinon kernels:
\begin{align}
    g_{xx}(\omega)=-\frac{\operatorname{Im}\Pi_f^L(\omega)}
    {\pi\omega^2},\quad \Omega_{xy}(\omega)=-\frac{2\operatorname{Re}\Pi_f^H(\omega)}
    {\pi\omega^2}\label{eq:Kf_to_geometry_letter}.
\end{align} 
To recover the spinon kernels from the measured
physical response, we need to circumvent the unknown charge contribution. 
In the circular basis, the absence of a chargon Hall response
gives us $\Pi_b^\pm=p_b \in \mathbb{R}$. Introducing the inverse responses
$R_\pm\equiv1/\Pi_{\rm phys}^\pm$, $r_\pm\equiv1/\Pi_f^\pm$, and
$q_b\equiv1/p_b$, we obtain
\begin{equation}
    R_\pm=q_b+r_\pm.
    \label{eq:inverse_circular_letter}
\end{equation}
The unknown chargon term identically enters in both channels and can be removed by taking their difference. 
In addition, since we only consider the frequency regime well below the chargon absorption threshold 
$\omega_b^{\rm th}$, taking the imaginary part of their sum removes the real $p_b$ as well. 
To express the spinon information isolated by these two combinations, 
we similarly define the longitudinal and Hall components of the inverse spinon kernel 
as $r_f^L=(r_++r_-)/2$ and $r_f^H=(r_+-r_-)/(2i)$. They satisfy
\begin{subequations}
\label{eq:inverse_observables_letter}
\begin{align}
    r_f^H(\omega)&=\frac{R_+(\omega)-R_-(\omega)}{2i},\\
    \operatorname{Im}r_f^L(\omega)
    &=\operatorname{Im}\frac{R_+(\omega)+R_-(\omega)}{2}.
\end{align}
\end{subequations}
These relations determine $r_f^H$ and $\operatorname{Im}[r_f^L]$ directly
from optical data. The inverse physical responses are obtained from
$R_\pm=i/(\omega\sigma_{\rm phys}^\pm)$. The four real conductivity
components entering $\sigma_{\rm phys}^\pm$ can be measured in the polarization-resolved reflection and transmission channels [Fig.~\ref{fig:model}]. Specifically, Kerr rotation and ellipticity in the reflection polarimetry encode ${\rm Re}[\sigma_{\rm phys}^{H}]$ and
${\rm Im}[\sigma_{\rm phys}^{H}]$, respectively. In transmission, the phase shift encodes ${\rm Im}[\sigma_{\rm phys}^{L}]$, whereas
${\rm Re}[\sigma_{\rm phys}^{L}]$ is inferred from the absorbed intensity
after correcting for reflection losses.

\begin{figure}[t]
    \centering
    \includegraphics[width=1\linewidth]{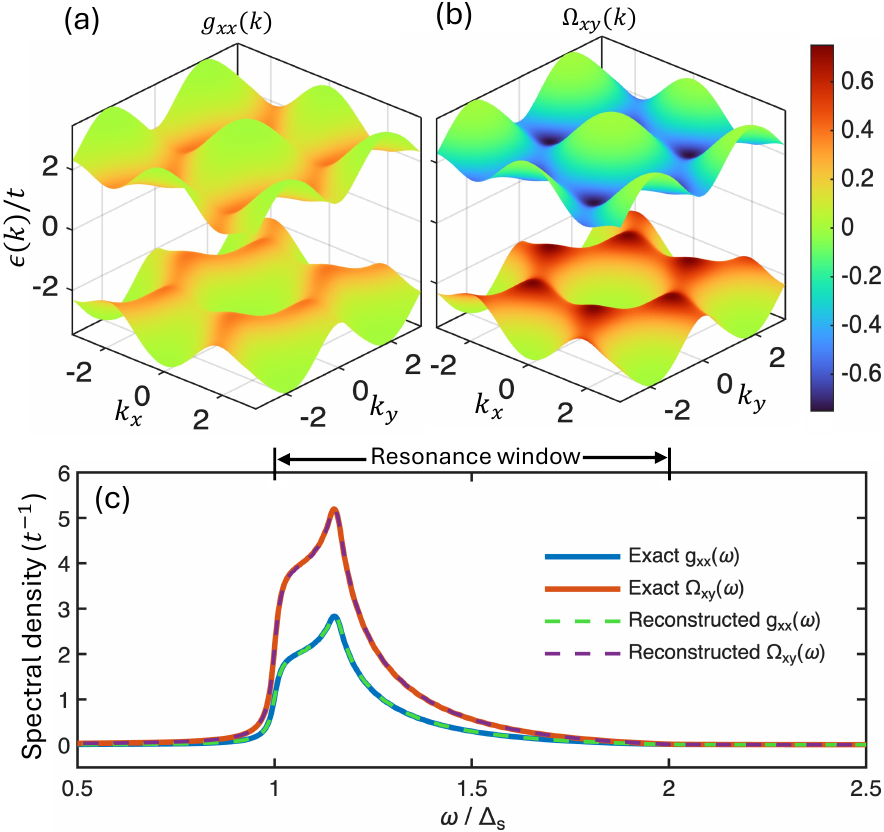}
    \caption{Spinon quantum geometry and its optical reconstruction for
    the triangular-lattice CSL. (a,b) Spinon bands colored by the quantum
    metric $g_{xx}(\bm k)$ and Berry curvature $\Omega_{xy}(\bm k)$,
    respectively. (c) Geometric spectral densities calculated directly
    (solid curves) and reconstructed using the inverse-response
    Kramers--Kronig protocol (dashed curves), at $U/t=15$ with small spinon broadening
    $\eta_s/t_f=0.05$.% The marked resonance window $\Delta_s\leq\omega\leq2\Delta_s$ corresponds to the spinon
    interband continuum.}
    \label{fig:metric_and_curvature}
\end{figure}

The last remaining unknown component, $\operatorname{Re}[r_f^L]$, 
is determined up to a subtraction constant by the Kramers-Kronig (KK) relation implied by causality:
\begin{equation}
    \operatorname{Re}\left[r_f^L(\omega)\right]
    =r_{f,\infty}^L+\frac{2}{\pi}\mathcal P
    \int_0^\infty d\omega'\,
    \frac{\omega'\operatorname{Im}r_f^L(\omega')}
    {\omega'^2-\omega^2}.
    \label{eq:inverse_KK_letter}
\end{equation}
Here $r_{f,\infty}^L=\lim_{|z|\to\infty}r_f^L(z)$ denotes the constant term at infinite frequency 
since $\Pi^L_f(\infty)$ is finite.
The Kramers-Kronig relation requires an analytic inverse kernel. 
$\Pi_f^\pm(z)$ must not have zeros on the upper half-plane, which is indeed the case in our model. 
Additionally, the opposite zero-frequency poles of $r_\pm$ cancel 
in $r_f^L(0)\propto(r_+(0) + r_-(0))$. The analytic conditions 
and numerical implementation are provided in the SM~\cite{SM}.

For the TLHH CSL ansatz, the constant term $r_{f,\infty}^L$ can be
calibrated from the spinon interband threshold. 
Carrying out the high-frequency expansion of Eq.~\eqref{eq:Pi_L} and Eq.~\eqref{eq:Pi_H} 
directly gives $r_{f,\infty}^L=(t_f\mathcal D_\triangle)^{-1}$, 
where $\mathcal D_\triangle\simeq1.38681$ is a dimensionless geometrical factor 
that depends only on the lattice structure and the number of spin flavors~\cite{SM}.
The spinon gap is given by $\Delta_s=2\sqrt3t_f$, which leads to
$r_{f,\infty}^L\simeq2.49788/\Delta_s$.
The interband factors $\delta[\omega-2d(\bm k)]$ imply that the
spinon response has no absorptive weight below $\Delta_s=2\min_{\bm k}d(\bm k)$. 
Thus, the measured leading interband edge $\omega^{th}_f$ of $\operatorname{Im}[r_f^L(\omega)]$ 
identifies $\Delta_s= \omega^{th}_f$~\cite{SM}. 
This determines $r_{f,\infty}^L$ and completes the KK reconstruction. 
Since $r^L_f(\omega)$ only has a measurable weight in the interband window $[\Delta_s,2\Delta_s]$, 
the KK integral can be evaluated with a finite cutoff $2\Delta_s <\Omega_{\rm cut} < 2\Delta_b$ in the integral~\eqref{eq:inverse_KK_letter}. We also provide another calibration method 
for $r_{f,\infty}^L$ based on the first moment of the dynamical spin structure factor~\cite{SM}.

%%%%%%

Once $r_f^L$ is completely determined, the spinon kernel follows from
\begin{equation}
    \Pi_f^\pm(\omega)
    =\bigl[r_f^L(\omega)\pm i r_f^H(\omega)\bigr]^{-1}.
    \label{eq:KK_spinon_recovery_letter}
\end{equation}
Recovering $\Pi_f^L=(\Pi_f^++\Pi_f^-)/2$ and
$\Pi_f^H=(\Pi_f^+-\Pi_f^-)/(2i)$ then yields the geometric spectra through
Eq.~\eqref{eq:Kf_to_geometry_letter}. We test this protocol using the self-consistent TLHH ansatz.
Figs.~\ref{fig:metric_and_curvature}(a,b) show the spinon bands colored
by their quantum metric and Berry curvature. The solid curves in
Fig.~\ref{fig:metric_and_curvature}(c) are obtained by directly integrating
these microscopic geometric quantities over momentum with the chosen
spinon broadening; they serve as the exact benchmark within the
mean-field model. For the dashed curves, we first combine the microscopic
spinon and chargon kernels through Eq.~\eqref{eq:Pi_phys} to generate
synthetic optical conductivities. We then treat these conductivities as the
measured inputs and apply the inverse-response KK protocol, with the
subtraction constant calibrated from $\Delta_s$, to recover the spinon
kernels and hence the geometric spectra. 
The reconstruction protocol itself uses no microscopic chargon kernel as input. 
Both calculations use the same broadening. 
The near coincidence of the solid and dashed curves 
in Fig.~\ref{fig:metric_and_curvature}(c) shows that the protocol accurately reproduces 
the spectral properties of the spinon quantum geometry, 
including the peak positions and heights, the relative spectral 
weights of the metric and curvature, and the broadened tails, 
thereby validating our inverse-response KK reconstruction theory. 
In practice, experimental reconstruction requires accurate measurements 
of the optical response and accurate calibration
of $\Delta_s$. The small chargon kernel comparing to the spinon kernel 
makes the inverse-response
extraction sensitive to measurement errors. In principle, the experimental accuracy 
depends on signal to noise and calibration precision~\cite{SM}.

% 加上一段讨论相关材料体系，moire, organics, 等等

\noindent\emph{Discussion.}---We have established a quantitative optical framework to extract the spinon Chern number and reconstruct the fractionalized quantum geometry within the topological Mott regime for a CSL. Virtual chargon polarization couples an optical probe to neutral spinons through the emergent gauge field. The resulting tensor Ioffe-Larkin response supports two complementary
measurements. A self-calibrated low-frequency conductivity ratio extracts
the spinon Chern number, while inverse-response Kramers--Kronig protocal
reconstructs the fractionalized quantum geometry. 
While our numerical calculations utilize the triangular-lattice Hofstadter-Hubbard (TLHH) model, which directly simulates the physics of moir\'e heterostructures under a strong orbital magnetic flux, the core physical insights and the tensor Ioffe-Larkin inversion protocol developed here 
hold broader relevance for a wide class of triangular-lattice Mott insulators
and other topological Mott insulators.

Potential applications include surfaces of the cluster magnet
Nb$_3$Br$_8$~\cite{Yao2026SurfaceCSL},
$1\text{T-TaS}_2$~\cite{Ruan2021,Manas-Valero2021,Law_Lee_2017},
and triangular-lattice organic spin liquids such as
$\kappa\text{-(BEDT-TTF)}_2\text{Cu}_2(\text{CN})_3$
~\cite{Shimizu_PRL_2003}, which have been proposed to host CSLs
or related fractionalized phases.
For a CSL realized without external magnetic flux $\bar{\bm{A}}=0$, however, both the spinons and chargons simultaneously experience a nonzero emergent gauge field. Specifically, the chargons couple to $\bar{\bm{A}}-\bar{\bm{a}}=-\bar{\bm{a}}\neq0$, which is exactly opposite to the spinon gauge field, which generates a subgap virtual Hall response, $\Pi_b^H\neq0$.
The present extraction protocols must therefore be extended
to account for the chargon Hall kernel, whose contribution
would require independent constraints from microscopic
calculations or additional magneto-optical measurements.

Our established framework also extends beyond two-dimensional CSLs to three-dimensional topological Mott insulators~\cite{PesinBalents2010} and spinon Weyl semimetals~\cite{FernandezLopez2024} where the nontrivial internal magnetoelectric response may be revealed by physical measurements~\cite{TangChen2026}. Additionally, recent studies of geometric effects in anyon binding and interaction-driven phases of fractionalized
quasiparticles~\cite{Li2026AnyonBinding,Pichler2026AnyonSC} 
motivate a further question, how does spinon quantum geometry
influence instabilities~\cite{BalentsLucile2026} driven by residual interactions? 
In particular, electronic anomalous Hall
crystals~\cite{Dong2024AHC,TanDevakul2024} motivate exploring
whether analogous topological crystalline phases can emerge
from interacting spinons such that conventional spin orders can coexist with fractionalization~\cite{PhysRevB.60.1654,PhysRevResearch.6.033031,PhysRevResearch.5.L032027,PhysRevB.63.134521}. 
These extensions and open questions provide interesting directions for future work.

\noindent\textit{Acknowledgments.---}We thank Jiahao Yang for fruitful discussions. 
This work was supported by the Quantum Science and Technology--National Science and Technology Major Project (Grant No. 2025ZD0300500), the NSFC (Grant Nos. 92565110 and 12574061), 
and the BJNSF (Grant No. F261004).   
 
\bibliography{references} 

\clearpage
\onecolumngrid
\setcounter{page}{1}
\renewcommand{\thepage}{S\arabic{page}}
\setcounter{section}{0}
\renewcommand{\thesection}{S\arabic{section}}
\renewcommand{\theHsection}{S\arabic{section}}
\setcounter{equation}{0}
\renewcommand{\theequation}{S\arabic{equation}}
\renewcommand{\theHequation}{S\arabic{equation}}
\setcounter{figure}{0}
\renewcommand{\thefigure}{S\arabic{figure}}
\renewcommand{\theHfigure}{S\arabic{figure}}
\setcounter{table}{0}
\renewcommand{\thetable}{S\arabic{table}}
\renewcommand{\theHtable}{S\arabic{table}}
\begin{center}
{\large\bf Supplementary Material for}\\[0.5em]
{\large\bf  Optical Probes of Spinon Quantum Geometry in Mott-Insulating Chiral Spin Liquids}\\[1.0em]

Junyu Tang$^{1}$, Hongquan Lv$^{1}$, Gang v. Chen$^{1,2,3}$\\[0.5em]

{\small
$^{1}$\textit{International Center for Quantum Materials, School of Physics,
Peking University, Beijing 100871, China}\\
$^{2}$\textit{Beijing Key Laboratory of Quantum Devices, Peking University, Beijing 100871, China}\\
$^{3}$\textit{Collaborative Innovation Center of Quantum Matter, 100871, Beijing, China}
}
\end{center}
\supplementarytableofcontents 
\writesupplementarycontents

\vspace{2cm}
\twocolumngrid 

\section{Slave-rotor structure of the Hofstadter--Hubbard model}
\label{app:hofstadter_hubbard_parton}

We give a microscopic realization of the parton structure used in the main
text by starting from the spin-$1/2$ Hofstadter--Hubbard model on the
triangular lattice,
\begin{equation}
    H_{\rm HH}
    =-t\sum_{\langle ij\rangle,\sigma}
    \left(e^{i\bar A_{ij}}c^\dagger_{i\sigma}c_{j\sigma}
    +\mathrm{h.c.}\right)
    +\frac{U}{2}\sum_i(n_i-1)^2 .
    \label{eq:HH_microscopic}
\end{equation}
Here, the site number
$n_i=\sum_\sigma c^\dagger_{i\sigma}c_{i\sigma}$ accounts for the two (pseudo)spin
flavors $\sigma=\{\uparrow,\downarrow\}$, and we work at half filling,
$\langle n_i\rangle=1$, i.e., one electron per site.  We specialize to the
commensurate Hofstadter flux
\begin{equation*}
    \sum_{\partial\triangle}\bar A_{ij}
    \equiv\Phi_\triangle=\frac{\pi}{2}
    \quad (\mathrm{mod}\ 2\pi)
\end{equation*}
through each oriented elementary triangle.  This is an externally chosen
parameter of the microscopic model, not a quantity determined by the parton
saddle.  More generally, the flux-matched $SU(N)$ construction takes
$\Phi_\triangle=\pi/N$.  Because a primitive triangular-lattice unit cell
contains two elementary triangles, its flux is then $2\pi/N$, so that the
magnetic unit cell contains $N$ sites. At one particle per site, a magnetic unit cell containing \(N\) sites accommodates \(N\) particles. In an \(SU(N)\)-symmetric state, this corresponds to one particle per flavor per magnetic unit cell. Since each magnetic subband contains one state per flavor per magnetic unit cell, the lowest \(C=1\) Hofstadter band of each flavor is completely filled. Thus, the total Chern number is $C=N$ for the filled lowest Hofstadter band of the $SU(N)$ model. The present $SU(2)$ problem is the $N=2$ member of
this construction (for \(N=2\), one particle per site corresponds to half filling). The bar on $\bar A_{ij}$ distinguishes this fixed
background from the weak electromagnetic field used below as an optical
probe.  Up to an additive constant, the interaction in
Eq.~\eqref{eq:HH_microscopic} is the usual on-site Hubbard repulsion at half
filling.

We then apply the standard slave-rotor decomposition~\cite{FlorensGeorges2004}
\begin{equation}
    c_{i\sigma}=b_i f_{i\sigma},
    \label{eq:slave_rotor_decomposition}
\end{equation}
where \(b_i=e^{i\theta_i}\) is the charged rotor, referred to as the chargon,
and \(f_{i\sigma}\) is a neutral spinon carrying spin $\sigma$.  With
$L_i=-i\partial_{\theta_i}$, the physical Hilbert space at half filling is
selected by
\begin{equation}
    L_i+n_i^f=1,
    \qquad
    n_i^f=\sum_\sigma f^\dagger_{i\sigma}f_{i\sigma}.
    \label{eq:slave_rotor_constraint}
\end{equation}
Thus a singly occupied site has $L_i=0$, whereas an empty (doubly occupied)
site has $L_i=+1$ ($-1$), and the Hubbard term becomes $U L_i^2/2$ in the
enlarged Hilbert space.  Natural units $e=\hbar=k_B=1$ are used throughout.

Substitution of Eq.~\eqref{eq:slave_rotor_decomposition} into the kinetic
energy gives the quartic parton hopping
\begin{equation}
    H_t=-t\sum_{\langle ij\rangle,\sigma}
    \left(e^{i\bar A_{ij}}b_i^\dagger b_j
    f_{i\sigma}^\dagger f_{j\sigma}+\mathrm{h.c.}\right).
    \label{eq:quartic_parton_hopping}
\end{equation}
Introducing the bond fields
\begin{subequations}\label{eq:bond_mean_fields}
    \begin{align}
    \chi^b_{ij}&\equiv\langle b_i^\dagger b_j\rangle=|\chi^b_{ij}|e^{i\phi^b_{ij}}
    \label{eq:chargon_bond_mean_field}\\
    \chi^f_{ij}&\equiv\sum_\sigma
    \langle f_{i\sigma}^\dagger f_{j\sigma}\rangle=|\chi^f_{ij}|e^{i\phi^f_{ij}}.
    \label{eq:spinon_bond_mean_field}
    \end{align}
\end{subequations}
After mean-field decoupling, the kinetic Hamiltonian~\eqref{eq:quartic_parton_hopping} becomes
\begin{subequations}\label{eq:explicit_mf_decoupling}
\begin{align}
H_f^{(0)}={}&-t\sum_{ij,\sigma}
e^{i\bar A_{ij}}\chi^b_{ij}
f^\dagger_{i\sigma}f_{j\sigma}
+\mathrm{h.c.}+h\sum_i n_i^f\\
H_b^{(0)}={}&-t\sum_{ij}
e^{i\bar A_{ij}}\chi^f_{ij}b_i^\dagger b_j
+\mathrm{h.c.}+\frac{U}{2}\sum_iL_i^2
+h\sum_i(L_i-1)
\end{align}
\end{subequations}
where the constant term is ignored. For the translation-invariant saddle, the Lagrange multiplier can be relaxed to
a uniform number $h$. The slave-rotor decomposition introduces an internal \(U(1)\) gauge redundancy
$f_{i\sigma}\to e^{i\alpha_i}f_{i\sigma}$ and
$b_i\to e^{-i\alpha_i}b_i$. Note that under the internal gauge transformation, the bond phases transform as
$\phi^b_{ij}\to\phi^b_{ij}+\alpha_i-\alpha_j$ and
$\phi^f_{ij}\to\phi^f_{ij}-\alpha_i+\alpha_j$.  The individual bond phases
are therefore gauge dependent, whereas their sum and the corresponding loop
fluxes are gauge invariant.  Before imposing a particular saddle, from Eq.~\eqref{eq:explicit_mf_decoupling} we define
the two static phases that actually enter the parton hopping Hamiltonians,
\begin{align}
    \vartheta^f_{ij}
    \equiv \bar A_{ij}+\phi^b_{ij},\quad
    \vartheta^b_{ij}
    \equiv \bar A_{ij}+\phi^f_{ij}.
    \label{eq:parton_effective_static_phases}
\end{align}
Thus, $\vartheta^f_{ij}$ is the effective hopping phase seen by the spinon,
whereas $\vartheta^b_{ij}$ is the effective hopping phase seen by the
chargon.  Equivalently,
\begin{align}
    t^f_{ij}e^{i\vartheta^f_{ij}}
    \equiv t e^{i\bar A_{ij}}\chi^b_{ij},\quad
    t^b_{ij}e^{i\vartheta^b_{ij}}
    \equiv t e^{i\bar A_{ij}}\chi^f_{ij},
    \label{eq:mf_hopping_definitions}
\end{align}
with $t^f_{ij}=t|\chi^b_{ij}|$ and
$t^b_{ij}=t|\chi^f_{ij}|$.  In particular,
$\bar A_{ij}+\phi^b_{ij}+\phi^f_{ij}$ controls the factorized physical bond
current, and a generic finite-field saddle may support equilibrium loop
currents.

The mean-field equations determine the bond fields $\chi_{ij}^{b/f}$ with fixed $\bar A_{ij}$;
they do not determine the external flux $\bar A_{ij}$.  Within this fixed
commensurate model, we adopt the translation-invariant flux-partition ansatz used for the
$SU(2)$ CSL in Refs.~\cite{ZhangLiuSong2026,Divic2025},
\begin{equation}
    \sum_{\partial\triangle}\vartheta^f_{ij}=\frac{\pi}{2},
    \qquad
    \sum_{\partial\triangle}\vartheta^b_{ij}=0
    \quad (\mathrm{mod}\ 2\pi).
    \label{eq:flux_partition_saddle}
\end{equation}
This ansatz is a self-consistent saddle branch rather than a consequence of
gauge redundancy or a unique solution for arbitrary external flux.  A
zero-flux rotor bond correlator leaves the spinon hopping with the microscopic
Hofstadter phase; conversely, the filled spinon Hofstadter band produces a
bond field whose phase cancels $\bar A_{ij}$ in the rotor hopping.  The
mean-field equations then determine the corresponding hopping magnitudes and
the remaining saddle parameters.

Because the chargon link configuration has zero flux, it is pure gauge in
the bulk.  Any pattern of $\pi$ link signs compatible with zero flux on
every contractible loop is likewise pure gauge and can be removed.  By
contrast, assigning a phase $\pi$ to every bond of the nonbipartite
triangular lattice produces a $\pi$ flux through each triangle and belongs
to a different flux sector.  We choose the convenient bond gauge
\begin{equation}
    \vartheta^f_{ij}=\bar A_{ij},
    \qquad
    \vartheta^b_{ij}=0,
    \label{eq:Hofstadter_saddle_effective_phases}
\end{equation}
which, by Eq.~\eqref{eq:parton_effective_static_phases}, is equivalent to
\begin{equation}
    \phi^b_{ij}=0,
    \qquad
    \phi^f_{ij}=-\bar A_{ij}.
    \label{eq:Hofstadter_saddle_bond_phases}
\end{equation}
The factor $e^{i\bar A_{ij}}$ in the chargon hopping is therefore canceled by
the spinon bond field $\chi^f_{ij}$, while the real rotor bond field
$\chi^b_{ij}$ leaves the spinon hopping with the full microscopic Peierls
phase.  The relation
$\bar A_{ij}+\phi^b_{ij}+\phi^f_{ij}=0$ and the vanishing factorized bond
current are properties of this adopted flux-partition branch.

We next identify the static internal link field and note that the static
relative field for chargon vanishes,
\begin{equation}
    \bar a_{ij}\equiv\vartheta^f_{ij}=\bar A_{ij},
    \qquad
    \vartheta^b_{ij}=\bar A_{ij}-\bar a_{ij}=0.
    \label{eq:static_internal_relative_fields}
\end{equation}
A fixed mean-field ansatz freezes the phases of the gauge-charged bond fields and therefore amounts to a choice of gauge. To retain the low-energy gauge fluctuations associated with the local parton redundancy, we allow these bond phases to fluctuate about the saddle and parameterize them by a dynamical internal link field \(a_{ij}\). The total internal field transforms as
\[
\bar a_{ij}+a_{ij}\rightarrow
\bar a_{ij}+a_{ij}+\alpha_i-\alpha_j,
\]
which makes the separate spinon and chargon Hamiltonians gauge covariant. We next restore the dynamical internal fluctuation
$a_{ij}=\int_i^j\bm a\cdot d\bm r$ and the optical probe
$A_{ij}=\int_i^j\bm A\cdot d\bm r$, and define
$\mathcal A_{ij}\equiv A_{ij}-a_{ij}$.  Since
$\bar a_{ij}=\bar A_{ij}$ at the adopted saddle, the full phase seen by the
chargon reduces to
\begin{equation*}
    (\bar A_{ij}+A_{ij})-(\bar a_{ij}+a_{ij})
    =A_{ij}-a_{ij}=\mathcal A_{ij}.
\end{equation*}
The two parton Hamiltonians expanded around this saddle are therefore
\begin{align}
H_f
=
&-\sum_{ij,\sigma}
t^f_{ij}e^{i(\bar A_{ij}+a_{ij})}
f^\dagger_{i\sigma}f_{j\sigma}
+\mathrm{h.c.}+h\sum_i n_i^f,
\label{eq:Hf_mf_revised}
\\
H_b
=
&-\sum_{ij}
t^b_{ij}e^{i\mathcal A_{ij}}
b_i^\dagger b_j
+\mathrm{h.c.}+
\frac{U}{2}\sum_i L_i^2+h\sum_i(L_i-1).
\label{eq:Hb_mf_revised}
\end{align}
where we have utilized condition~\eqref{eq:static_internal_relative_fields}. Here, $h$ enforces Eq.~\eqref{eq:slave_rotor_constraint} on average; its
half-filled saddle-point value vanishes in the particle--hole-symmetric
rotor problem ($h=0$).  From the above two Hamiltonian we see that the spinon inherits the full microscopic Hofstadter flux from the static external magnetic field. Additionally, the unbar fields $a_{ij}$ and $A_{ij}$ can now be understood as the fluctuating fields with respect to the static background in a unified manner ($A_{ij}$ corresponds to the probe field, which is also a kind of perturbation to the vacuum). The spinon therefore couples to $\bar a+a = \bar A + a$, whereas the chargon couples only to the fluctuating relative
field $\mathcal A=A-a$ (its static relative background vanishes). The product $c=bf$ is invariant under the internal gauge transformations and has the
physical electromagnetic charge.  In the microscopic saddle above, the two
parton sectors have different topology because they experience different
static fluxes, while their hopping magnitudes $t^f$ and $t^b$ are fixed by
different bond expectation values.

The difference in static backgrounds is precisely what allows the two
parton sectors to have different topology.  The spinons retain
$\sum_{\partial\triangle}\bar a_{ij}=\pi/2$ and form a spin-singlet Chern
insulator with $C_s=1$ per spin flavor, hence total spinon Chern number
$\mathcal C_f=2$.  The chargon hopping, by contrast, has a real zero-flux
background and is topologically trivial at the mean-field level.

% \begin{table}[t]
%     \centering
%     \begin{tabular}{c|c c}
%          & $U_{\rm EM}(1)\qquad$ & $U_{\rm int}(1)$ \\
%         \hline
%         \(c=bf\) & $+1$ & $0$ \\
%         \(b\) & $+1$ & $-1$ \\
%         \(f\) & $0$ & $+1$
%     \end{tabular}
%     \caption{Gauge charges in the slave-rotor representation.  The chargon
%     carries $q_A=+1$ and $q_a=-1$ while the spinon carries only $q_a=+1$. Note that the
%     physical hole $c=bf$ (electron annihilation) carries a physical charge $q_A=+1$.}
%     \label{tab:gauge_couplings}
% \end{table}

We finally clarify the role of the Zeeman coupling.  Equation~\eqref{eq:HH_microscopic}
defines the spin-independent, or orbital-only, Hofstadter--Hubbard model: the
static magnetic background enters through its Peierls phase, while the two
$SU(2)$ flavors remain exactly degenerate.  This is the appropriate effective
description when the flavors are Zeeman-insensitive pseudospins (such as
layer or valley labels) or when the orbital flux is generated synthetically.
If $\sigma=\uparrow,\downarrow$ instead denotes the physical electron spin in
a real magnetic field, the symmetry-allowed microscopic Hamiltonian also
contains
\begin{equation}
    H_Z=-\frac{\Delta_Z}{2}\sum_i
    \left(n_{i\uparrow}-n_{i\downarrow}\right),
    \qquad
    \Delta_Z=g_{\rm eff}\mu_B|\bar{\bm B}| .
    \label{eq:Zeeman_coupling}
\end{equation}
Here the orbital flux is controlled by the perpendicular component of
$\bar{\bm B}$, whereas $\Delta_Z$ is controlled by the total field and the
effective $g$ factor.  These are consequently distinct parameters of the
low-energy lattice model.

The Zeeman term does not couple directly to the rotor as in the on-site terms, chargon phase cancels each other and the Zeeman term  becomes
\begin{equation}
    H_Z=-\frac{\Delta_Z}{2}\sum_i
    \left(f^\dagger_{i\uparrow}f_{i\uparrow}
    -f^\dagger_{i\downarrow}f_{i\downarrow}\right),
    \label{eq:Zeeman_spinon}
\end{equation}
In the magnetic-sublattice basis, the Hamiltonian of each spin flavor is
therefore
\begin{equation*}
    h_{f,\sigma}(\bm k)
    =\bm d(\bm k)\cdot\bm\tau
    -\frac{\sigma\Delta_Z}{2}\tau_0.
\end{equation*}
The Zeeman term changes only the scalar \(\tau_0\) component: it does not
change the vector \(\bm d(\bm k)\), its magnitude \(d(\bm k)\), or the
current vertex \(\partial_{k_i}h_{f,\sigma}\).  It merely shifts the two
spinon spectra as
$E_{\lambda\sigma}(\bm k)=E_\lambda(\bm k)-\sigma\Delta_Z/2$.
The Zeeman field consequently reduces both the mean-field single-spinon gap
to the chemical potential and the energy of a spin-flip excitation.  These
gaps should not, however, be identified with the interband threshold entering
the present optical kernel.  The orbital current vertex is spin conserving,
\(J_i=\partial_{k_i}h_f\otimes\mathbf 1_\sigma\), and therefore connects only
\((-,\sigma)\) to \((+,\sigma)\).  For every such transition,
\begin{align*}
    &E_{+\sigma}(\bm k)-E_{-\sigma}(\bm k)\\
    &=\left[E_+(\bm k)-\frac{\sigma\Delta_Z}{2}\right]
     -\left[E_-(\bm k)-\frac{\sigma\Delta_Z}{2}\right]=2d(\bm k).
\end{align*}
Thus the Zeeman shift cancels between the initial and final states.  As long
as $\Delta_Z$ is smaller than the zero-field spin-flip (or indirect spinon)
gap, one lower
$C_s=1$ band remains completely occupied for each spin flavor.  There is then
no transfer of occupation between the two flavors, the spin-summed bond field
$\chi^f_{ij}$ is unchanged, and the chargon Hamiltonian remains the real,
zero-flux Hamiltonian in Eq.~\eqref{eq:Hb_mf_revised}.  At zero temperature,
the Fermi occupation factors, current matrix elements, and spin-conserving
transition energies entering $\Pi_f^{ij}(\omega)$ are consequently identical
to their $\Delta_Z=0$ values within this subcritical regime.  In the
interacting language the corresponding condition is
$\Delta_Z<\Delta_{\rm spin}$, with $\Delta_{\rm spin}$ the zero-field gap to
the lowest spinful excitation.  The state then
remains adiabatically connected to the Kalmeyer--Laughlin CSL, although a
nonzero $\Delta_Z$ explicitly reduces spin $SU(2)$ to $U(1)$.

Once the Zeeman splitting closes the spin gap, spinon Fermi pockets,
magnetization, and a reconstruction of the internal-flux saddle can occur;
such field-induced phases have been analyzed in
Ref.~\cite{Yang_PRL_2026}.  They lie outside the gapped, spin-degenerate
$\mathcal C_f=2$ regime considered here.  Throughout this work we therefore
set $\Delta_Z=0$, while the response theory also applies without modification
for $\Delta_Z<\Delta_{\rm spin}$ under the conditions just stated.  In
particular, the Zeeman term neither acts on the chargon directly nor changes
the chargon antiunitary constraint discussed later.

\section{Tensor Ioffe--Larkin rule}
\label{app:tensor_IL}

We now derive the physical electromagnetic response around the saddle
constructed in the preceding section.  The static backgrounds $\bar A$ and
$\bar a=\bar A$ are already included in the mean-field band
Hamiltonians.  Throughout this section, the unbarred fields $A_\mu$ and
$a_\mu$ denote fluctuations about those backgrounds, and 
\begin{equation}
    \mathcal A_\mu\equiv A_\mu-a_\mu
    \label{eq:relative_fluctuating_field}
\end{equation}
is the fluctuating relative field minimally coupled to the chargon.

In the Mott phase the rotor is uncondensed,
$\langle b_i\rangle=\langle e^{i\theta_i}\rangle=0$, and the chargon sector
is gapped.  Its response to $\mathcal A_\mu$ is defined by
\begin{equation}
    e^{-S_b^{\rm eff}[\mathcal A]}
    =
    \int \mathcal D b\,\mathcal D b^\ast\,
    e^{-S_b[b;\mathcal A]}.
\end{equation}
Expanding about the stationary zero-probe saddle, the first nontrivial term
is quadratic,
\begin{equation}
    S_b^{\rm eff}[\mathcal A]
    =
    \frac{1}{2}\sum_q
    \mathcal A_\mu(-q)\Pi_b^{\mu\nu}(q)\mathcal A_\nu(q)
    +\cdots,
    \label{eq:rotor_effective_action}
\end{equation}
where $q=(i\Omega_n,\bm q)$ and
\begin{equation}
    \Pi_b^{\mu\nu}(q)
    =
    \left.
    \frac{\delta^2 S_b^{\rm eff}[\mathcal A]}
    {\delta \mathcal A_\mu(-q)\delta \mathcal A_\nu(q)}
    \right|_{\mathcal A=0}
\end{equation}
is the irreducible chargon polarization kernel.  Because the chargon is
gapped, this kernel is analytic at low frequency and long wavelength and
describes virtual charge polarization, including virtual doublon--holon
fluctuations, of the Mott insulator.

The gapped spinons form a Chern insulator and couple minimally to $a_\mu$.
Integrating them out gives
\begin{equation}
    S_f^{\rm eff}[a]
    =
    \frac{1}{2}\sum_q
    a_\mu(-q)\Pi_f^{\mu\nu}(q)a_\nu(q)
    +\cdots.
    \label{eq:spinon_effective_action}
\end{equation}
The kernel $\Pi_f^{\mu\nu}$ contains both the paramagnetic current--current
correlator and the diamagnetic contribution.  In the CSL it also contains
the Chern--Simons response and the finite-frequency interband response of the
spinon Chern bands.

For completeness, the temporal component of the internal gauge field is the
fluctuation of the Lagrange multiplier that enforces
Eq.~\eqref{eq:slave_rotor_constraint}.  Writing
\begin{equation}
    \lambda_i(\tau)=h+a_{0,i}(\tau)
\end{equation}
completes the dynamical field $a_\mu=(a_0,\bm a)$: the spatial components
are the bond-phase fluctuations discussed in the preceding section, while
$a_0$ enforces the local constraint.  The half-filled
particle--hole-symmetric rotor saddle has
$h=0$.  In the uniform finite-frequency optical limit, we choose temporal
gauge,
\begin{equation}
    A_0=a_0=0,
\end{equation}
so that the density sector decouples and only $i,j\in\{x,y\}$ remain. The total effective action up to quadratic order then reads
\begin{align}
    &S^{\rm eff}[A,a]= S^{\rm eff}_b + S^{\rm eff}_f =
    \frac{1}{2}
    \sum_{\Omega_n}
    \mathcal{A}_i
    \Pi_b^{ij}
    \mathcal{A}_j
    +
    \frac{1}{2}\sum_{\Omega_n}
    a_{i}
    \Pi_f^{ij}
    a_{j},
    \label{eq:spatial_quadratic_action}
\end{align}
where the frequency argument \(\Omega_n\) is suppressed for notational simplicity. 
Since \(a_i\) is an internal gauge fluctuation, it should also be integrated out
from the physical response.  At the Gaussian level this is equivalent to solving
its saddle-point equation $\delta S^{\rm eff}/\delta a_i=0$, which gives
\begin{equation}
    \bm{a}
    =
    \left[\Pi_b+\Pi_f\right]^{-1}
    \Pi_b \bm{A}
    \label{eq:a_A_relation}
\end{equation}
This matrix equation gives the internal gauge field induced by the physical
vector potential; spatial indices and frequency arguments are suppressed.
Substituting the saddle-point solution back into
Eq.~\eqref{eq:spatial_quadratic_action} gives the effective action for the
physical electromagnetic field, which we denote as \(S_{\rm phys}[A]\):
\begin{equation}
    S_{\rm phys}[A]
    =
    \frac{1}{2}
    \sum_{\Omega_n}
    A_i(-\Omega_n)
    \Pi_{\rm phys}^{ij}(i\Omega_n)
    A_j(\Omega_n).
    \label{eq:physical_effective_action}
\end{equation}
Here the physical electromagnetic response kernel is
\begin{equation}
    \Pi_{\rm phys}
    =
    \Pi_b
    -
    \Pi_b
    \left[
        \Pi_b+\Pi_f
    \right]^{-1}
    \Pi_b.
    \label{eq:sm_Pi_phys}
\end{equation}
All three kernels are matrices in spatial indices.  To make contact with the
standard tensor Ioffe--Larkin rule, we first rewrite Eq.~\eqref{eq:sm_Pi_phys} as
\begin{align}
    \Pi_{\rm phys}
    &=\Pi_b\left[I-(\Pi_b+\Pi_f)^{-1}\Pi_b\right]\notag\\
    &=\Pi_b\left[(\Pi_b+\Pi_f)^{-1}(\Pi_b+\Pi_f)
    -(\Pi_b+\Pi_f)^{-1}\Pi_b\right]\notag\\
    &=\Pi_b(\Pi_b+\Pi_f)^{-1}\Pi_f.
    \label{eq:Pi_phys_factorized}
\end{align}
When the relevant kernels are invertible, taking the matrix inverse gives
\begin{align}
    \Pi_{\rm phys}^{-1}
    &=\Pi_f^{-1}(\Pi_b+\Pi_f)\Pi_b^{-1}\notag\\
    &=\Pi_f^{-1}+\Pi_b^{-1}.
\end{align}
Thus, the physical response obeys the well-known tensor Ioffe--Larkin
composition rule
\begin{equation}
    \Pi_{\rm phys}^{-1}
    =\Pi_b^{-1}+\Pi_f^{-1}
    \label{eq:Pi_phys_inverse_IL}
\end{equation}
without requiring \(\Pi_b\) and \(\Pi_f\) to commute.  At frequencies where an individual kernel is singular, Eq.~\eqref{eq:sm_Pi_phys} remains the more direct form, while Eq.~\eqref{eq:Pi_phys_inverse_IL} is understood by a regulated limit. The measured response is therefore neither the bare spinon kernel nor the bare chargon kernel, but their gauge-constrained composition.

We now show why the chargon kernel has no antisymmetric, or Hall,
component at the adopted saddle.  Because the static relative flux vanishes,
the hopping amplitudes in $H_b$ are real.  At half filling $h=0$,
and complex conjugation in the rotor basis defines an antiunitary operation
$\Theta_b$ satisfying
\begin{align}
    \Theta_b b_i\Theta_b^{-1}&=b_i^\dagger,\notag\\
    \Theta_b L_i\Theta_b^{-1}&=-L_i,\notag\\
    \Theta_b H_b[\mathcal A]\Theta_b^{-1}&=H_b[-\mathcal A].
    \label{eq:chargon_antiunitary_transformation}
\end{align}
Thus $H_b[0]$ and its equilibrium density matrix are invariant under
$\Theta_b$.  The uniform chargon current
$j_b^i=-\partial H_b/\partial\mathcal A_i|_{\mathcal A=0}$ is odd,
\begin{equation}
    \Theta_b j_b^i\Theta_b^{-1}=-j_b^i,
\end{equation}
whereas the diamagnetic contact tensor, obtained from the second derivative
of $H_b$ with respect to $\mathcal A$, is even and symmetric in its spatial
indices.

To see the resulting reciprocity explicitly, consider the spectral
representation of the retarded paramagnetic kernel,
\begin{equation}
    \Pi_{b,{\rm para}}^{R,ij}(\omega)
    =\sum_{m,n}
    \frac{p_m-p_n}
    {\omega+E_m-E_n+i0^+}
    \langle m|j_b^i|n\rangle
    \langle n|j_b^j|m\rangle,
    \label{eq:chargon_kernel_spectral_representation}
\end{equation}
where $p_m=e^{-\beta E_m}/Z$.  Antiunitarity and the odd transformation of
the current imply
\begin{equation}
    \langle \Theta_b m|j_b^i|\Theta_b n\rangle
    =-\langle m|j_b^i|n\rangle^\ast.
\end{equation}
The states $|m\rangle$ and $\Theta_b|m\rangle$ have the same energy and
Boltzmann weight.  Relabeling every state in
Eq.~\eqref{eq:chargon_kernel_spectral_representation} by its antiunitary
partner therefore exchanges the two current matrix elements.  Since both
currents are odd, their two minus signs cancel, giving the Onsager relation
\begin{equation}
    \Pi_b^{R,ij}(\omega)=\Pi_b^{R,ji}(\omega).
    \label{eq:chargon_onsager_relation}
\end{equation}
The symmetric diamagnetic contact term obeys the same relation, so it applies to the full kernel and it holds for the Matsubara
kernel before analytic continuation.  Hence, in the uniform optical limit,
\begin{equation}
    \Pi_b^H(\omega)
    \equiv
    \frac{\Pi_b^{R,xy}(\omega)-\Pi_b^{R,yx}(\omega)}{2}=0.
    \label{eq:chargon_Hall_forbidden}
\end{equation}
This antiunitary symmetry belongs to the real, zero-relative-flux
chargon Hamiltonian.  It is not a time-reversal symmetry of the microscopic
Hofstadter--Hubbard model, whose external flux
$\Phi_\triangle=\pi/2$ explicitly breaks time reversal.  Importantly, the
argument forbids the entire finite-frequency chargon Hall kernel, not only
its dc chargon Hall kernel.

\section{Gaussian chargon response from the slave-rotor action}
\label{app:gaussian_chargon_kernel}

We now derive the chargon kernel directly from the Gaussian slave-rotor
theory with half filling,
where the uniform saddle has \(h=0\) and temporal gauge
\(\mathcal A_0=0\) is used.  Starting from chargon Hamiltonian [Eq.~\eqref{eq:Hb_mf_revised}], the canonical
Euclidean rotor action is
\begin{align}
    S_b[L,\theta;\mathcal A]
    =\int_0^\beta d\tau\bigg[
    &\sum_i\left(
        iL_i\partial_\tau\theta_i+\frac{U}{2}L_i^2
    \right)\notag\\
    &-\sum_{\langle ij\rangle}t^b_{ij}
    \left(
        e^{i\mathcal A_{ij}}e^{-i\theta_i}e^{i\theta_j}
        +\mathrm{c.c.}
    \right)\bigg].
    \label{eq:canonical_rotor_action}
\end{align}
The first term is the canonical, or Berry-phase-like term of the rotor. Since \(\theta_i\) and \(L_i\) are conjugate variables, \([\theta_i,L_j]=i\delta_{ij}\), so the phase-space path
integral for \(Z=\mathrm{Tr}\,e^{-\beta H_b}\) has the Euclidean form
\(S_E=\int d\tau\,[iL_i\partial_\tau\theta_i+H_b]\). Within the Gaussian rotor approximation the integer-valued \(L_i\) is treated as a continuous field.  Completing the square
\begin{equation*}
    iL_i\partial_\tau\theta_i+\frac{U}{2}L_i^2
    =
    \frac{U}{2}
    \left(L_i+\frac{i}{U}\partial_\tau\theta_i\right)^2
    +\frac{1}{2U}(\partial_\tau\theta_i)^2.
\end{equation*}
and integrating it out gives
the action~\cite{FlorensGeorges2004}
\begin{align}
    S_b[\theta;\mathcal A]
    =\int_0^\beta d\tau\bigg[
    &\frac{1}{2U}\sum_i(\partial_\tau\theta_i)^2\notag\\
    &-\sum_{\langle ij\rangle}t^b_{ij}
    \left(
        e^{i\mathcal A_{ij}}e^{-i\theta_i}e^{i\theta_j}
        +\mathrm{c.c.}
    \right)\bigg].
    \label{eq:phase_only_rotor_action}
\end{align}
Because \(\theta_i\) is an angular variable, \(\theta_i\) and
\(\theta_i+2\pi\) describe the same rotor.  An exact finite-temperature path
integral must therefore also sum over integer winding sectors satisfying
\(\theta_i(\beta)=\theta_i(0)+2\pi m_i\).  Here we retain the smooth Gaussian
fluctuations and ignore the nonperturbative winding events. To obtain a quadratic theory, define \(X_i=e^{i\theta_i}\).  For the exact
unit-modulus field we have
\begin{equation}
    |X_i|^2=1,
    \qquad
    |\partial_\tau X_i|^2=(\partial_\tau\theta_i)^2,
    \label{eq:unit_rotor_identity}
\end{equation}
The soft-rotor approximation then relaxes
the pointwise condition \(|X_i|^2=1\) to an average condition and imposes it
with a uniform multiplier \(\lambda\).  The resulting quadratic action is
\begin{align}
    S_b^{G}[X;\mathcal A]
    =\int_0^\beta d\tau\bigg[
    &\frac{1}{2U}\sum_i|\partial_\tau X_i|^2
    +\lambda\sum_i(|X_i|^2-1)\notag\\
    &-\sum_{\langle ij\rangle}t^b_{ij}
    \left(e^{i\mathcal A_{ij}}X_i^*X_j+\mathrm{c.c.}\right)
    \bigg].
    \label{eq:gaussian_rotor_action}
\end{align}
% Although \(\lambda\) may be called as bosonic chemical
% potential, here it is more precisely the saddle-point mass that enforces the
% soft-rotor normalization.  It is distinct from the physical electron chemical
% potential and from \(h\), which couples to the rotor charge \(L_i\) and
% enforces the parton constraint.  We therefore retain the symbol \(\lambda\)
% rather than \(\mu\).

For a translation-invariant, real zero-flux hopping pattern, define
\begin{equation}
    \varepsilon_b(\bm k)
    =2\sum_{\bm\delta}t^b_{\bm\delta}
    \cos(\bm k\cdot\bm\delta),
    \qquad
    \xi_b(\bm k)=\lambda-\varepsilon_b(\bm k)>0,
    \label{eq:gaussian_rotor_dispersion}
\end{equation}
where one vector from each undirected bond pair is included in the
\(\bm\delta\) sum. The $\xi_b(\bm k)$ term is positive definite, and the
rotor is uncondensed, when
\(\lambda>\max_{\bm k}\varepsilon_b(\bm k)\) or equivalently \(\xi_b(\bm k)>0\). At equality the lowest rotor mode becomes gapless and can condense.  Thus the Mott saddle has the
single-chargon gap \(\Delta_b=\min_{\bm k}E_b(\bm k)>0\). At zero probe field, use the bosonic Fourier transform
\begin{equation}
    X_i(\tau)
    =
    \frac{1}{\sqrt{\beta N}}
    \sum_{\bm k,\nu_n}
    e^{i\bm k\cdot\bm r_i-i\nu_n\tau}
    X(\bm k,i\nu_n),
    \quad
    \nu_n=2\pi nT.
    \label{eq:gaussian_rotor_fourier_transform}
\end{equation}
Substitution into Eq.~\eqref{eq:gaussian_rotor_action} diagonalizes the
quadratic action:
\begin{equation}
    S_b^G[\mathcal A=0]
    =
    \sum_{\bm k,\nu_n}
    X^*(\bm k,i\nu_n)
    \left[
        \frac{\nu_n^2}{2U}+\xi_b(\bm k)
    \right]
    X(\bm k,i\nu_n)
    -\beta N\lambda.
    \label{eq:gaussian_rotor_momentum_action}
\end{equation}
For a Gaussian complex field, its two-point Green's function is the inverse
of the quadratic coefficient.  Before evaluating it, we implement the
atomic-limit calibration of the soft-rotor approximation.  For the exact
compact rotor, \(H_{\rm at}=UL^2/2\), the \(L=0\to L=\pm1\) excitation costs
\(U/2\).  By contrast, relaxing \(|X_i|=1\) to the average constraint in
Eq.~\eqref{eq:gaussian_rotor_constraint} would place the atomic pole at \(U\)
if the same \(U\) were used directly in the Gaussian propagator.  We therefore
replace the inertial parameter of the soft theory by
\(U_s\equiv U/2\), while \(U\) continues to denote the physical Hubbard
interaction in the exact rotor Hamiltonian and phase-only action above.
Equivalently, the \(U\) in the soft kinetic term of
Eq.~\eqref{eq:gaussian_rotor_momentum_action} is understood as \(U_s\) when
constructing the Gaussian propagator and all response functions
below.  Thus,
\begin{align}
    G_b(\bm k,i\nu_n)
    &\equiv
    \langle X(\bm k,i\nu_n)X^*(\bm k,i\nu_n)\rangle\notag\\
    &={}
    \frac{1}{\nu_n^2/(2U_s)+\xi_b(\bm k)}
    =
    \frac{2U_s}{\nu_n^2+E_b^2(\bm k)},\\
    E_b(\bm k)
    &={}
    \sqrt{2U_s\xi_b(\bm k)}.
    \label{eq:gaussian_rotor_propagator}
\end{align}
Varying the Gaussian free energy with respect to \(\lambda\) restores the
relaxed unit-modulus condition on average:
\begin{equation}
    1=\langle |X_i|^2\rangle
    =
    \frac{T}{N}\sum_{\bm k,\nu_n}G_b(\bm k,i\nu_n).
    \label{eq:gaussian_rotor_constraint}
\end{equation}
This constraint also makes the atomic-limit calibration explicit.  When
\(t_b=0\) ($\epsilon_b=0$), the rotor mode is momentum independent,
\(E_b(\bm k)=E_0=\sqrt{2U_s\lambda}\).  At zero temperature,
the elementary bosonic frequency sum gives
\begin{equation}
\begin{aligned}
    1&=T\sum_{\nu_n}G_b(i\nu_n)=\frac{U_s}{E_0}\rightarrow E_0=U_s,\ \lambda=\frac{U_s}{2}
\end{aligned}
    \label{eq:gaussian_rotor_atomic_gap} 
\end{equation}
Hence the atomic single-chargon gap is
\(\Delta_b=E_0=U_s=U/2\), not \(U_s/2\).  The latter is the atomic value
of the soft constraint multiplier \(\lambda\), rather than an excitation
energy.  At large but finite \(U/t_b\), the constraint has no correction
linear in \(t_b\) to \(\lambda\), because the Brillouin-zone average of
\(\varepsilon_b(\bm k)\) vanishes.  Using
\(\Delta_b=\sqrt{2U_s(\lambda-6t_b)}\) therefore gives
\begin{equation}
    \Delta_b
    =U_s-6t_b+O(t_b^2/U_s)
    =\frac{U}{2}-6t_b+O(t_b^2/U).
    \label{eq:gaussian_rotor_large_U_gap}
\end{equation}
Thus \(U/2\) is the leading atomic-limit value, while the experimentally
calibrated gap should include the finite-bandwidth renormalization.

In a fully self-consistent treatment,
Eq.~\eqref{eq:gaussian_rotor_constraint}, together with the bond saddle
equations, determines \(\lambda\) and \(t^b_{ij}\).  In the present work, we instead take an uncondensed CSL saddle as input and focus on probing the spinon QGT in the main text.

We next calculate the uniform optical response.  A spatially uniform,
time-dependent relative vector potential enters by the Peierls substitution
\(\xi_b(\bm k)\rightarrow\xi_b(\bm k-\bm{\mathcal A})\).  Introduce the
vertices
\begin{equation}
    v_i(\bm k)=\partial_{k_i}\xi_b(\bm k),
    \qquad
    m_{ij}(\bm k)=\partial_{k_i}\partial_{k_j}\xi_b(\bm k).
    \label{eq:gaussian_rotor_vertices}
\end{equation}
For completeness, the expansion leading to the response kernel is
\begin{equation}
    \xi_b(\bm k-\bm{\mathcal A})
    =
    \xi_b(\bm k)-v_i(\bm k)\mathcal A_i
    +\frac{1}{2}m_{ij}(\bm k)\mathcal A_i\mathcal A_j
    +O(\mathcal A^3).
    \label{eq:gaussian_rotor_peierls_expansion}
\end{equation}
Writing
\(G_b^{-1}[\mathcal A]=G_b^{-1}+V_1+V_2+\cdots\), where \(V_1\) and \(V_2\)
are respectively the terms linear and quadratic in \(\mathcal A\), integrating
out the complex field gives
\begin{align}
    S_{\mathrm{eff}}[\mathcal A]
    ={}&
    \mathrm{Tr}\ln G_b^{-1}
    +\mathrm{Tr}(G_bV_1)
    +\mathrm{Tr}(G_bV_2)\notag\\
    &-\frac{1}{2}\mathrm{Tr}(G_bV_1G_bV_1)
    +O(\mathcal A^3).
    \label{eq:gaussian_rotor_trace_log_expansion}
\end{align}
The first-order term vanishes after the Brillouin-zone sum.  The
\(\mathrm{Tr}(G_bV_2)\) term produces the diamagnetic contact vertex
\(m_{ij}\), whereas the term with two \(V_1\)'s produces the paramagnetic
bubble with vertices \(v_i v_j\).  Defining
\(S_{\mathrm{eff}}^{(2)}
=\frac{1}{2}\sum_{\Omega_n}\mathcal A_i(-i\Omega_n)
\Pi_{b,E}^{ij}(i\Omega_n)\mathcal A_j(i\Omega_n)\) then gives
\begin{align}
    \Pi_{b,E}^{ij}(i\Omega_n)
    =&
    \int_{\bm k}T\sum_{\nu_n}
    \big[
    m_{ij}(\bm k)G_b(\bm k,i\nu_n)\notag\\
    &-v_i(\bm k)v_j(\bm k)
    G_b(\bm k,i\nu_n)
    G_b(\bm k,i\nu_n+i\Omega_n)
    \big],
    \label{eq:gaussian_chargon_matsubara_kernel}
\end{align}
where \(\int_{\bm k}\equiv N^{-1}\sum_{\bm k}\).  This expression is already
the complete Gaussian chargon kernel; omitting the first term would violate
gauge invariance.  Indeed, periodicity of the Brillouin zone gives
\begin{equation}
    \Pi_{b,E}^{ij}(0)
    =
    \int_{\bm k}T\sum_{\nu_n}
    \partial_{k_i}
    \left[v_j(\bm k)G_b(\bm k,i\nu_n)\right]
    =0.
    \label{eq:gaussian_chargon_ward_identity}
\end{equation}
Thus a static uniform vector potential produces no current in the Mott
insulator.

At zero temperature the Matsubara frequency sums are
\begin{align}
    T\sum_{\nu_n}G_b
    &={}
    \frac{U_s}{E_b},\notag\\
    T\sum_{\nu_n}G_b(i\nu_n)G_b(i\nu_n+i\Omega_n)
    &={}
    \frac{4U_s^2}{E_b(\Omega_n^2+4E_b^2)}.
    \label{eq:gaussian_rotor_frequency_sums}
\end{align}
An integration by parts converts the contact term according to
\begin{equation}
    \int_{\bm k}\frac{U_s}{E_b}m_{ij}
    =
    \int_{\bm k}\frac{U_s^2}{E_b^3}v_i v_j.
    \label{eq:gaussian_rotor_contact_identity}
\end{equation}
Consequently, the Euclidean kernel takes the manifestly gauge-invariant form
\begin{equation}
    \Pi_{b,E}^{ij}(i\Omega_n)
    =
    \int_{\bm k}
    \frac{U_s^2v_i(\bm k)v_j(\bm k)}{E_b^3(\bm k)}
    \frac{\Omega_n^2}{\Omega_n^2+4E_b^2(\bm k)}.
    \label{eq:gaussian_chargon_euclidean_closed}
\end{equation}
Writing \(z=\omega+i0^+\), the same analytic continuation used for the
spinon kernel gives
\begin{equation}
    \Pi_b^{R,ij}(\omega)
    =
    \int_{\bm k}
    \frac{U_s^2v_i(\bm k)v_j(\bm k)}{E_b^3(\bm k)}
    \frac{z^2}
    {z^2-4E_b^2(\bm k)}.
    \label{eq:gaussian_chargon_retarded_kernel}
\end{equation}
The optical threshold is therefore \(2\Delta_b\): below it the clean
Gaussian kernel is real and describes only virtual chargon-antichargon
polarization, whereas above it the imaginary part resolves the continuum of
two-chargon excitations.

% For \(|\omega|\ll2\Delta_b\), Eq.~\eqref{eq:gaussian_chargon_retarded_kernel}
% reduces to
% \begin{align}
%     \Pi_b^{R,ij}(\omega)
%     &={}
%     -\chi_b^{ij}z^2+O(z^4),\notag\\
%     \chi_b^{ij}
%     &={}
%     \int_{\bm k}
%     \frac{U_s^2v_i(\bm k)v_j(\bm k)}
%     {4E_b^5(\bm k)}.
%     \label{eq:gaussian_chargon_polarizability}
% \end{align}
% The polarizability is thus determined by the Gaussian saddle rather than
% introduced as an independent phenomenological constant.
We now apply the result to the triangular lattice with three
undirected nearest-neighbor vectors $\bm\delta_1=a(1,0)$, $\bm\delta_2=a(1/2,\sqrt{3}/2)$, $\bm\delta_3=a(-1/2,\sqrt{3}/2)$. Although the physical system is placed in an external orbital field, the
Hofstadter--Hubbard saddle chosen above obeys
\(\bar{\mathcal A}_{ij}=\bar A_{ij}-\bar a_{ij}=0\).  Its hopping is therefore
real and translation invariant in the original one-site unit cell.  The
doubled magnetic unit cell associated with the spinon Hofstadter problem does
not enter the chargon dispersion.  For a uniform nearest-neighbor amplitude
\(t_b>0\), Eqs.~\eqref{eq:gaussian_rotor_dispersion} and
\eqref{eq:gaussian_rotor_propagator} become
\begin{align}
    \gamma_\triangle(\bm k)
    &=\cos(ak_x)
    +2\cos\left(\frac{ak_x}{2}\right)
      \cos\left(\frac{\sqrt{3}ak_y}{2}\right)
    ,\notag\\
    \varepsilon_b(\bm k)
    &=2t_b\gamma_\triangle(\bm k),\notag\\
    \xi_b(\bm k)
    &=r+2t_b[3-\gamma_\triangle(\bm k)],\notag\\
    E_b(\bm k)
    &=\sqrt{2U_s\xi_b(\bm k)},\notag
\end{align}
where $r\equiv\lambda-6t_b>0$ and $\Delta_b=\sqrt{2U_sr}$. Note that \(\varepsilon_b\) is maximal at \(\Gamma\), while
\(\varepsilon_{b,\min}=-3t_b\) at the \(K\) points.  Thus
\(E_{b,\max}=\sqrt{\Delta_b^2+18U_st_b}\).

Equation~\eqref{eq:gaussian_chargon_retarded_kernel} generally contains a
continuum of pair poles.  If the chargon band is narrow compared with its
gap, \(E_b(\bm k)\simeq\Delta_b\) over the optically weighted momenta, it
reduces to an effective single pole.  To see this directly, first rewrite
each term in Eq.~\eqref{eq:gaussian_chargon_retarded_kernel} exactly as
\begin{align}
    \Pi_b^{R,ij}(z)
    &=-z^2\int_{\bm k}w_{ij}(\bm k)
    \frac{4E_b^2(\bm k)}
    {4E_b^2(\bm k)-z^2},\notag\\
    w_{ij}(\bm k)
    &\equiv
    \frac{U_s^2v_i(\bm k)v_j(\bm k)}
    {4E_b^5(\bm k)},
    \qquad
    \chi_b^{ij}=\int_{\bm k}w_{ij}(\bm k).
    \label{eq:gaussian_chargon_weighted_pole}
\end{align}
Writing \(E_b(\bm k)=\Delta_b+\delta E_b(\bm k)\) and denoting the chargon bandwidth by \(W_b\), the narrow-band condition is \(W_b/\Delta_b\ll1\) or \(\delta E_b(\bm k)/\Delta_b\ll 1\).  Thus, we have the approximation
\begin{equation}
    \frac{4E_b^2(\bm k)}
    {4E_b^2(\bm k)-z^2}
    \approx
    \frac{4\Delta_b^2}{4\Delta_b^2-z^2}.
    \label{eq:gaussian_chargon_narrow_band_expansion}
\end{equation}
Note that the chargon band gap is larger than spinon band gap and when on-shell resonance of spinon kernel occurs, the spinon frequency is below the chargon threshold, away from the chargon pole. Therefore, under the narrow-band condition,we have 
\begin{equation}
    \Pi_b^{R,ij}(\omega)
    \simeq
    -\chi_b^{ij}z^2
    \frac{4\Delta_b^2}
    {4\Delta_b^2-z^2}.
    \label{eq:gaussian_chargon_effective_pole}
\end{equation}
The approximation is controlled throughout a subgap spinon spectroscopy
window provided both \(W_b/\Delta_b\ll1\) and
\(|4\Delta_b^2-z^2|\) remains large compared with the bandwidth correction.
It necessarily fails parametrically close to the chargon-pair threshold,
where the individual continuum poles can no longer be replaced by one pole.
Eq.~\eqref{eq:gaussian_chargon_effective_pole} has the same form of the response kerneal of a simple anisotropically polarizable medium after identifying its optical pole \(\Omega_b=2\Delta_b\). The
minus sign in Eq.~\eqref{eq:gaussian_chargon_effective_pole} is the retarded vector-potential-kernel convention obtained by analytic continuation of the Euclidean action; it must be used consistently when comparing with the phenomenological expression.

Importantly, the leading low-frequency response does not require a narrow
chargon band.  Expanding the full Gaussian continuum in
Eq.~\eqref{eq:gaussian_chargon_retarded_kernel} gives
\begin{align}
    \Pi_b^{R,ij}(z)
    &=-\chi_b^{ij}z^2-\gamma_b^{ij}z^4+O(z^6)\\
    \gamma_b^{ij}
    &=\int_{\bm k}
    \frac{U_s^2v_i(\bm k)v_j(\bm k)}
    {16E_b^7(\bm k)}.
    \label{eq:gaussian_chargon_low_frequency_expansion}
\end{align}
By comparison, Eq.~\eqref{eq:gaussian_chargon_effective_pole} yields
\(\Pi_b^{R,ij}=-\chi_b^{ij}z^2-
\chi_b^{ij}z^4/(4\Delta_b^2)+O(z^6)\).  The narrow-band condition
\(E_b(\bm k)\simeq\Delta_b\) makes
\(\gamma_b^{ij}\simeq\chi_b^{ij}/(4\Delta_b^2)\) and hence controls the
single-pole form beyond leading order.  Without that condition, the
continuum still has the same universal
\(-\chi_b^{ij}z^2\) scaling at low frequency. Thus, the low-frequency polarizability $\chi_b^{ij}$ is a robust property and can still be extracted from the low-frequency longitudinal response.

We finally separate the Hall question from spatial isotropy.  Apparently, each term in
Eqs.~\eqref{eq:gaussian_chargon_matsubara_kernel} and
\eqref{eq:gaussian_chargon_retarded_kernel} is symmetric under
\(i\leftrightarrow j\), so
\begin{equation}
    \Pi_b^{R,ij}(\omega)=\Pi_b^{R,ji}(\omega),
    \qquad
    \Pi_b^{H}(\omega)
    \equiv\frac{\Pi_b^{R,xy}-\Pi_b^{R,yx}}{2}=0.
    \label{eq:gaussian_chargon_no_hall}
\end{equation}
This explicitly reproduces the antiunitary-symmetry result of the preceding
section.  On a completely general anisotropic lattice, however, a symmetric
off-diagonal component
\(\Pi_b^{R,xy}=\Pi_b^{R,yx}\) can be present in a nonprincipal coordinate
basis (but it is not a Hall response).  The tensor Ioffe--Larkin rule itself
requires neither diagonality nor isotropy.  The scalar reduction used in the
main text follows from the unbroken triangular-lattice rotation symmetry.  In
the uniform optical limit, let
\begin{equation*}
    R_3=
    \begin{pmatrix}
        -1/2&-\sqrt{3}/2\\
        \sqrt{3}/2&-1/2
    \end{pmatrix}
\end{equation*}
represent a rotation by \(2\pi/3\).  Invariance under \(C_3\) imposes the
tensor constraint
\begin{equation*}
    \bm\Pi_b^R(z)=R_3\bm\Pi_b^R(z)R_3^T.
\end{equation*}
Equation~\eqref{eq:gaussian_chargon_no_hall} makes this tensor symmetric, so
we may write it as
\(\bm\Pi_b^R=\left(\begin{smallmatrix}a&c\\c&b\end{smallmatrix}\right)\).
Substitution into the constraint above gives
\(a=b\) and \(c=0\).  Thus \(C_3\) makes the two longitudinal responses equal,
while the absence of a Hall component removes the only other rotationally
invariant tensor structure.  We therefore define the scalar chargon channel
and its nonnegative optical weight by
\begin{subequations}\label{eq:Pi_b_def_appendix}
\begin{align}
    \Pi_b^{R,ij}(z)&=\Pi_b(z)\delta^{ij},\quad \chi_b^{ij}=\chi_b\delta^{ij},\\
    \Pi_b(z) &\equiv\frac{1}{2}\delta_{ij}\Pi_b^{R,ij}(z)=-z^2\int_{\bm k}w(\bm k)
    \frac{4E_b^2(\bm k)}{4E_b^2(\bm k)-z^2},\\
    w(\bm k)&\equiv\frac{1}{2}\delta^{ij}w_{ij}(\bm k)=\frac{U_s^2[v_x^2(\bm k)+v_y^2(\bm k)]}{8E_b^5(\bm k)}\geq0,\\
    \chi_b&\equiv\int_{\bm k}w(\bm k).
\end{align}
\end{subequations}

For any scalar function \(O(\bm k)\), the corresponding normalized optical
average is
\begin{equation}
    \langle O\rangle_w
    \equiv
    \frac{\int_{\bm k}w(\bm k)O(\bm k)}
    {\int_{\bm k}w(\bm k)}
    =\frac{1}{\chi_b}\int_{\bm k}w(\bm k)O(\bm k).
    \label{eq:sm_normalized_optical_average}
\end{equation}
\(C_3\) requires isotropy only after the Brillouin-zone integral. Specifically, although \(w_{ij}(\bm k)\neq w_{ij}(\bm k)\), after integration over the Brillouin zone, we still have \(\chi_b^{xx}=\chi_b^{yy}\) and \(\chi_b^{xy}=0\).

\section{Spinon Hamiltonian of the Hofstadter–Hubbard model}
In this section we derive the spinon Hamiltonian used in the kernel calculation from the
triangular-lattice Hofstadter saddle of Eq.~\eqref{eq:Hf_mf_revised}. 
\begin{figure}
    \centering
    \includegraphics[width=0.7\linewidth]{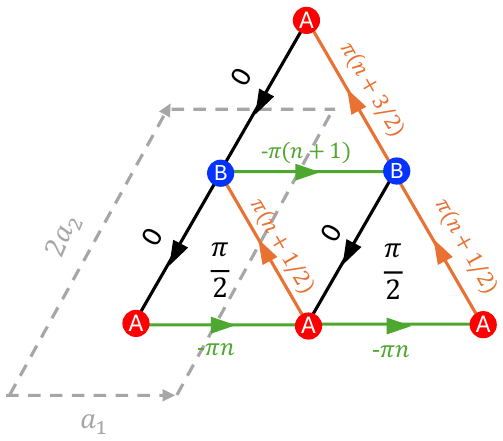}
    \caption{The magnetic gauge for the \(\pi/2\)-flux state on the triangular lattice. The grey dashed lines indicate the magnetic unit cell. Hopping along different bonds with different colors acquires different phases as indicated by the number near the bonds. The magnetic flux threading each trangular plaquette gives a phase \(\pi/2\) when traversing the plaquette in a counterclockwise direction.}
\end{figure}
We work in the uniform optical limit, set the
temporal gauge fluctuation to zero, and take a uniform nearest-neighbor
spinon hopping \(t_f\).  Write the lattice sites as
\(\bm r_{m,n}=m\bm a_1+n\bm a_2\), with
\begin{equation}
    \bm a_1=a(1,0),
    \qquad
    \bm a_2=a(1/2,\sqrt{3}/2).
\end{equation}
For the \(\pi/2\)-flux state, a convenient Landau-type magnetic gauge,
gauge-equivalent to other commonly used choices for the triangular-lattice
Hofstadter problem~\cite{ZhangLiuSong2026,Divic2025}, is
\begin{align}
    \bar a_{\bm r,\bm r+\bm a_1}&=-\pi n,\notag\\
    \bar a_{\bm r,\bm r+\bm a_2}&=0,\notag\\
    \bar a_{\bm r,\bm r+\bm a_2-\bm a_1}
    &=+\pi\left(n+\frac{1}{2}\right)
    \quad (\mathrm{mod}\ 2\pi).
    \label{eq:spinon_pi_over_two_gauge}
\end{align}
This choice depends only on the row coordinate \(n\), which makes its
two-row magnetic periodicity explicit.  It is useful first to verify the
flux.  Orienting the boundary of each
triangle counterclockwise, the upper triangle
\(\bm r\to\bm r+\bm a_1\to\bm r+\bm a_2\to\bm r\) gives
\begin{align}
    \Phi_{\triangle_\uparrow}^{\rm ccw}
    &=\bar a_{\bm r,\bm r+\bm a_1}
      +\bar a_{\bm r+\bm a_1,\bm r+\bm a_2}
      +\bar a_{\bm r+\bm a_2,\bm r}\notag\\
    &=-\pi n+\pi\left(n+\frac12\right)-0
      =+\frac{\pi}{2},
    \label{eq:upper_triangle_flux_check}
\end{align}
where the second edge is along \(\bm a_2-\bm a_1\).  For the lower triangle
\(\bm r\to\bm r+\bm a_2\to
\bm r+\bm a_2-\bm a_1\to\bm r\), one similarly obtains
\begin{align}
    \Phi_{\triangle_\downarrow}^{\rm ccw}
    &=0+\pi(n+1)-\pi\left(n+\frac12\right)
      =+\frac{\pi}{2}.
    \label{eq:lower_triangle_flux_check}
\end{align}
Thus every elementary triangle carries the same uniform flux, with the
counterclockwise convention giving \(+\pi/2\), consistently with
\(\partial\triangle\) above.  Complex conjugating all bond amplitudes
reverses the physical chirality, leaves the dispersion unchanged, and
reverses the band Chern number.  The gauge in
Eq.~\eqref{eq:spinon_pi_over_two_gauge} is periodic under \(n\to n+2\), so
the magnetic primitive vectors may be chosen as
\(\bm a_1^m=\bm a_1\) and \(\bm a_2^m=2\bm a_2\).  The magnetic unit cell
therefore contains two sites, denoted \(A\) (\(n\) even) and \(B\) (\(n\)
odd), in agreement with the general \(N=2\) Hofstadter counting discussed
above.

To see explicitly how the Bloch matrix follows from this gauge, retain one
orientation of the three nearest-neighbor bonds,
\(\bm\delta_1=\bm a_1\), \(\bm\delta_2=\bm a_2\), and
\(\bm\delta_3=\bm a_2-\bm a_1\), and write
\begin{equation}
    H_f^{(0)}
    =-t_f\sum_{\bm r,\sigma}\sum_{\nu=1}^{3}
    \left(
    e^{i\bar a_{\bm r,\bm r+\bm\delta_\nu}}
    f_{\bm r\sigma}^\dagger
    f_{\bm r+\bm\delta_\nu,\sigma}
    +\mathrm{h.c.}\right).
    \label{eq:spinon_oriented_bond_hamiltonian}
\end{equation}
Let
\(\bm R=m\bm a_1+2\ell\bm a_2\) label a magnetic unit cell, with
\(\bm s_A=0\) and \(\bm s_B=\bm a_2\), and use the full-position Fourier
convention
\begin{equation}
    f_{\bm R+\bm s_\alpha,\sigma}
    =\frac{1}{\sqrt{N_m}}
    \sum_{\bm k\in\mathrm{MBZ}}
    e^{i\bm k\cdot(\bm R+\bm s_\alpha)}
    f_{\bm k\alpha\sigma}.
    \label{eq:spinon_magnetic_fourier_transform}
\end{equation}
Introduce \(k_1=\bm k\cdot\bm a_1\),
\(k_2=\bm k\cdot\bm a_2\), and the magnetic-sublattice spinor
\(\Psi_{\bm k\sigma}=(f_{\bm kA\sigma},f_{\bm kB\sigma})^{T}\).
For brevity, define \(\gamma_1=\cos k_1\),
\(\gamma_2=\cos k_2\), and
\(\gamma_3=\cos(k_2-k_1)\).
The \(\bm a_1\) bonds have hopping phases \(+1\) on the \(A\) rows and
\(-1\) on the \(B\) rows; the \(\bm a_2\) bonds have phase \(+1\);
and the \(\bm a_2-\bm a_1\) bonds have phases \(+i\) and \(-i\) when
they start, respectively, on \(A\) and \(B\).  Adding each bond to its
Hermitian conjugate therefore gives
\begin{align}
    h_f^{(\bm a_1)}(\bm k)&=-2t_f\gamma_1\tau_z,\\
    h_f^{(\bm a_2)}(\bm k)&=-2t_f\gamma_2\tau_x,\\
    h_f^{(\bm a_2-\bm a_1)}(\bm k)&=+2t_f\gamma_3\tau_y.
    \label{eq:spinon_three_bond_contributions}
\end{align}
At the half-filled saddle \(h=0\), their sum is
\begin{align}
    h_f(\bm k)
    &=-2t_f
    \begin{pmatrix}
    \gamma_1&\gamma_2+i\gamma_3\\
    \gamma_2-i\gamma_3&-\gamma_1
    \end{pmatrix}=\bm d(\bm k)\cdot\bm\tau\\
    \bm d(\bm k)
    &=-2t_f
    \begin{pmatrix}
    \gamma_2, &
    -\gamma_3, &
    \gamma_1
    \end{pmatrix}^T.
    \label{eq:sm_spinon_two_band_hamiltonian}
\end{align}
Here the Pauli matrices \(\bm\tau\) act on the two magnetic sublattices,
not on the physical \(SU(2)\) spin.  The two magnetic bands are
\begin{equation}
    \varepsilon_\pm(\bm k)
    =\pm d(\bm k)
    =\pm2t_f
    \sqrt{
    \cos^2k_1+\cos^2k_2+\cos^2(k_2-k_1)}.
    \label{eq:triangular_hofstadter_spinon_dispersion}
\end{equation}
The matrix form of \(h_f(\bm k)\) is gauge dependent, whereas this
dispersion is not.  Indeed, a lattice gauge transformation
\begin{equation}
    f_{\bm r\sigma}\to e^{i\chi_{\bm r}}f_{\bm r\sigma},
    \qquad
    \bar a_{\bm r\bm r'}
    \to\bar a_{\bm r\bm r'}+\chi_{\bm r}-\chi_{\bm r'}
    \label{eq:spinon_lattice_gauge_change}
\end{equation}
changes the magnetic-sublattice representation of the Bloch Hamiltonian by
a unitary basis transformation, possibly together with a relabeling or
folding of magnetic momentum.  It therefore cannot change the eigenvalues
or the gap.  The Chern number and the response kernels are likewise gauge
invariant when the states and current vertices are transformed consistently,
although the displayed matrix entries and the components of
\(\bm d(\bm k)\) generally change.
The square root ranges from \(\sqrt{3}/2\) to \(\sqrt{3}\), so the direct
spinon band gap is
\(\Delta_s=\min_{\bm k}[\varepsilon_+(\bm k)-\varepsilon_-(\bm k)]
=2\sqrt{3}t_f\).  At one spinon per site there are two spinons per magnetic
unit cell.  Consequently, for each of the two degenerate \(SU(2)\) spin
flavors the lower magnetic band is completely filled and the upper band is
empty.  The occupied band has \(\lvert C_s\rvert=1\) per spin flavor; we
choose the orientation convention \(C_s=+1\), giving total
\(\mathcal C_f=N_\sigma C_s=2\).

Thus this Hofstadter spinon problem is still a two-band
\(\bm d\cdot\bm\tau\) model for each spin flavor, and the general response
derivation below applies without change.  Note that all momentum integrals run over the
magnetic Brillouin zone (denoted simply by BZ below).  Spatial derivatives
\(\partial_{k_i}\) continue to refer to the Cartesian components
\(i=x,y\).  In this section \(g_{ij}\) and \(\Omega_{ij}\) denote the quantum
metric and Berry curvature of the occupied lower Hofstadter band; we suppress
the band label ``\(-\)'' for brevity.  Since \(h_f\) is independent of the
physical spin label, tracing over the \(N_\sigma=2\) degenerate copies
produces the overall factor \(N_\sigma\) used below.

\section{Spinon response kernel from the imaginary-time path integral}
\label{app:spinon_kernel_path_integral}

\subsection{Quadratic action for the internal gauge field}

The Peierls coupling of the spinons to a spatially uniform internal gauge
field is implemented by
\begin{equation}
    h_f(\bm k)\longrightarrow h_f(\bm k+\bm a).
\end{equation}
Expanding to second order gives
\begin{equation}
    h_f(\bm k+\bm a)
    =h_f(\bm k)+a_i \mathcal{J}_i(\bm k)
    +\frac{1}{2}a_i a_j\tau_{ij}(\bm k)+O(a^3),
    \label{eq:sm_h_expansion}
\end{equation}
where
\begin{equation}
    \mathcal{J}_i(\bm k)=\partial_{k_i}h_f(\bm k),
    \qquad
    \tau_{ij}(\bm k)=\partial_{k_i}\partial_{k_j}h_f(\bm k).
    \label{eq:sm_current_vertices}
\end{equation}
Here \(\mathcal{J}_i\) is the paramagnetic current vertex and \(\tau_{ij}\) is the
diamagnetic vertex, or contact vertex.

The imaginary-time spinon action is
\begin{equation}
    S_f[\bar f,f,a]
    =
    \int_0^\beta d\tau
    \sum_{\bm k,\sigma}
    \bar f_{\bm k\sigma}
    \left[\partial_\tau+h_f(\bm k+\bm a)\right]
    f_{\bm k\sigma}.
    \label{eq:sm_spinon_euclidean_action}
\end{equation}
We Fourier transform the fermion and gauge fields, with $\partial_\tau \to -i\omega_n$, according to
\begin{equation}
    f(\tau)\propto\sum_{i\omega_n}e^{-i\omega_n\tau}f(i\omega_n),
    \qquad
    a_i(\tau)\propto\sum_{i\Omega_m}e^{-i\Omega_m\tau}a_i(i\Omega_m),
    \label{eq:sm_fourier_convention}
\end{equation}
where the fermionic and bosonic Matsubara frequencies are \(\omega_n=(2n+1)\pi T\) and \(\Omega_m=2m\pi T\), respectively. Specifically, the gauge field $a$ carries a bosonic Matsubara frequency \(\Omega_m\), while \(\omega_n\) for the spinon is fermionic. The overall Fourier normalization factors are absorbed into the frequency sums below. Using the expansion in Eq.~\eqref{eq:sm_h_expansion}, in Matsubara space we write the action as
\begin{equation}
    S_f=\bar f\left(G_0^{-1}+V\right)f=\bar f\left(G_0^{-1}+V_1+V_2+\cdots\right)f,
\end{equation}
with the free spinon Green's function and the linear and quadratic vertices
\begin{align}
    &G_0^{-1}(i\omega_n,\bm k)=-i\omega_n+h_f(\bm k),\notag\\
    &V_1=a_i \mathcal{J}_i,
    \qquad
    V_2=\frac{1}{2}a_i a_j\tau_{ij}.
    \label{eq:sm_G0_and_vertices}
\end{align}
In the absence of the gauge perturbation, the free Green's function is diagonal in frequency,
momentum, and physical spin:
\begin{equation}
    [G_0]_{(i\omega_n,\bm k,\sigma),(i\omega_n',\bm k',\sigma')}
    =
    \delta_{\omega_n,\omega_n'}
    \delta_{\bm k,\bm k'}
    \delta_{\sigma,\sigma'}
    G_0(i\omega_n,\bm k),
\end{equation}
where $G_0(i\omega_n,\bm k)$ is now a two-by-two matrix in the orbital/sublattice pseudospin space.

To see how the frequency indices of the linear vertex arise, consider the
term in the imaginary-time action that is first order in the gauge field,
\begin{equation}
    S_{f,1}
    =\int_0^\beta d\tau\sum_{\bm k,\sigma}
    \bar f_{\bm k\sigma}(\tau)
    a_i(\tau)\mathcal{J}_i(\bm k)
    f_{\bm k\sigma}(\tau).
    \label{eq:sm_Sf1_time_domain}
\end{equation}
Substituting the Fourier expansions in
Eq.~\eqref{eq:sm_fourier_convention}, and suppressing the overall Fourier
normalization factors, gives
\begin{align}
    S_{f,1}\propto
    &\sum_{n',n,m}\sum_{\bm k,\sigma}
    \bar f_{\bm k\sigma}(i\omega_{n'})
    a_i(i\Omega_m)\mathcal{J}_i(\bm k)
    f_{\bm k\sigma}(i\omega_n)\notag\\
    &\times\int_0^\beta d\tau\,
    e^{i(\omega_{n'}-\omega_n-\Omega_m)\tau}.
    \label{eq:sm_Sf1_frequency_expansion}
\end{align}
The imaginary-time integral imposes frequency conservation, 
\begin{equation}
    \int_0^\beta d\tau\,
    e^{i(\omega_{n'}-\omega_n-\Omega_m)\tau}
    =\beta\,
    \delta_{\omega_{n'},\,\omega_n+\Omega_m}.
    \label{eq:sm_V1_frequency_delta}
\end{equation}
Thus a gauge-field mode with bosonic Matsubara frequency $i\Omega_m$
connects an incoming spinon state at $i\omega_n$ to an outgoing state at
$i\omega_n+i\Omega_m$.  In the uniform limit, the corresponding nonzero
matrix element of $V_1$ is therefore
\begin{equation}
    [V_1]_{i\omega_n+i\Omega_m,\,i\omega_n}
    =
    a_i(i\Omega_m)\mathcal{J}_i(\bm k),
    \label{eq:V1_matrix_element}
\end{equation}
where the momentum $\bm k$, spin \(\sigma\), and orbital indices \(\alpha\) are implicit (omitted for brevity). Its complete form reads
\begin{align}
    &[V_1]_{(i\omega_n',\bm k,\alpha,\sigma),
    (i\omega_n,\bm k,\beta,\sigma')}
    \notag\\
    &\quad=\delta_{\sigma\sigma'}
    \sum_{i\Omega_m}a_i(i\Omega_m)
    [\mathcal{J}_i(\bm k)]_{\alpha\beta}
    \delta_{\omega_n',\omega_n+\Omega_m},
    \label{eq:sm_V1_matrix_element}
\end{align}
Thus, $V_1$ changes the spinon frequency from $i\omega_n$ to
$i\omega_n+i\Omega_m$.  Similarly, the quadratic vertex $V_2$ carries two bosonic frequencies from the two gauge fields.  Its nonzero matrix element is
\begin{equation}
    [V_2]_{i\omega_n+i\Omega_1+i\Omega_2,\,i\omega_n}
    =
    \frac{1}{2}
    a_i(i\Omega_1)a_j(i\Omega_2)
    \tau_{ij}(\bm k).
    \label{eq:V2_matrix_element}
\end{equation}
Again, the momentum, spin, and orbital indices are omitted for brevity. Its complete form reads
\begin{align}
    &[V_2]_{(i\omega_n',\bm k,\alpha,\sigma),
    (i\omega_n,\bm k,\beta,\sigma')}
    \notag\\
    &\quad=\frac{1}{2}\delta_{\sigma\sigma'}
    \sum_{i\Omega_1,i\Omega_2}
    a_i(i\Omega_1)a_j(i\Omega_2)
    [\tau_{ij}(\bm k)]_{\alpha\beta}
    \delta_{\omega_n',\omega_n+\Omega_1+\Omega_2}.
    \label{eq:sm_V2_matrix_element}
\end{align}

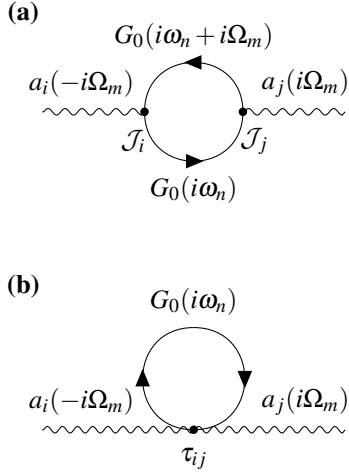
\begin{figure}[t]
    \renewcommand{\theHfigure}{S\arabic{figure}}
    \centering
    \begin{tikzpicture}
            % Paramagnetic bubble
            \coordinate (al) at (-2.0,0);
            \coordinate (v1) at (-0.65,0);
            \coordinate (v2) at (0.65,0);
            \coordinate (ar) at (2.0,0);
            \draw[/tikzfeynman/photon] (al) -- (v1);
            \draw[/tikzfeynman/fermion] (v2)
                .. controls (0.65,0.87) and (-0.65,0.87) .. (v1);
            \draw[/tikzfeynman/fermion] (v1)
                .. controls (-0.65,-0.87) and (0.65,-0.87) .. (v2);
            \draw[/tikzfeynman/photon] (v2) -- (ar);
            \fill (v1) circle (1.6pt);
            \fill (v2) circle (1.6pt);
            \node at (-1.48,0.38) {$a_i(-i\Omega_m)$};
            \node at (1.48,0.38) {$a_j(i\Omega_m)$};
            \node at (-0.8,-0.38) {$\mathcal{J}_i$};
            \node at (0.8,-0.38) {$\mathcal{J}_j$};
            \node at (0,1) {$G_0(i\omega_n+i\Omega_m)$};
            \node at (0,-1.02) {$G_0(i\omega_n)$};
            \node at (-2.25,1.3) {\textbf{(a)}};

            % Diamagnetic contact diagram
            \coordinate (bl) at (-2.0,-4.2);
            \coordinate (vc) at (0,-4.2);
            \coordinate (bu) at (0,-2.85);
            \coordinate (br) at (2.0,-4.2);
            \draw[/tikzfeynman/photon] (bl) -- (vc);
            \draw[/tikzfeynman/fermion] (vc)
                .. controls (-0.90,-4.2) and (-0.90,-2.85) .. (bu);
            \draw[/tikzfeynman/fermion] (bu)
                .. controls (0.90,-2.85) and (0.90,-4.2) .. (vc);
            \draw[/tikzfeynman/photon] (vc) -- (br);
            \fill (vc) circle (1.6pt);
            \node at (-1.48,-3.82) {$a_i(-i\Omega_m)$};
            \node at (1.48,-3.82) {$a_j(i\Omega_m)$};
            \node at (0,-4.58) {$\tau_{ij}$};
            \node at (0,-2.5) {$G_0(i\omega_n)$};
            \node at (-2.25,-2.3) {\textbf{(b)}};
    \end{tikzpicture}
    \caption{Diagrammatic representation of the quadratic spinon response.
    (a) Paramagnetic current--current bubble generated by the two linear
    vertices in $\frac{1}{2}\Tr(G_0V_1G_0V_1)$.  (b) Diamagnetic contact
    contribution generated by the quadratic vertex in $-\Tr(G_0V_2)$.
    Solid directed lines denote spinon propagators, and wavy lines denote
    external legs of the internal gauge field.}
    \label{fig:sm_spinon_response_diagrams}
\end{figure}

The effective action is further defined by the Grassmann functional integral
\begin{equation}
    e^{-S_f^{\rm eff}[a]}
    =\int\mathcal D\bar f\,\mathcal Df\,
    e^{-S_f[\bar f,f,a]}.
    \label{eq:sm_define_spinon_effective_action}
\end{equation}
Using \(\int\mathcal D\bar f\,\mathcal Df\,
e^{-\bar f M f}=\det M\) and \(\ln\det M=\Tr\ln M\), we find
\begin{align}
    S_f^{\rm eff}[a]
    &=-\Tr\ln\left(G_0^{-1}+V_1+V_2+\cdots\right)\notag\\
    &=\text{const.}-\Tr(G_0V_2)
    +\frac{1}{2}\Tr(G_0V_1G_0V_1)+O(a^3).
    \label{eq:sm_trace_log_expansion}
\end{align}
The two quadratic contributions in Eq.~\eqref{eq:sm_trace_log_expansion}
are represented diagrammatically in Fig.~\ref{fig:sm_spinon_response_diagrams}.
Here \(\ \Tr:= \sum_{i\omega_n} \sum_{\bm k}\sum_\sigma \mathrm{tr}_\alpha\) includes all compound indices and frequency-momentum sums,
whereas \(\tr\) below acts only in the two-component orbital space.  The
expansion follows from
\begin{equation}
    -\Tr\ln(1+G_0V)
    =-\Tr(G_0V)+\frac{1}{2}\Tr(G_0VG_0V)+\cdots.
\end{equation}
The term linear in \(a_i\) vanishes because we expand about a saddle point
with no equilibrium transport current. Since $G_0$ is diagonal in fermionic
Matsubara frequency, we have
\begin{align}
    \Tr(G_0V_1G_0V_1)
    &=
    \sum_{\omega_0,\omega_2}
    G_0(i\omega_0)
    [V_1]_{\omega_0,\omega_2} 
    G_0(i\omega_2)
    [V_1]_{\omega_2,\omega_0}.
\end{align}

We define the quadratic kernel by
\begin{equation}
    S_f^{(2)}[a]
    =\frac{1}{2}\sum_{i\Omega_m}
    a_i(-i\Omega_m)\Pi_f^{ij}(i\Omega_m)a_j(i\Omega_m).
    \label{eq:sm_K_definition}
\end{equation}
For the paramagnetic contribution $\Tr(G_0V_1G_0V_1)$, the coefficient of $a_i(-i\Omega_m)a_j(i\Omega_m)$ is obtained by taking $\omega_0=\omega_n$ and $\omega_2=\omega_n+\Omega_m$.  The first vertex then carries frequency $-i\Omega_m$ and contributes $a_i(-i\Omega_m)\mathcal{J}_i$,
while the second vertex carries frequency $+i\Omega_m$ and contributes
$a_j(i\Omega_m)\mathcal{J}_j$. For the diamagnetic term $-\Tr(G_0V_2)$, since $V_2$ already
contains two powers of the gauge field, the trace contains only one spinon
Green's function. Frequency conservation requires the two external
gauge fields to carry opposite frequencies. This is equivalent to the standard Feynman diagram of the paramagnetic bubble and diamagnetic contact terms, as shown in Fig.~\ref{fig:sm_spinon_response_diagrams}.

For the bubble, one chooses an initial fermion frequency \(i\omega_n\).  One
linear vertex carries \(+i\Omega_m\) and takes the fermion to
\(i\omega_n+i\Omega_m\); the second carries \(-i\Omega_m\) and closes the
trace.  Hence the coefficient of
\(a_i(-i\Omega_m)a_j(i\Omega_m)\) is
\begin{align}
    &T\sum_{i\omega_n}\int_{\rm BZ}\frac{d^2k}{(2\pi)^2}
    \tr\bigl[
    G_0(i\omega_n,\bm k)\mathcal{J}_i(\bm k)
    \notag\\[-0.2em]
    &\hspace{8em}\times
    G_0(i\omega_n+i\Omega_m,\bm k)\mathcal{J}_j(\bm k)
    \bigr].
    \label{eq:sm_trace_frequency_flow}
\end{align}
For the diamagnetic contact term, frequency conservation instead imposes
\(i\Omega_1+i\Omega_2=0\) inside the single quadratic vertex.  Comparing
the resulting two terms with Eq.~\eqref{eq:sm_K_definition} gives the
paramagnetic bubble and the frequency-independent contact term as
\begin{align}
    \Pi_{f,{\rm para}}^{ij}(i\Omega_m)
    =&N_\sigma T\sum_{i\omega_n}
    \int_{\rm BZ}\frac{d^2k}{(2\pi)^2}
    \tr\bigl[
    G_0(i\omega_n,\bm k)\mathcal{J}_i(\bm k)\notag\\
    &\hspace{5.7em}\times
    G_0(i\omega_n+i\Omega_m,\bm k)\mathcal{J}_j(\bm k)
    \bigr],
    \label{eq:sm_paramagnetic_bubble}\\
    \Pi_{f,{\rm dia}}^{ij}
    =&-N_\sigma T\sum_{i\omega_n}
    \int_{\rm BZ}\frac{d^2k}{(2\pi)^2}
    \tr\left[G_0(i\omega_n,\bm k)\tau_{ij}(\bm k)\right].
    \label{eq:sm_diamagnetic_term}
\end{align}
The trace in these expressions is over the two-component orbital space.

\subsection{Matsubara sum and interband response}

Using the band decomposition
\begin{equation}
    G_0(i\omega_n,\bm k)
    =\sum_{s=\pm}
    \frac{|u_{s\bm k}\rangle\langle u_{s\bm k}|}
    {-i\omega_n+\varepsilon_s(\bm k)},
    \label{eq:sm_G_band_decomposition}
\end{equation}
the contact term can now be evaluated directly.  Defining
\(\tau_{ij}^{ss}=\langle u_s|\tau_{ij}|u_s\rangle\) and using
\begin{equation}
    T\sum_{i\omega_n}\frac{1}{-i\omega_n+\varepsilon_s}
    =-n_F(\varepsilon_s),
\end{equation}
Eq.~\eqref{eq:sm_diamagnetic_term} becomes
\begin{equation}
    \Pi_{f,{\rm dia}}^{ij}
    =N_\sigma\int_{\rm BZ}\frac{d^2k}{(2\pi)^2}
    \sum_{s=\pm}n_F(\varepsilon_s)\tau_{ij}^{ss}(\bm k).
    \label{eq:sm_diamagnetic_band_form}
\end{equation}
For the paramagnetic bubble we use the standard Matsubara sum
\begin{align}
    &T\sum_{i\omega_n}
    \frac{1}{-i\omega_n+\varepsilon_s}
    \frac{1}{-i(\omega_n+\Omega_m)+\varepsilon_{s'}}\notag\\
    &\hspace{4em}=
    \frac{n_F(\varepsilon_s)-n_F(\varepsilon_{s'})}
    {i\Omega_m+\varepsilon_s-\varepsilon_{s'}},
    \label{eq:sm_matsubara_sum}
\end{align}
we obtain
\begin{align}
    \Pi_{f,{\rm para}}^{ij}(i\Omega_m)
    =N_\sigma\int_{\rm BZ}\frac{d^2k}{(2\pi)^2}
    \sum_{s,s'=\pm}
    \frac{n_F(\varepsilon_s)-n_F(\varepsilon_{s'})}
    {i\Omega_m+\varepsilon_s-\varepsilon_{s'}}
    \mathcal{J}_i^{ss'}\mathcal{J}_j^{s's},
    \label{eq:sm_para_band_sum}
\end{align}
where \(\mathcal{J}_i^{ss'}=\langle u_s|\mathcal{J}_i|u_{s'}\rangle\).  At zero temperature
\(n_F(\varepsilon_-)=1\) and \(n_F(\varepsilon_+)=0\).  The intraband terms
therefore vanish, and the remaining interband bubble is
\begin{align}
    \Pi_{f,{\rm para}}^{ij}(i\Omega_m)
    =&N_\sigma\int_{\rm BZ}\frac{d^2k}{(2\pi)^2}\notag\\
    &\times
    \left[
    \frac{M_{ij}(\bm k)}{i\Omega_m-2d(\bm k)}
    -\frac{M_{ji}(\bm k)}{i\Omega_m+2d(\bm k)}
    \right],
    \label{eq:sm_interband_matsubara_kernel}
\end{align}
with
\begin{equation}
    M_{ij}(\bm k)
    =\langle u_-|\mathcal{J}_i|u_+\rangle
    \langle u_+|\mathcal{J}_j|u_-\rangle.
    \label{eq:sm_Mij_definition}
\end{equation}
The first term in Eq.~\eqref{eq:sm_interband_matsubara_kernel} is the
\(s=-,s'=+\) contribution.  It carries the occupation difference
\(n_F(\varepsilon_-)-n_F(\varepsilon_+)=1\) and the excitation energy
\(\varepsilon_+-\varepsilon_-=2d\).  The second is the
\(s=+,s'=-\) contribution; its occupation difference is \(-1\), which
accounts for the explicit minus sign and gives the negative-frequency
counterpart required by the current commutator.

The relation to the QGT follows directly by differentiating the band
eigenvalue equation.  For two different bands,
\begin{equation}
    \langle u_{s'}|\partial_{k_i}h_f|u_s\rangle
    =(\varepsilon_s-\varepsilon_{s'})
    \langle u_{s'}|\partial_{k_i}u_s\rangle.
    \label{eq:sm_interband_velocity_identity}
\end{equation}
In particular,
\begin{align}
    \langle u_+|\mathcal{J}_i|u_-\rangle
    &=-2d\langle u_+|\partial_{k_i}u_-\rangle,\notag\\
    \langle u_-|\mathcal{J}_i|u_+\rangle
    &=-2d\langle\partial_{k_i}u_-|u_+\rangle,
    \label{eq:sm_velocity_matrix_elements}
\end{align}
where the second line also uses the derivative of
\(\langle u_-|u_+\rangle=0\).  Therefore
\begin{align}
    M_{ij}(\bm k)
    =[2d(\bm k)]^2
    \langle\partial_{k_i}u_-|u_+\rangle
    \langle u_+|\partial_{k_j}u_-\rangle.
    \label{eq:sm_Mij_geometric_matrix_element}
\end{align}
Since in a two-band model
\(1-|u_-\rangle\langle u_-|=|u_+\rangle\langle u_+|\), the last two
matrix elements are precisely the occupied-band QGT\@. Hence
\begin{equation}
    M_{ij}(\bm k)
    =[2d(\bm k)]^2Q_{ij}(\bm k),
    \qquad
    Q_{ij}=g_{ij}-\frac{i}{2}\Omega_{ij}.
    \label{eq:sm_Mij_QGT_relation}
\end{equation}

\subsection{Gauge-invariant kernel and analytic continuation}

We now analytically continue \(i\Omega_m\to z\), where
\(z=\omega+i0^+\).  A finite numerical broadening is included by taking
\(z=\omega+i\eta_s\).  To display the metric and curvature contributions
separately, define
\begin{equation}
    S_{ij}=\frac{M_{ij}+M_{ji}}{2},
    \qquad
    A_{ij}=\frac{M_{ij}-M_{ji}}{2}.
    \label{eq:sm_SA_definition}
\end{equation}
Since \(Q_{ji}=Q_{ij}^*=g_{ij}+i\Omega_{ij}/2\),
Eq.~\eqref{eq:sm_Mij_QGT_relation} gives
\begin{equation}
    S_{ij}=4d^2g_{ij},
    \qquad
    A_{ij}=-2id^2\Omega_{ij}.
    \label{eq:sm_SA_geometry}
\end{equation}
The analytically continued paramagnetic bubble can then be rearranged as
\begin{align}
    \Pi_{f,{\rm para}}^{ij}(z)
    =&N_\sigma\int_{\rm BZ}\frac{d^2k}{(2\pi)^2}
    \Bigg\{
    S_{ij}\left[\frac{1}{z-2d}-\frac{1}{z+2d}\right]
    \notag\\
    &\hspace{5em}+A_{ij}
    \left[\frac{1}{z-2d}+\frac{1}{z+2d}\right]
    \Bigg\}.
    \label{eq:sm_para_SA_form}
\end{align}
Here and in the next few equations the momentum arguments of \(d\),
\(S_{ij}\), and \(A_{ij}\) are implicit.  Using
\begin{align}
    \frac{1}{z-2d}-\frac{1}{z+2d}
    &=\frac{4d}{z^2-(2d)^2},\notag\\
    \frac{1}{z-2d}+\frac{1}{z+2d}
    &=\frac{2z}{z^2-(2d)^2},
    \label{eq:sm_denominator_identities}
\end{align}
we obtain
\begin{align}
    \Pi_{f,{\rm para},S}^{ij}(z)
    =&N_\sigma\int_{\rm BZ}\frac{d^2k}{(2\pi)^2}
    \frac{16d^3(\bm k)}{z^2-[2d(\bm k)]^2}
    g_{ij}(\bm k),
    \label{eq:sm_para_symmetric}\\
    \Pi_{f,{\rm para},A}^{ij}(z)
    =&-iN_\sigma\int_{\rm BZ}\frac{d^2k}{(2\pi)^2}
    \frac{4z d^2(\bm k)}{z^2-[2d(\bm k)]^2}
    \Omega_{ij}(\bm k).
    \label{eq:sm_para_antisymmetric}
\end{align}
The paramagnetic bubble alone is not gauge invariant: its symmetric part is
nonzero for a static, uniform vector potential.  The contact term in
Eq.~\eqref{eq:sm_diamagnetic_term} cancels this static response (f-sum rule).  Equivalently,
the full symmetric kernel can be written as the static subtraction
\begin{equation}
    \Pi_{f,S}^{ij}(z)
    =\Pi_{f,{\rm para},S}^{ij}(z)
    -\Pi_{f,{\rm para},S}^{ij}(0).
    \label{eq:sm_static_subtraction}
\end{equation}
This identity can be verified explicitly from the band expression for the
contact term.  At zero temperature, Eq.~\eqref{eq:sm_diamagnetic_band_form}
contains only \(\tau_{ij}^{--}\).  Differentiating the lower-band energy
twice gives the effective-mass identity
\begin{align}
    \partial_{k_i}\partial_{k_j}\varepsilon_-
    =&\tau_{ij}^{--}
    +\frac{M_{ij}+M_{ji}}{\varepsilon_--\varepsilon_+}\notag\\
    =&\tau_{ij}^{--}-4d\,g_{ij},
    \label{eq:sm_effective_mass_identity}
\end{align}
where Eqs.~\eqref{eq:sm_Mij_QGT_relation} and
\eqref{eq:sm_SA_geometry} were used in the second line.  Therefore
\begin{align}
    \Pi_{f,{\rm dia}}^{ij}
    =&N_\sigma\int_{\rm BZ}\frac{d^2k}{(2\pi)^2}
    \left(\partial_{k_i}\partial_{k_j}\varepsilon_-+4d\,g_{ij}\right)
    \notag\\
    =&4N_\sigma\int_{\rm BZ}\frac{d^2k}{(2\pi)^2}d\,g_{ij}.
    \label{eq:sm_dia_metric_form}
\end{align}
The total derivative integrates to zero over the periodic Brillouin zone.
On the other hand, Eq.~\eqref{eq:sm_para_symmetric} gives
\begin{equation}
    \Pi_{f,{\rm para},S}^{ij}(0)
    =-4N_\sigma\int_{\rm BZ}\frac{d^2k}{(2\pi)^2}d\,g_{ij}.
    \label{eq:sm_para_static_metric_form}
\end{equation}
Equations~\eqref{eq:sm_dia_metric_form} and
\eqref{eq:sm_para_static_metric_form} establish the static subtraction,
or the corresponding \(f\)-sum rule, without leaving the contact term
implicit.
The antisymmetric bubble already vanishes at \(z=0\), and the contact vertex
is symmetric in \(i,j\), so the Hall part is unchanged.

Using
\begin{equation}
    \frac{16d^3}{z^2-(2d)^2}+4d
    =\frac{4dz^2}{z^2-(2d)^2},
\end{equation}
we arrive at the full retarded spinon kernel
\begin{align}
    \Pi_{f,S}^{ij}(z)
    =&N_\sigma\int_{\rm BZ}\frac{d^2k}{(2\pi)^2}
    \frac{4d(\bm k)z^2}{z^2-[2d(\bm k)]^2}
    g_{ij}(\bm k),
    \label{eq:sm_full_spinon_symmetric_kernel}\\
    \Pi_{f,A}^{ij}(z)
    =&-iN_\sigma\int_{\rm BZ}\frac{d^2k}{(2\pi)^2}
    \frac{4z d^2(\bm k)}{z^2-[2d(\bm k)]^2}
    \Omega_{ij}(\bm k).
    \label{eq:sm_full_spinon_antisymmetric_kernel}
\end{align}
For the \(C_3\)-symmetric triangular-lattice Hofstadter model considered here,
\begin{equation}
    \Pi_f^{ij}=\Pi_f^L\delta^{ij}+\Pi_f^H\epsilon^{ij},
    \qquad \epsilon^{xy}=1,
\end{equation}
and \(C_3\) symmetry gives \(\Pi_{f,S}^{xx}=\Pi_{f,S}^{yy}\).  Consequently,
\begin{align}
    \Pi_f^{L}(z)=N_\sigma\int_{\rm BZ}\frac{d^2k}{(2\pi)^2}
    \frac{4d(\bm k)z^2}{z^2-[2d(\bm k)]^2}
    g_{xx}(\bm k),
    \label{eq:sm_final_KfL}\\
    \Pi_f^{H}(z)=-iN_\sigma\int_{\rm BZ}\frac{d^2k}{(2\pi)^2}
    \frac{4z d^2(\bm k)}{z^2-[2d(\bm k)]^2}
    \Omega_{xy}(\bm k).
    \label{eq:sm_final_KfH}
\end{align}
At nonzero Zeeman splitting, the corresponding band-sum derivation replaces
the factor \(N_\sigma\) in these equations by the momentum-dependent
occupation sum
\begin{equation*}
    N_\sigma
    \longrightarrow
    \sum_{\sigma=\uparrow,\downarrow}
    \left[
    n_F(E_{-\sigma}(\bm k))-n_F(E_{+\sigma}(\bm k))
    \right],
\end{equation*}
but leaves every occurrence of \(d(\bm k)\) unchanged.  At zero temperature
and below the band-reoccupation threshold, the lower band of each flavor is
filled and the upper band is empty, so the bracket equals one for each
\(\sigma\) and the sum reduces exactly to \(N_\sigma=2\).  Thus the extra
gap- and frequency-dependent factors displayed above are also unchanged; at
finite temperature, or after Zeeman-induced band reoccupation, the occupation
sum is no longer constant and the kernels do change.
The static subtraction forces \(\Pi_f^L\propto z^2\), while the antisymmetric response is
linear in \(z\) and therefore yields a finite spinon Hall conductivity in
the dc limit.  In particular, below the interband threshold,
\begin{align}
    \Pi_f^L(\omega)&=-\chi_f^L\omega^2+O(\omega^4),\notag\\
    \chi_f^L&=N_\sigma\int_{\rm BZ}\frac{d^2k}{(2\pi)^2}
    \frac{g_{xx}(\bm k)}{d(\bm k)}>0,
\end{align}
in agreement with the convention adopted in the main text.

We now separate the real and imaginary parts of
Eqs.~\eqref{eq:sm_final_KfL} and \eqref{eq:sm_final_KfH}.  For the retarded
response, we set \(z=\omega+i0^+\).  At positive frequency,
\((\omega+i0^+)^2-(2d)^2\) may be written as
\(\omega^2-(2d)^2+i0^+\), with the positive factor multiplying \(0^+\)
absorbed into its definition.  The distribution identity
\begin{equation}
    \frac{1}{x+i0^+}
    =
    \mathcal P\frac{1}{x}
    -i\pi\delta(x)
\end{equation}
therefore gives
\begin{equation}
    \frac{1}{(\omega+i0^+)^2-(2d)^2}
    =
    \mathcal P\frac{1}{\omega^2-(2d)^2}
    -i\pi\delta\!\left[\omega^2-(2d)^2\right].
\end{equation}
For \(\omega>0\) and \(d>0\), only the positive-frequency root contributes,
so that
\begin{equation}
    \delta\!\left[\omega^2-(2d)^2\right]
    =
    \frac{1}{2\omega}\delta(\omega-2d).
    \label{eq:sm_positive_frequency_delta}
\end{equation}
Using this result in Eq.~\eqref{eq:sm_final_KfL}, and introducing
\(\int[d\bm k]\equiv
N_\sigma\int_{\rm BZ}d^2k/(2\pi)^2\), we obtain
\begin{align}
    \operatorname{Re}\Pi_f^L(\omega)
    &=
    \mathcal P\int[d\bm k]\,
    \frac{4d(\bm k)\omega^2}
    {\omega^2-[2d(\bm k)]^2}
    g_{xx}(\bm k),
    \label{eq:sm_Re_KfL}\\
    \operatorname{Im}\Pi_f^L(\omega)
    &=
    -\pi\int[d\bm k]\,
    4d(\bm k)\omega^2 g_{xx}(\bm k)
    \delta\!\left[\omega^2-[2d(\bm k)]^2\right]\notag\\
    &=
    -\pi\omega^2\int[d\bm k]\,
    g_{xx}(\bm k)\delta[\omega-2d(\bm k)].
    \label{eq:sm_Im_KfL}
\end{align}
In the last line, the delta function sets \(d(\bm k)=\omega/2\).
Combining these two parts reproduces Eq.(8) of the main text.

For the Hall kernel, the overall factor \(-i\) in
Eq.~\eqref{eq:sm_final_KfH} interchanges the dispersive and absorptive
parts.  We find
\begin{align}
    \operatorname{Re}\Pi_f^H(\omega)
    &=
    -\pi\int[d\bm k]\,
    4\omega d^2(\bm k)\Omega_{xy}(\bm k)
    \delta\!\left[\omega^2-[2d(\bm k)]^2\right]\notag\\
    &=
    -\frac{\pi\omega^2}{2}\int[d\bm k]\,
    \Omega_{xy}(\bm k)\delta[\omega-2d(\bm k)],
    \label{eq:sm_Re_KfH}\\
    \operatorname{Im}\Pi_f^H(\omega)
    &=
    -\mathcal P\int[d\bm k]\,
    \frac{4\omega d^2(\bm k)\Omega_{xy}(\bm k)}
    {\omega^2-[2d(\bm k)]^2}.
    \label{eq:sm_Im_KfH}
\end{align}
Here we again used \(d(\bm k)=\omega/2\) on the support of the delta
function.  These two contributions combine to give
Eq.(9) of the main text.

\section{Chargon-independent reconstruction from Kramers--Kronig relation}
\label{app:inverse_KK_reconstruction}

Direct inversion of Eq.(5) in the main text requires the full
chargon kernel.  We now show that, under a different set of experimentally
testable conditions, the spinon kernel can instead be reconstructed without
specifying the frequency dependence of the chargon response.  The essential
ingredients are the absence of a chargon Hall response, the absence of
chargon absorption in the spinon spectroscopy window, and causality.

We work with retarded kernels at a complex frequency \(z\) and define the
inverse circular responses
\begin{equation}
    R_\pm(z)\equiv\frac{1}{\Pi_{\rm phys}^{\pm}(z)},
    \quad
    r_\pm(z)\equiv\frac{1}{\Pi_f^{\pm}(z)},\quad q_b(z)\equiv\frac{1}{\Pi_b(z)}
    \label{eq:sm_inverse_circular_definitions}
\end{equation}
The circular Ioffe--Larkin rule then becomes additive,
\begin{equation}
    R_\pm(z)=q_b(z)+r_\pm(z),
    \label{eq:sm_inverse_IL_circular}
\end{equation}
where the same scalar \(q_b\) enters both circular channels because
\(\Pi_b^H=0\).  It is useful to introduce the longitudinal and Hall
components of the inverse spinon kernel,
\begin{subequations}\label{eq:sm_inverse_spinon_components}
\begin{align}
    r_f^L
    &\equiv\frac{r_++r_-}{2}
    =\frac{\Pi_f^L}{(\Pi_f^L)^2+(\Pi_f^H)^2},\\
    r_f^H
    &\equiv\frac{r_+-r_-}{2i}
    =-\frac{\Pi_f^H}{(\Pi_f^L)^2+(\Pi_f^H)^2}.
\end{align}
\end{subequations}
Likewise, \(R_{\rm phys}^L=(R_++R_-)/2\) and
\(R_{\rm phys}^H=(R_+-R_-)/(2i)\). Note that chargon kernel does not have a Hall component, so \(q_b^{\pm}=q_b\). Then, form Equation~\eqref{eq:sm_inverse_IL_circular} we know that
\begin{equation}
    R_{\rm phys}^H=r_f^H,\qquad R_{\rm phys}^L=q_b+r_f^L.
    \label{eq:sm_inverse_IL_LH} 
\end{equation}
Thus, the difference of the two measured inverse circular responses removes
the chargon contribution exactly. Equivalently, the Hall part of the inverse physical response is purely the Hall part of the inverse spinon kernel. On the other hand, the longitudinal part of the inverse physical response is the sum of the chargon and spinon contributions, but below the onset of the chargon continuum, denoted by
\(\omega_b^{\rm th}\), the clean chargon response is reactive.  Away from a
zero of \(\Pi_b\), both \(\Pi_b(\omega)\) and \(q_b(\omega)\) are then real.
For \(0<\omega<\omega_b^{\rm th}\), Eq.~\eqref{eq:sm_inverse_IL_LH}
consequently implies
\begin{subequations}\label{eq:sm_direct_inverse_observables}
\begin{align}
    r_f^H(\omega)
    &=\frac{1}{2i}
    \left[
    \frac{1}{\Pi_{\rm phys}^+(\omega)}
    -\frac{1}{\Pi_{\rm phys}^-(\omega)}
    \right],\\
    \operatorname{Im}r_f^L(\omega)
    &=\operatorname{Im}\frac{1}{2}
    \left[
    \frac{1}{\Pi_{\rm phys}^+(\omega)}
    +\frac{1}{\Pi_{\rm phys}^-(\omega)}
    \right].
\end{align}
\end{subequations}
These quantities can be formed directly from the measured circular
conductivities using
\(\Pi_{\rm phys}^{\pm}=-i\omega\sigma_{\rm phys}^{\pm}\), or
\(R_\pm=i/(\omega\sigma_{\rm phys}^{\pm})\).  

The only part that remains unknown is \(\operatorname{Re}r_f^L\). This missing part is fixed by causality up to one real subtraction constant.
Additional care is required since the inverse of a causal response is not
automatically causal.  Besides analyticity of \(\Pi_f^{\pm}(z)\), we require
\begin{equation}
    \Pi_f^+(z)\Pi_f^-(z)
    =(\Pi_f^L)^2+(\Pi_f^H)^2\ne0,
    \qquad \operatorname{Im}z>0,
    \label{eq:sm_inverse_zero_free_condition}
\end{equation}
such that \(r_f^L(z)\) is analytic in the upper half-plane.  Any isolated zero
would instead produce a pole whose residue must be included explicitly in
the dispersion relation.  The zero-frequency behavior of a Chern insulator
does not by itself invalidate Eq.~\eqref{eq:sm_inverse_zero_free_condition}.
Indeed, \(\Pi_f^\pm\) vanish linearly with opposite signs as \(z\to0\); their
inverse poles cancel in \((r_++r_-)/2\), leaving \(r_f^L(0)\) finite. And we have checked that \(r_f^L(z)\) is indeed analytic in the upper half-plane.

We then have the following Kramers--Kronig relation to obtain the \(\operatorname{Re}r_f^L\):
\begin{equation}
    \operatorname{Re}r_f^L(\omega)
    =r_{f,\infty}^L
    +\frac{2}{\pi}\mathcal P\int_0^\infty d\omega'
    \frac{\omega'\operatorname{Im}r_f^L(\omega')}
    {\omega'^2-\omega^2},
    \label{eq:sm_inverse_KK_relation}
\end{equation}
where \(r_{f,\infty}^L\equiv
\lim_{|z|\to\infty}r_f^L(z)\) is real. For numerical evaluation with an upper cutoff \(\Omega_{\rm cut}\), the principal-value singularity at \(\omega'=\omega\) can be removed
analytically. Define \(f_r(\omega')\equiv\omega'\operatorname{Im}r_f^L(\omega')\).
For \(0<\omega<\Omega_{\rm cut}\), subtracting and adding \(f_r(\omega)\)
gives the exact identity
\begin{align}
    \mathcal P\int_0^{\Omega_{\rm cut}}
    \frac{f_r(\omega')}{\omega'^2-\omega^2}\,d\omega'
    &=\int_0^{\Omega_{\rm cut}}
    \frac{f_r(\omega')-f_r(\omega)}{\omega'^2-\omega^2}\,d\omega'
    \notag\\
    &\quad+\frac{f_r(\omega)}{2\omega}
    \ln\left|\frac{\Omega_{\rm cut}-\omega}
    {\Omega_{\rm cut}+\omega}\right|.
    \label{eq:sm_KK_principal_value_subtraction}
\end{align}
If \(f_r\) is continuously differentiable near \(\omega\), as in our
finite-broadening calculation, the integrand on the right has the finite
limit
\begin{equation}
    \lim_{\omega'\to\omega}
    \frac{f(\omega')-f(\omega)}{\omega'^2-\omega^2}
    =\frac{f'(\omega)}{2\omega}.
    \label{eq:sm_KK_removed_pole_limit}
\end{equation}
The logarithmic term is the analytically integrated principal value of
the subtracted part and accounts for the cancellation of the two-sided
singularity. In our numerical implementation, we use the above limit at
the coincident grid point, evaluate the regular integral with the
trapezoidal rule, and add back the logarithmic term. The endpoint
\(\omega=\Omega_{\rm cut}\) is excluded, and the cutoff is chosen above
the frequency range of interest; any omitted spectral tail remains a
separate truncation error.

Basing on the mean-field kernel in Eqs.~\eqref{eq:sm_final_KfL} and \eqref{eq:sm_final_KfH}, we know
\begin{equation}
    r_{f,\infty}^L
    =\left[
    N_\sigma\int_{\rm BZ}\frac{d^2k}{(2\pi)^2}
    4d(\bm k)g_{xx}(\bm k)
    \right]^{-1}.
    \label{eq:sm_inverse_KK_constant}
\end{equation}
In practice, ``infinity'' denotes the asymptotic regime of the low-energy spinon theory---frequencies above the spinon bandwidth but still below the scale at which the parton description itself ceases to apply. Note that at $z\to \infty$, \(\Pi_f^H(z)\) approaches to zero while \(\Pi_f^L(z)\) approaches the quantity inside the square brackets of above equation. Thus, the inverse constant $(r_{f,\infty}^L)^{-1}$ is related to the quantum metric as
\begin{equation}
    D_f\equiv\lim_{|z|\to\infty}\Pi_f^L(z) = (r_{f,\infty}^L)^{-1}
    =N_\sigma\int_{\rm BZ}\frac{d^2k}{(2\pi)^2}
    4d(\bm k)g_{xx}(\bm k).
\end{equation}

For the triangular Hofstadter ansatz, $ D_f$ can be determined by
the measured spinon interband threshold. Note that the spinon eigenvectors and hence
\(g_{xx}\) are independent of \(t_f\), so we can write
\begin{equation}
    D_f=t_f\mathcal D_\triangle,
    \qquad
    \mathcal D_\triangle\equiv
    N_\sigma\int_{\rm BZ}\frac{d^2k}{(2\pi)^2}
    4\widetilde d(\bm k)g_{xx}(\bm k).
\end{equation}
where the spinon hopping parameter $t_f$ is seperated out from $\widetilde d(\bm k)$ as \(d(\bm k)=t_f\widetilde d(\bm k)\). Thus, $\mathcal D_\triangle$ is a dimensionless quantity that only depends on the lattice geometry instead of spinon hopping parameter. Numerical integration over the magnetic Brillouin zone gives
\(\mathcal D_\triangle=1.38681422\) for \(N_\sigma=2\).
The spinon hopping \(t_f\) can then be inferred from the measured
interband threshold, \(\Delta_s=2\sqrt{3}t_f\). Consequently, the subtraction constant is
calibrated as
\begin{equation}
    r_{f,\infty}^L=\frac{1}{D_f}
    =\frac{2.49788441}{\Delta_s}.
\end{equation}
Although it relies on the spinon ansatz on triangular-lattice, this calibration is independent of the chargon model.

\begin{figure}[!h]
    \includegraphics[width=1\linewidth]{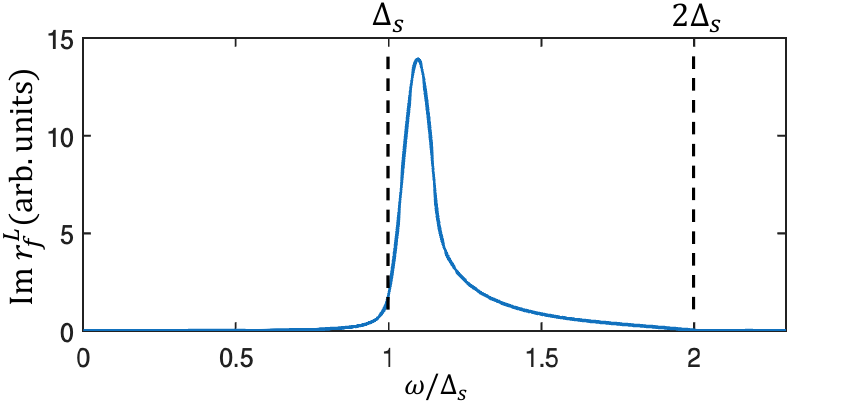}
    \caption{Distribution of \(\operatorname{Im}r_f^L\) (arbitary units) over the spinon interband continuum with $U/t=15$ and $\eta_s/t_f=0.05$. The dashed vertical lines indicate the spinon interband threshold \(\Delta_s\) and the upper edge \(2\Delta_s\).} \label{fig:rLf}
\end{figure}

An isolated spinon is gauge charged, so a local spectroscopic probe does not
directly measure \(\Delta_f=\Delta_s/2\).  What is measured is the gauge-invariant
particle--hole threshold \(\Delta_s\).  Experimentally, the two complex
circular conductivities determine
\begin{equation}
    Y_L(\omega)\equiv
    \operatorname{Im}\frac{1}{2}
    \left[\frac{i}{\omega\sigma_{\rm phys}^+(\omega)}
          +\frac{i}{\omega\sigma_{\rm phys}^-(\omega)}\right]
    =\operatorname{Im}r_f^L(\omega)
\end{equation} 
below the chargon absorption threshold.  In the clean limit, \(Y_L\) has no
absorptive support below \(\Delta_s\), and its first interband edge fixes
\(\Delta_s\). Thus, \(\Delta_s\) is
obtained by fitting the broadened leading edge (first dashed line in Fig.~\ref{fig:rLf}).  The present dispersion further predicts support
over \(\Delta_s\leq\omega\leq2\Delta_s\), so the upper edge (second dashed line in Fig.~\ref{fig:rLf}) also provides an
internal check of the assumed band shape. 

There is also a spectral-separation requirement.  Equation~\eqref{eq:sm_inverse_KK_relation}
needs \(\operatorname{Im}r_f^L\) over its full absorptive support, whereas
Eq.~\eqref{eq:sm_direct_inverse_observables} isolates it only where the
chargon is lossless.  A strictly chargon-independent closure is therefore
possible when the entire spinon interband continuum relevant to the
reconstruction lies below \(\omega_b^{\rm th}\), so that
\(\operatorname{Im}r_f^L\) has vanished before chargon absorption begins.
If the two continua overlap, the unmeasured high-frequency part of the
Kramers--Kronig integral must be supplied by an additional model.

For the present finite-band ansatz the clean spinon support ends at
\(2\Delta_s\).  Hence, if
\(2\Delta_s<\Omega_{\rm cut}<\omega_b^{\rm th}\), the upper limit in
Eq.~\eqref{eq:sm_inverse_KK_relation} can be replaced by
\(\Omega_{\rm cut}\) exactly: an infinite experimental scan is unnecessary.
Weak high-energy tails or overlap with the chargon continuum must instead be
included as a finite-window systematic uncertainty.

After completing \(r_f^L\), the circular and Cartesian spinon kernels are
obtained algebraically:
\begin{subequations}\label{eq:sm_inverse_recovery_of_Kf}
\begin{align}
    r_\pm&=r_f^L\pm i r_f^H,
    &\Pi_f^\pm&=\frac{1}{r_\pm},\\
    \Pi_f^L&=\frac{\Pi_f^++\Pi_f^-}{2},
    &\Pi_f^H&=\frac{\Pi_f^+-\Pi_f^-}{2i}.
\end{align}
\end{subequations}
Substitution into Eq.(19) in the main text finally yields the
quantum-metric and Berry-curvature spectral densities.  This construction
does not require a microscopic frequency-dependent chargon model.  It does
require a vanishing chargon Hall channel, negligible chargon absorption over
the spinon support, an upper-half-plane zero-free inverse spinon kernel (or
explicit pole terms), and the scalar normalization fixed from the measured
interband threshold.

\subsection{Alternative method: first moment of the dynamical spin
structure factor}

As an independent experimental cross-check for $r^L_{f,\infty}$, one may utilize the small-momentum
first moment of the dynamical spin structure factor (DSSF).  At zero temperature,
with \(S^z=(n_\uparrow-n_\downarrow)/2\) and
\(\omega_{n0}=E_n-E_0>0\), we define
\begin{equation}
    S^{zz}(\bm q,\omega)
    \equiv\sum_n
    |\langle n|S^z_{\bm q}|0\rangle|^2
    \delta(\omega-\omega_{n0}).
\end{equation}
If \(S^z\) is conserved, its long-wavelength continuity equation is
\begin{equation}
    \partial_tS^z_{\bm q}+iqJ_L^z(\bm q)=O(q^2),
\end{equation}
where \(J_L^z=\hat{\bm q}\cdot\bm J^z\).  Taking a matrix element between
the ground state and an excited state gives
\begin{equation}
    \omega_{n0}\langle n|S^z_{\bm q}|0\rangle
    =-q\langle n|J_L^z(\bm q)|0\rangle+O(q^2).
\end{equation}
The sign depends on the Fourier-transform convention and drops out after
taking the modulus squared.  Introducing the positive-frequency
longitudinal spin-current spectrum
\begin{equation}
    \mathcal S_{JJ}^{L}(\bm q,\omega)
    \equiv\sum_n
    |\langle n|J_L^z(\bm q)|0\rangle|^2
    \delta(\omega-\omega_{n0}),
\end{equation}
the continuity equation therefore yields
\begin{equation}
    \lim_{q\to0}\mathcal S_{JJ}^{L}(\bm q,\omega)
    =\lim_{q\to0}\frac{\omega^2}{q^2}
    S^{zz}(\bm q,\omega).
\end{equation} 
 
To connect this spectrum to an instantaneous response weight, we write the
retarded physical spin-current kernel as
\begin{align} 
    \Pi_{f,\mathrm{spin}}^L(z)=\mathcal{D}_f+\chi_{JJ}^R(z),
\end{align}
which should be distinguished from the internal-gauge current kernel
\(\Pi_f^L\) in Eq.~\eqref{eq:sm_final_KfL}.  Here \(\mathcal{D}_f\) is the
frequency-independent diamagnetic/contact term.  The paramagnetic term has
the Lehmann representation
\begin{equation}
    \chi_{JJ}^R(z)=\int_0^\infty d\omega'\,
    \mathcal S_{JJ}^{L}(\bm 0,\omega')
    \left(\frac{1}{z-\omega'}-\frac{1}{z+\omega'}\right).
\end{equation}
The gapped spin liquid has no spin stiffness, so the static uniform kernel
vanishes, \(\Pi_{f,\mathrm{spin}}^L(0)=0\).  It follows that
\begin{align}
    \mathcal{D}_f&=-\chi_{JJ}^R(0)
    =2\int_0^\infty\frac{d\omega}{\omega}
    \mathcal S_{JJ}^{L}(\bm 0,\omega)\notag\\
    &=2\lim_{q\to0}\frac{1}{q^2}
    \int_0^\infty d\omega\,\omega S^{zz}(\bm q,\omega).
\end{align}
Before comparing \(\mathcal D_f\) with the internal-gauge weight \(D_f\), we
make the origin of the single-flavor quantity \(D_0\) explicit.  Removing
the overall factor \(N_\sigma\) from
Eqs.~\eqref{eq:sm_para_symmetric} and \eqref{eq:sm_dia_metric_form}, the
longitudinal kernel of one flavor with unit probe charge is
\begin{align}
    \Pi_0^L(z)&=D_0+\Pi_{0,{\rm para}}^L(z),\\
    D_0&=\int_{\rm BZ}\frac{d^2k}{(2\pi)^2}
    4d(\bm k)g_{xx}(\bm k),\\
    \Pi_{0,{\rm para}}^L(z)
    &=\int_{\rm BZ}\frac{d^2k}{(2\pi)^2}
    \frac{16d^3(\bm k)}{z^2-[2d(\bm k)]^2}g_{xx}(\bm k).
\end{align}
The last line decays as \(z^{-2}\) for \(|z|\to\infty\), whereas the
contact term is frequency independent.  Therefore
\begin{equation}
    \lim_{|z|\to\infty}\Pi_0^L(z)=D_0.
\end{equation}
Equivalently, this follows by setting \(N_\sigma=1\) in
Eq.~\eqref{eq:sm_final_KfL} before taking the high-frequency limit.  Thus
\(D_0\) is not an additional phenomenological parameter: it is the
single-flavor diamagnetic weight derived from the same spinon Hamiltonian.

It remains to account for the probe charges carried by the two flavors. Both
flavors carry unit internal gauge charge, whereas their physical \(S^z\)
charges are \(+1/2\) and \(-1/2\).  Since the two contributions are equal
and cross-flavor correlators vanish,
\begin{subequations}
\begin{align}
    D_f&=(1^2+1^2)D_0=2D_0,\\
    \mathcal{D}_f&=\left[\left(\frac12\right)^2
    +\left(-\frac12\right)^2\right]D_0
    =\frac{D_0}{2}=\frac{D_f}{4}.
\end{align}
\end{subequations}
Consequently,
\begin{equation}
    D_f=8\lim_{q\to0}\frac{1}{q^2}
    \int_0^\infty d\omega\,\omega S^{zz}(\bm q,\omega).
\end{equation}
The right-hand side can in principle be determined from absolutely
normalized, polarization-resolved inelastic neutron scattering.  This test
requires the full magnetic continuum, a controlled small-\(q\) extrapolation,
and an approximately conserved spin component.  The continuity equation is general under these conditions, whereas \(D_f=4\mathcal{D}_f\) additionally assumes the present two-flavor mean-field saddle and can receive interaction or vertex corrections.

\section{Self-consistent mean-field method}
\label{app:mean_field_method}

We first clarify which quantities are
independent inputs and which are outputs of the parton saddle.  This
distinction is important because the parameters appearing in the two parton
Hamiltonians cannot all be varied independently if they are to represent one
microscopic Hofstadter--Hubbard model.

The triangular lattice, half filling, and the number of flavors
\(N_\sigma=2\) are fixed throughout the work apparently.  The bare electron hopping \(t\) sets the
microscopic energy unit.  The orbital flux
\(\Phi_\triangle=\pi/2\) is imposed externally rather than obtained from a
mean-field equation, and \(\bar A_{ij}\) is a gauge representative of this
fixed flux.  Likewise, \(U/t\) is a tunable microscopic control parameter, not
a saddle-point output.  It may be chosen within the CSL interval established
for the same model, preferably away from either phase boundary
\cite{ZhangLiuSong2026,Divic2025}.  We work at \(T=0\) for a clean exposure of the spinon quantum geometry (otherwise, there will be additional distribution factors in the optical kernels). We ignore the Zeeman splitting.  A nonzero Zeeman splitting below the spin gap does not modify the spin-conserving optical kernels at zero temperature before the
band occupations change, as explained in the discussion following
Eq.~\eqref{eq:Zeeman_spinon}.

The gauge-invariant flux ansatz of the adopted saddle is the flux partition
\begin{equation*}
    \Phi_f=\sum_{\partial\triangle}\bar a_{ij}=\frac{\pi}{2},
    \qquad
    \Phi_b=\sum_{\partial\triangle}(\bar A_{ij}-\bar a_{ij})=0.
\end{equation*}
This uniform CSL branch, including its two-site spinon magnetic unit cell and
topological band filling, can be adopted from the self-consistent and
many-body studies\cite{ZhangLiuSong2026,Divic2025}; The bond-by-bond choices
\(\bar a_{ij}=\bar A_{ij}\), \(\phi^b_{ij}=0\), and
\(\phi^f_{ij}=-\bar A_{ij}\) are gauge conventions. The unbarred field \(a_\mu\) represents a dynamical gauge
fluctuation around this saddle.  It is integrated out in the tensor
Ioffe--Larkin rule and is neither an externally chosen number nor an
additional mean-field fitting parameter.  The weak field
\(A_\mu\) is the optical source used to probe the response.

The Lagrangian multiplier \(\lambda\) can be determined once \(U_s\)
and \(t_b\) are fixed.  Performing the Matsubara sum in
Eq.~\eqref{eq:gaussian_rotor_constraint} gives
\begin{equation}
    1=\frac{1}{N}\sum_{\bm k}
    \frac{U_s}{E_b(\bm k)}
    \coth\!\left[\frac{E_b(\bm k)}{2T}\right].
    \label{eq:sm_finite_temperature_lambda_constraint}
\end{equation}
Hence at zero temperature the triangular-lattice equation reduces to
\begin{equation}
    1=\int_{\bm k}
    \frac{U_s}
    {\sqrt{2U_s[\lambda-2t_b\gamma_\triangle(\bm k)]}},
    \qquad \lambda>6t_b.
    \label{eq:sm_zero_temperature_lambda_constraint}
\end{equation}
This is a one-dimensional root-finding problem for \(\lambda\).  Its uncondensed solution is
unique whenever the selected parameters lie on the Mott side of the rotor
condensation transition.  For \(U_s\gg t_b\), using
\(\langle\varepsilon_b\rangle_{\bm k}=0\) and
\(\langle\varepsilon_b^2\rangle_{\bm k}=6t_b^2\), one has the expsion:
\begin{equation*}
    \lambda
    =\frac{U_s}{2}+\frac{9t_b^2}{U_s}
    +O(t_b^3/U_s^2).
\end{equation*}

We now evaluate directly the bond-field definitions in
Eqs.~\eqref{eq:bond_mean_fields}.  In the gauge of
Eq.~\eqref{eq:Hofstadter_saddle_bond_phases}, they read
\begin{equation*}
    \chi^b_{ij}=|\chi^b_{ij}|,
    \qquad
    \chi^f_{ij}=|\chi^f_{ij}|e^{-i\bar A_{ij}}.
\end{equation*}
Thus \(\chi^f_{ij}\) itself is generally complex.  Magnetic translations and the (projective) \(C_3\)
symmetry make the magnitudes identical on all nearest-neighbor bonds.  We may
therefore suppress the bond indices on the magnitudes and write
\(|\chi^b_{ij}|=|\chi^b|\) and
\(|\chi^f_{ij}|=|\chi^f|\), while retaining the bond-dependent Peierls phase
explicitly.  These bond fields should not be confused with the
low-frequency chargon polarizability \(\chi_b\), whose subscript carries a lower
parton label.  From Eq.~\eqref{eq:bond_mean_fields} and Eq.~\eqref{eq:mf_hopping_definitions}, the two hopping amplitudes satisfy
\begin{subequations}\label{eq:sm_bond_amplitude_self_consistency}
\begin{align}
    \frac{t_b}{t}=|\chi^f|
    &=\frac{N_\sigma}{6}
      \big\langle\rho(\bm k)\big\rangle_{\rm MBZ},
      \label{eq:sm_tb_bond_self_consistency}\\
    \frac{t_f}{t}=|\chi^b|
    &=\int_{\bm k}\cos(\bm k\!\cdot\!\bm\delta_\nu)
      \frac{U_s}{E_b(\bm k)}
     =\frac{1}{3}\int_{\bm k}\gamma_\triangle(\bm k)
      \frac{U_s}{E_b(\bm k)}.
      \label{eq:sm_tf_bond_self_consistency}
\end{align}
\end{subequations}
Here \(N_\sigma=2\),
\(\rho(\bm k)=
\sqrt{\cos^2 k_1+\cos^2 k_2+\cos^2(k_2-k_1)}\), and
\(E_b(\bm k)=\sqrt{2U_s[\lambda-2t_b\gamma_\triangle(\bm k)]}\).
The \(\lambda\) entering the second line is not an additional unresolved
quantity: once the first line fixes \(t_b\), it is obtained numerically from
the zero-temperature constraint in
Eq.~\eqref{eq:sm_zero_temperature_lambda_constraint} above.

We now clarify how Eq.~\eqref{eq:sm_tb_bond_self_consistency} can be derived from
Eq.~\eqref{eq:spinon_bond_mean_field}.  Let \(N_m\) be the number of two-site
magnetic unit cells, so that the number of lattice sites is \(N_s=2N_m\),
and define
\(\langle O\rangle_{\rm MBZ}=N_m^{-1}
\sum_{\bm k\in{\rm MBZ}}O(\bm k)\).  At half filling the lower spinon band
is completely filled for each of the \(N_\sigma=2\) spin flavors.  Using
\(\epsilon_-(\bm k)=-2t_f\rho(\bm k)\), its ground-state energy per site is
\begin{equation*}
    \frac{E_f^{(0)}}{N_s}
    =\frac{N_\sigma}{2N_m}
      \sum_{\bm k\in{\rm MBZ}}\epsilon_-(\bm k)
    =-N_\sigma t_f\langle\rho(\bm k)\rangle_{\rm MBZ}
    .
\end{equation*}
The same expectation value evaluated from the real-space spinon Hamiltonian
is
\begin{align*}
    E_f^{(0)}
    &=-t_f\sum_{\langle ij\rangle}
      \left[e^{i\bar A_{ij}}\chi^f_{ij}
      +\mathrm{c.c.}\right]=-2t_f|\chi^f|\,(3N_s)
\end{align*}
Here the phase in
\(\chi^f_{ij}=|\chi^f|e^{-i\bar A_{ij}}\) cancels the explicit Peierls
phase.  The factor \(3N_s\) is the number of undirected nearest-neighbor
bonds on a triangular lattice: the coordination number is six, and each bond
is shared by two sites.  Equating the momentum- and real-space expressions
gives
\(|\chi^f|=(N_\sigma/6)\langle\rho\rangle_{\rm MBZ}\), which is
Eq.~\eqref{eq:sm_tb_bond_self_consistency}.  This is also the
Hellmann--Feynman evaluation of the occupied-band projector.  Because the
lower band is completely filled, its eigenvectors and \(\rho\) do not depend
on the overall scale \(t_f\), so this equation directly determines
\(t_b=t|\chi^f|\) without an initial guess for \(t_f\).

We next derive Eq.~\eqref{eq:sm_tf_bond_self_consistency} from
Eq.~\eqref{eq:chargon_bond_mean_field}.  Since \(b_i=X_i=e^{i\theta_i}\), the
zero-flux gauge gives, for a representative bond
\(j=i+\bm\delta_\nu\),
\begin{align*}
    |\chi^b|
    &=\langle X_i^*X_{i+\bm\delta_\nu}\rangle\\
    &=\frac{T}{N_s}\sum_{\bm k,\nu_n}
      e^{i\bm k\cdot\bm\delta_\nu}
      G_b(\bm k,i\nu_n).
\end{align*}
Because \(E_b(\bm k)=E_b(-\bm k)\), the sine part of the exponential
integrates to zero.  Substituting
\(G_b=2U_s/[\nu_n^2+E_b^2(\bm k)]\) from
Eq.~\eqref{eq:gaussian_rotor_propagator} and using the bosonic sum
\begin{equation*}
    T\sum_{\nu_n}\frac{1}{\nu_n^2+E^2}
    =\frac{1}{2E}\coth\!\left(\frac{E}{2T}\right)
\end{equation*}
yields
\begin{equation*}
    |\chi^b|
    =\int_{\bm k}\cos(\bm k\!\cdot\!\bm\delta_\nu)
      \frac{U_s}{E_b(\bm k)}
      \coth\!\left[\frac{E_b(\bm k)}{2T}\right].
\end{equation*}
At \(T=0\), the hyperbolic cotangent tends to one.  Moreover, \(C_3\)
symmetry makes this correlator equal for the three bond directions.  Summing
over \(\nu=1,2,3\) and using
\(\gamma_\triangle(\bm k)=\sum_{\nu=1}^3
\cos(\bm k\cdot\bm\delta_\nu)\) then gives the two equivalent forms in
Eq.~\eqref{eq:sm_tf_bond_self_consistency}.
 
\begin{figure}[t]
    \centering
    \includegraphics[width=0.8\linewidth]{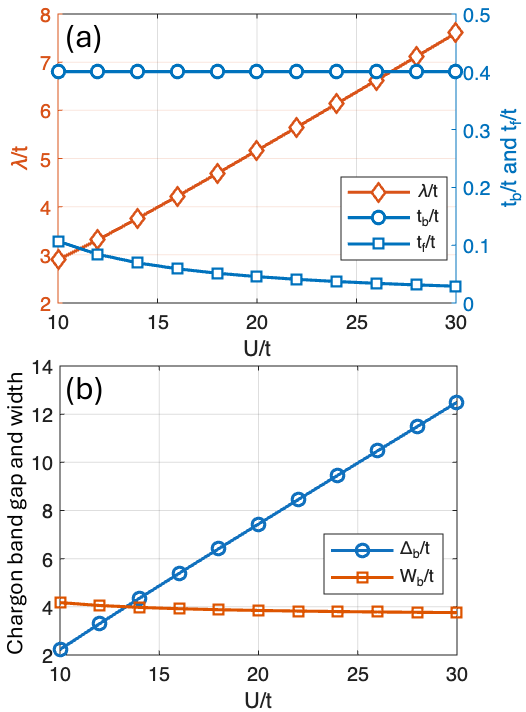}
    \caption{(a) Mean-field solutions for $\lambda$, $t_b$, and $t_f$ as a function of \(U/t\). (b) The chargon band gap and bandwidth calculated from Eq.~\eqref{eq:sm_derived_chargon_scales}.}
    \label{fig:sm_mf_solutions}
\end{figure}

Consequently, the actual sequence of self-consistent calculations is:  solve
Eq.~\eqref{eq:sm_zero_temperature_lambda_constraint} for the multiplier \(\lambda\) for the soft mode constraint, evaluate the first line of
Eq.~\eqref{eq:sm_bond_amplitude_self_consistency} to obtain \(t_b\), and evaluate the second line of Eq.~\eqref{eq:sm_bond_amplitude_self_consistency} to obtain \(t_f\). The mean-field solutions for each \(U/t\) are shown in Fig.~\ref{fig:sm_mf_solutions}. 

% No coupled nonlinear iteration remains for this special filled-band ansatz.  A
% simultaneous iteration is required only for a more general ansatz in which
% the spinon occupations or eigenvectors also change with the bond amplitudes.
% The other uniform constraint field, \(h\), is fixed to zero at the
% half-filled saddle.  The soft-rotor multiplier \(\lambda\) is distinct both
% from \(h\) and from the reconstruction control parameter \(\lambda^\pm\)
% introduced later.

Existing works can only be used directly for the existence and flux
structure of the CSL branch and for locating a reasonable interval of
\(U/t\). Previous works~\cite{ZhangLiuSong2026,Divic2025} do not tabulate a compatible pair \((t_b,t_f)\) for
our atomic-limit calibration \(U_s=U/2\); in particular, results from a
strong-coupling spinon-only saddle do not determine the Gaussian rotor bond
in Eq.~\eqref{eq:sm_bond_amplitude_self_consistency}.  Therefore the flux branch
can be adopted from the literature, whereas a quantitatively microscopic
kernel at a selected \(U/t\) should use Eq.~\eqref{eq:sm_bond_amplitude_self_consistency}
together with the already given \(\lambda\) constraint to recompute its two
bond amplitudes in our convention.

Once \(U_s\), \(t_b\), and \(\lambda\) are known, the gap and band width of Gaussian chargon can be derived from
\begin{align} 
    \Delta_b&=\sqrt{2U_s(\lambda-6t_b)},\\
    W_b&=\sqrt{\Delta_b^2+18U_st_b}-\Delta_b,
    \label{eq:sm_derived_chargon_scales}
\end{align}
Adiditionally, \(\chi_b\), \(\gamma_b\), and the full \(\Pi_b(\omega)\) through
the Brillouin-zone integrals can also be derived.  Similarly, onece \(t_f\) is known, the energy scale of the spinon band and spinon kernel \(\Pi_f^{ij}(\omega)\) can be determined. Note that the spinon QGT is independent of the overall scale \(t_f\).

\section{Derivation of the quantum geometric tensor}

In this appendix we derive the quantum geometric tensor (QGT) for a generic
two-band Hamiltonian using the projector formalism.  Throughout this appendix
we use the convention
\begin{equation}
    Q_{ij}
    =
    g_{ij}
    -
    \frac{i}{2}\Omega_{ij},
\end{equation}
where \(g_{ij}\) is the quantum metric and \(\Omega_{ij}\) is the Berry
curvature.

\subsection{Projector representation of the QGT}
\label{app:qgt_projector}

Consider a single nondegenerate Bloch band with normalized eigenstate
\(|u(\bm k)\rangle\).  The band projector is
\begin{equation}
    P(\bm k)
    =
    |u(\bm k)\rangle\langle u(\bm k)|.
\end{equation}
The quantum geometric tensor is defined as
\begin{equation}
    Q_{ij}
    \equiv
    \langle \partial_i u|
    (1-P)
    |\partial_j u\rangle ,
    \qquad
    \partial_i\equiv \partial_{k_i}\label{eq:QGT_def}.
\end{equation}
This form is invariant under the local \(U(1)\) gauge transformation
\(|u(\bm k)\rangle\rightarrow e^{i\varphi(\bm k)}|u(\bm k)\rangle\), because
the projector \(1-P\) removes the component of \(|\partial_j u\rangle\)
parallel to \(|u\rangle\).

We now show that the same tensor can be written in terms of projectors.
Taking the derivative of \(P=|u\rangle\langle u|\), one obtains
\begin{equation}
    \partial_i P
    =
    |\partial_i u\rangle\langle u|
    +
    |u\rangle\langle \partial_i u|.
\end{equation}
Using the normalization condition
\(\langle u|u\rangle=1\), one has
\begin{equation}
    \langle \partial_i u|u\rangle
    =
    -\langle u|\partial_i u\rangle.
\end{equation}
It follows that
\begin{equation}
    (\partial_j P) |u\rangle
    =
    (1-P)|\partial_j u\rangle ,
\end{equation}
and
\begin{equation}
    \langle u|(\partial_i P)
    =
    \langle \partial_i u|(1-P).
\end{equation}
Therefore
\begin{align}
    \Tr\left[
    P\partial_i P\partial_j P
    \right]
    &=
    \langle u|
    \partial_i P\partial_j P
    |u\rangle
    \nonumber\\
    &=
    \langle \partial_i u|
    (1-P)^2
    |\partial_j u\rangle
    \nonumber\\
    &=
    \langle \partial_i u|
    (1-P)
    |\partial_j u\rangle
    \nonumber\\
    &=
    Q_{ij}.
\end{align}
Similarly,
\begin{equation}
    \Tr\left[
    P\partial_j P\partial_i P
    \right]
    =
    Q_{ji}.
\end{equation}

The antisymmetric part gives the Berry curvature.  From
\begin{align}
    \Tr\left[
    P[\partial_iP,\partial_jP]
    \right]
    &=
    \Tr\left[
    P\partial_iP\partial_jP
    \right]
    -
    \Tr\left[
    P\partial_jP\partial_iP
    \right]
    \nonumber\\
    &=
    Q_{ij}-Q_{ji},
\end{align}
and using the convention \(Q_{ij}=g_{ij}-i\Omega_{ij}/2\), we find
\begin{equation}
    Q_{ij}-Q_{ji}
    =
    -i\Omega_{ij}.
\end{equation}
Thus the Berry curvature can be written as
\begin{equation}
    \Omega_{ij}
    =
    i\Tr\left[
    P[\partial_iP,\partial_jP]
    \right].
\end{equation}

The symmetric part gives the quantum metric.  Since \(Q_{ji}=Q_{ij}^*\),
\begin{equation}
    g_{ij}
    =
    \mathrm{Re}\,Q_{ij}
    =
    \frac{1}{2}
    \left(
    Q_{ij}+Q_{ji}
    \right).
\end{equation}
From the definition of the QGT in Eq.~\eqref{eq:QGT_def}, one has
\begin{align}
    Q_{ij}+Q_{ji}=&\braket{\partial_iu|\partial_ju}+\braket{\partial_ju|\partial_iu}\nonumber\\
     &+ \braket{\partial_i u|u}\braket{\partial_j u|u} +\braket{\partial_j u|u}\braket{\partial_i u|u}.
\end{align}
On the other hand, using the identity ${\rm Tr}(\ket{a}\bra{b})=\braket{a|b}$, we have
\begin{align}
    {\rm Tr}[\partial_i P\partial_j P]=&\braket{\partial_iu|\partial_ju}+\braket{\partial_ju|\partial_iu}\nonumber\\
     &+ \braket{\partial_i u|u}\braket{\partial_j u|u} +\braket{ u| \partial_j u}\braket{ u|\partial_i u}.
\end{align}
Note that $\braket{u|\partial_j u}\braket{u|\partial_i u}=\braket{\partial_j u|u}\braket{\partial_i u|u}$. Therefore, based on the preceding two equations, one obtains
\begin{equation}
    g_{ij}
    =
    \frac{1}{2}
    \Tr\left[
    \partial_i P\partial_j P
    \right].
\end{equation}

Hence, for the convention used here,
\begin{equation}
    g_{ij}
    =
    \frac{1}{2}
    \Tr\left[
    \partial_iP\partial_jP
    \right],
    \qquad
    \Omega_{ij}
    =
    i\Tr\left[
    P[\partial_iP,\partial_jP]
    \right].
\end{equation}

\subsection{Application to triangular-lattice Hofstadter spinons}

Now consider a generic two-band Hamiltonian
\begin{equation}
    h(\bm k)
    =
    \bm d(\bm k)\cdot\bm \tau ,
\end{equation}
where \(\bm \tau=(\tau_x,\tau_y,\tau_z)\) are Pauli matrices in the two-band
space.  We define
\begin{equation}
    d(\bm k)=|\bm d(\bm k)|,
    \qquad
    \hat{\bm d}(\bm k)=\frac{\bm d(\bm k)}{d(\bm k)}.
\end{equation}
The Hamiltonian can be written as
\begin{equation}
    h(\bm k)
    =
    d(\bm k)\,
    \hat{\bm d}(\bm k)\cdot\bm \tau.
\end{equation}
Since
\begin{equation}
    \left(
    \hat{\bm d}\cdot\bm\tau
    \right)^2
    =
    1,
\end{equation}
the eigenvalues of \(\hat{\bm d}\cdot\bm\tau\) are \(\pm1\), and the band
energies are
\begin{equation}
    \varepsilon_\pm(\bm k)=\pm d(\bm k).
\end{equation}
The projectors onto the upper and lower bands are therefore
\begin{equation}
    P_\pm(\bm k)
    =\ket{u_\pm}\bra{u_\pm}=
    \frac{1}{2}
    \left[
    1\pm \hat{\bm d}(\bm k)\cdot\bm\tau
    \right].
\end{equation}
For the lower band,
\begin{equation}
    P_-(\bm k)
    =
    \frac{1}{2}
    \left[
    1-\hat{\bm d}(\bm k)\cdot\bm\tau
    \right].
\end{equation}
One can verify that \(P_-^2=P_-\), $P_-\ket{u_-}=\ket{u_-}$, and $P_-\ket{u_+}=0$. Taking a momentum derivative gives
\begin{equation}
    \partial_iP_-
    =
    -\frac{1}{2}
    \left(
    \partial_i\hat{\bm d}
    \right)
    \cdot\bm\tau.
\end{equation}

We first compute the quantum metric.  In 2D, using
\begin{equation}
    (\bm a\cdot\bm\tau)(\bm b\cdot\bm\tau)
    =
    (\bm a\cdot\bm b)\mathbb{I}
    +
    i(\bm a\times\bm b)\cdot\bm\tau ,
\end{equation}
and
\begin{equation}
    \Tr\mathbb{I}=2,
    \qquad
    \Tr\,\tau_a=0,
\end{equation}
one finds
\begin{align}
    \Tr\left[
    \partial_iP_-\partial_jP_-
    \right]
    &=
    \frac{1}{4}
    \Tr\left[
    \left(
    \partial_i\hat{\bm d}\cdot\bm\tau
    \right)
    \left(
    \partial_j\hat{\bm d}\cdot\bm\tau
    \right)
    \right]
    \nonumber\\
    &=
    \frac{1}{2}
    \partial_i\hat{\bm d}\cdot\partial_j\hat{\bm d}.
\end{align}
Therefore
\begin{equation}
    g_{ij}^-
    =
    \frac{1}{4}
    \partial_i\hat{\bm d}\cdot\partial_j\hat{\bm d}.
\end{equation}

Next we compute the Berry curvature. Using the identity
\begin{equation}
    [\bm a\cdot\bm\tau,\bm b\cdot\bm\tau]
    =
    2i(\bm a\times\bm b)\cdot\bm\tau        
\end{equation}
the commutator of the projector
derivatives is
\begin{align}
    [\partial_iP_-,\partial_jP_-]
    &=
    \frac{1}{4}
    \left[
    \partial_i\hat{\bm d}\cdot\bm\tau,
    \partial_j\hat{\bm d}\cdot\bm\tau
    \right]
    \nonumber\\
    &=
    \frac{i}{2}
    \left(
    \partial_i\hat{\bm d}\times\partial_j\hat{\bm d}
    \right)
    \cdot\bm\tau.
\end{align}
Using
\begin{equation}
    P_-
    =
    \frac{1}{2}
    \left[
    1-\hat{\bm d}\cdot\bm\tau
    \right],
\end{equation}
we obtain
\begin{align}
    \Tr\left[
    P_-
    \left(
    \partial_i\hat{\bm d}\times\partial_j\hat{\bm d}
    \right)
    \cdot\bm\tau
    \right]
    &=
    -\hat{\bm d}\cdot
    \left(
    \partial_i\hat{\bm d}\times\partial_j\hat{\bm d}
    \right).
\end{align}
Thus
\begin{align}
    \Tr\left[
    P_-[\partial_iP_-,\partial_jP_-]
    \right]
    &=
    -\frac{i}{2}
    \hat{\bm d}\cdot
    \left(
    \partial_i\hat{\bm d}\times\partial_j\hat{\bm d}
    \right).
\end{align}
With the convention
\(\Omega_{ij}=i\Tr\{P[\partial_iP,\partial_jP]\}\), this gives
\begin{equation}
    \Omega_{ij}^-
    =
    \frac{1}{2}
    \hat{\bm d}\cdot
    \left(
    \partial_i\hat{\bm d}\times\partial_j\hat{\bm d}
    \right).
\end{equation}
Combining the metric and Berry curvature, the lower-band QGT is
\begin{equation}
    Q_{ij}^-
    =
    \frac{1}{4}
    \left[
    \partial_i\hat{\bm d}\cdot\partial_j\hat{\bm d}
    -
    i\,
    \hat{\bm d}\cdot
    \left(
    \partial_i\hat{\bm d}\times\partial_j\hat{\bm d}
    \right)
    \right].
\end{equation}

It is often useful to rewrite these expressions directly in terms of
\(\bm d\).  Since
\begin{equation}
    \partial_i\hat{\bm d}
    =
    \frac{\partial_i\bm d}{d}
    -
    \frac{
    \bm d
    \left(
    \bm d\cdot\partial_i\bm d
    \right)
    }{d^3},
\end{equation}
the quantum metric becomes
\begin{equation}
    g_{ij}^-
    =
    \frac{1}{4}
    \left[
    \frac{
    \partial_i\bm d\cdot\partial_j\bm d
    }{d^2}
    -
    \frac{
    \left(
    \bm d\cdot\partial_i\bm d
    \right)
    \left(
    \bm d\cdot\partial_j\bm d
    \right)
    }{d^4}
    \right].
\end{equation}
Similarly,
\begin{equation}
    \hat{\bm d}\cdot
    \left(
    \partial_i\hat{\bm d}\times\partial_j\hat{\bm d}
    \right)
    =
    \frac{
    \bm d\cdot
    \left(
    \partial_i\bm d\times\partial_j\bm d
    \right)
    }{d^3},
\end{equation}
and therefore
\begin{equation}
    \Omega_{ij}^-
    =
    \frac{
    \bm d\cdot
    \left(
    \partial_i\bm d\times\partial_j\bm d
    \right)
    }{
    2d^3
    }.
\end{equation}

We now apply these expressions to the triangular-lattice Hofstadter spinon
Hamiltonian in Eq.~\eqref{eq:sm_spinon_two_band_hamiltonian}.  Define
\begin{equation*}
    k_1=ak_x,
    \quad
    k_2=\frac{a}{2}(\sqrt{3}k_y+k_x),
    \quad
    k_3\equiv k_2-k_1=\frac{a}{2}(\sqrt{3}k_y-k_x),
\end{equation*}
together with
\begin{equation*}
    c_n=\cos k_n,
    \qquad
    s_n=\sin k_n,
    \qquad
    \rho^2=c_1^2+c_2^2+c_3^2.
\end{equation*}
For the \(+\pi/2\)-flux chirality chosen above,
\begin{equation*}
    \bm d(\bm k)
    =-2t_f(c_2,-c_3,c_1),
    \qquad
    d(\bm k)\equiv |\bm{d}(\bm{k})| =2t_f\rho.
\end{equation*}
The Cartesian momentum derivatives are
\begin{align*}
    \partial_{k_x}\bm d
    &=t_fa(s_2,s_3,2s_1),\\
    \partial_{k_y}\bm d
    &=\sqrt{3}t_fa(s_2,-s_3,0).
\end{align*}
For compactness, introduce
\begin{align*}
    A_x&=c_2s_2-c_3s_3+2c_1s_1,
    &
    A_y&=c_2s_2+c_3s_3\\
    B_x&=s_2^2+s_3^2+4s_1^2,
    &
    B_y&=s_2^2+s_3^2.
\end{align*}
Substitution into the general projector formulas gives the Cartesian quantum
metric
\begin{align}
    g_{xx}^-
    &=\frac{a^2}{16}
    \left(
    \frac{B_x}{\rho^2}
    -\frac{A_x^2}{\rho^4}
    \right),\\
    g_{yy}^-
    &=\frac{3a^2}{16}
    \left(
    \frac{B_y}{\rho^2}
    -\frac{A_y^2}{\rho^4}
    \right),\\
    g_{xy}^-
    &=\frac{\sqrt{3}a^2}{16}
    \left[
    \frac{s_2^2-s_3^2}{\rho^2}
    -\frac{A_xA_y}{\rho^4}
    \right].
\end{align}
The corresponding Berry curvature simplifies to
\begin{equation}
    \Omega_{xy}^-(\bm k)
    =
    \frac{\sqrt{3}a^2}{4\rho^3}
    \left(1-c_1c_2c_3\right).
    \label{eq:hofstadter_berry_curvature_explicit}
\end{equation}
Here we used
\(\bm d\cdot(\partial_{k_x}\bm d\times\partial_{k_y}\bm d)
=4\sqrt{3}t_f^3a^2(1-c_1c_2c_3)\).
Integrating Eq.~\eqref{eq:hofstadter_berry_curvature_explicit} over the
magnetic Brillouin zone gives \(C_s=+1\) for the occupied band in the
chirality convention adopted here.  Complex conjugating all hopping
amplitudes changes the second component of \(\bm d\) from \(-c_3\) to
\(+c_3\), and hence reverses \(\Omega_{xy}^-\) and \(C_s\), while leaving
\(d\) and \(g_{ij}^-\) unchanged.  The above derivations use the convention
\(Q_{ij}^-=g_{ij}^- - i\Omega_{ij}^-/2\).

\section{Further analysis of reconstruction protocol}
\label{app:Further_analysis} 

\subsection{Magnitude estimation of physical conducivity}

We first discuss the scale of the measured optical signal.  The conductivity
in our two-dimensional calculation is a genuine sheet conductance, and its
dimensionless numerical value is converted to SI units by multiplying by $e^2/\hbar=2.434\times10^{-4}\ {\rm S}$. For the representative parameters used in our numerical calculation, the
subgap response is typically of order
\begin{align*}
    |\operatorname{Im}\sigma_{\rm phys}^L|
    &\sim10^{-4}\frac{e^2}{\hbar}
    \simeq2.4\times10^{-8}\ {\rm S},\\
    |\operatorname{Re}\sigma_{\rm phys}^H|
    &\sim10^{-8}\frac{e^2}{\hbar}
    \simeq2.4\times10^{-12}\ {\rm S}.
\end{align*}

\begin{figure}[!h]
    \includegraphics[width=1\linewidth]{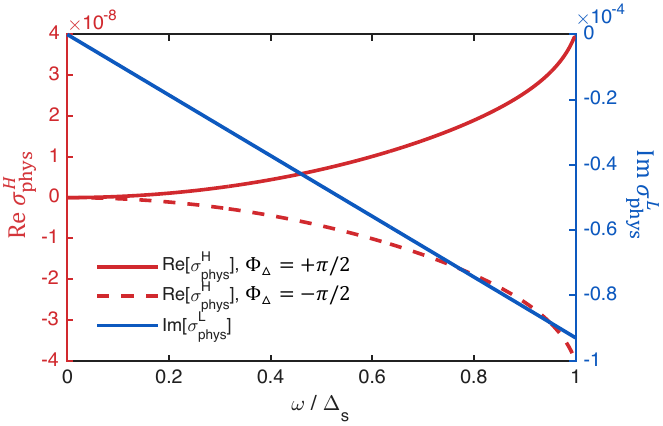}
    \caption{The 2D longitudinal and Hall conductivities (plotted in units of $e^2/\hbar$) as a function of frequency with $U/t=15$. Their ratio $[\operatorname{Im}\sigma_{\rm phys}^L]^2/\operatorname{Im}\sigma_{\rm phys}^H$ is proportional to the spinon Chern number. The longitudinal conductivity is invariant under the sign change of the flux, while the Hall conductivity changes sign.}
    \label{figs:conductivity}
\end{figure}

Consequently, in the strict single-layer limit, the Hall sheet conductance is only a
few picosiemens.  Since the polarization rotation of an optically thin sheet
is parametrically of order \(Z_0\sigma_{\rm phys}^H\), with
\(Z_0\simeq377\,\Omega\), the corresponding single-layer rotation is of
order \(10^{-9}\) rad or smaller.  Such a signal is difficult to
resolve with conventional broadband polarimetry, even though the larger
longitudinal response may remain accessible to phase-sensitive or
resonantly enhanced measurements.

A comparison with experiment must therefore account for the finite thickness
of a realistic layered sample.  If adjacent layers, separated by a distance
\(d\), realize the same CSL chirality, their long-wavelength conductivities
contribute to the total response.  The equivalent three-dimensional conductivity and the effective sheet
conductance of a sample of thickness \(L=Nd\) are
\begin{equation*}
    \sigma_{\rm 3D}^{L,H}
    =\frac{\sigma_{\rm 2D}^{L,H}}{d},
    \qquad
    G_{\rm eff}^{L,H}\simeq L\sigma_{\rm 3D}^{L,H}
    =N\sigma_{\rm 2D}^{L,H},
\end{equation*}
where the last relation applies in the optically thin, coherently probed
limit.  A sufficiently thick stack can consequently enhance the absolute
Hall signal by many orders of magnitude. Additionally, increasing the number of aligned layers improves the absolute
signal-to-noise ratio but does not change the intrinsic hierarchy ratio
\(|\sigma_{\rm phys}^H/\sigma_{\rm phys}^L|\sim10^{-4}\).

For a concrete estimate, consider an illustrative interlayer spacing
\(d=1\,\mathrm{nm}\).  A \(1\,\mu\mathrm{m}\)-thick sample then contains
approximately \(N=10^3\) contributing layers, giving
\begin{equation*}
    |G_{\rm eff}^L|\sim2.4\times10^{-5}\ {\rm S},
    \qquad
    |G_{\rm eff}^H|\sim2.4\times10^{-9}\ {\rm S}.
\end{equation*}
The Hall response is therefore enhanced from the picosiemens scale to the
nanosiemens scale, while \(Z_0|G_{\rm eff}^H|\sim9\times10^{-7}\) corresponds
to a polarization rotation of order one microradian, up to refractive-index
and substrate-dependent factors.  For \(L=10\,\mu\mathrm{m}\), the same
linear estimate gives \(|G_{\rm eff}^H|\sim2.4\times10^{-8}\,\mathrm{S}\)
and a rotation of order \(10^{-5}\,\mathrm{rad}\).  Thus realistic sample
thicknesses can improve the absolute Hall signal by three to four orders of
magnitude relative to a single layer.

\subsection{Finite spinon broadening}
\begin{figure}[!t]
    \includegraphics[width=1\linewidth]{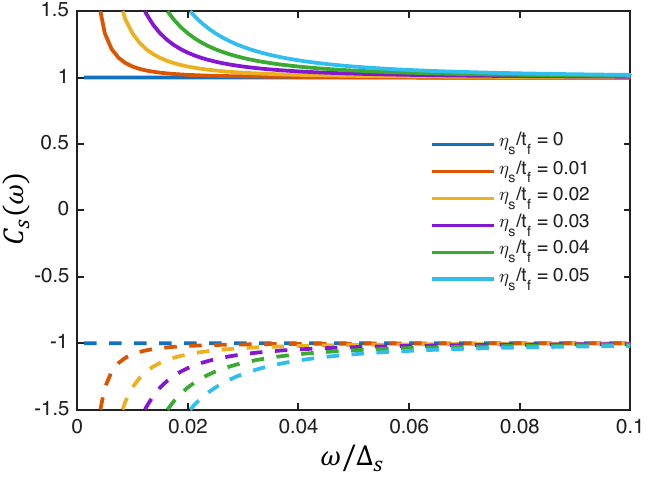}
    \caption{Extracted spinon Chern number \(C_s\) with finite spinon broadening at $U/t=15$. The dashed lines corresponds to the negative flux $-\pi/2$ result, which yield opposite sign of the Chern number.}
    \label{fig:Cs_eta_s}
\end{figure}

We next discuss the low-frequency divergence of the Chern-number estimator
with finite spinon broadening, as illustrated in Fig.~\ref{fig:Cs_eta_s}.
Here the broadening is implemented by evaluating the full spinon kernels at
\(z=\omega+i\eta_s\), while the subgap chargon kernel is kept unbroadened.
For \(\omega,\eta_s\ll\Delta_s\), Eqs.~\eqref{eq:sm_final_KfL}
and \eqref{eq:sm_full_spinon_antisymmetric_kernel} give
\begin{equation*}
    \Pi_f^L\simeq-\chi_f^L z^2,\quad
    \Pi_f^H\simeq i\kappa_s z,\quad
    \Pi_b\simeq-\chi_b\omega^2,
    \quad \kappa_s\equiv\frac{N_\sigma C_s}{2\pi},
\end{equation*}
where \(C_s=\pm1\) denotes the band Chern number.  At sufficiently small
\(\omega\) and \(\eta_s\), the Hall term dominates the Ioffe--Larkin
denominator, \((\Pi_b+\Pi_f^L)^2+(\Pi_f^H)^2\simeq-\kappa_s^2z^2\).
The physical kernels consequently reduce to
\(\Pi_{\rm phys}^L\simeq-\chi_b\omega^2\) and
\(\Pi_{\rm phys}^H\simeq-i\chi_b^2\omega^4/(\kappa_s z)\).
Using \(\sigma_{\rm phys}^{L,H}=i\Pi_{\rm phys}^{L,H}/\omega\), with
\(\omega>0\), we obtain
\begin{equation}
    \operatorname{Im}\sigma_{\rm phys}^L\simeq-\chi_b\omega,
    \qquad
    \operatorname{Re}\sigma_{\rm phys}^H\simeq
    \frac{\chi_b^2}{\kappa_s}
    \frac{\omega^4}{\omega^2+\eta_s^2}.
    \label{eq:sm_broadened_low_frequency_conductivities}
\end{equation}
Thus the finite-frequency estimator becomes
\begin{equation}
    C_s^{\rm est}(\omega)\equiv\frac{2\pi}{N_\sigma}
    \frac{[\operatorname{Im}\sigma_{\rm phys}^L]^2}
    {\operatorname{Re}\sigma_{\rm phys}^H}
    \simeq C_s\left(1+\frac{\eta_s^2}{\omega^2}\right).
    \label{eq:sm_broadened_chern_estimator}
\end{equation}
Finite broadening therefore suppresses the Hall denominator more strongly
than the longitudinal numerator, producing a signed divergence as
\(\omega\to0\) at fixed nonzero \(\eta_s\).  The scale
\(\omega\sim\eta_s\) marks the onset of an appreciable deviation, rather
than a pole at \(\omega=\eta_s\). The ideal result in main-text Fig.~2 takes
\(\eta_s=0\) before the low-frequency limit.  For small finite broadening,
the appropriate window for estimating the quantized value is instead
\(\eta_s\ll\omega\ll\Delta_s\).

\subsection{Reconstruction error analysis}

Although the inverse-response protocol removes the need to model the
frequency-dependent chargon kernel, it remains sensitive to experimental
errors. According to our mean-field results, we have \(|\Pi_b|\ll|\Pi_f^\pm|\). The inverse
Ioffe--Larkin rule then gives
\begin{equation}
    R_\pm=\frac{1}{\Pi_{\rm phys}^\pm}
    =\frac{1}{\Pi_b}
    +\frac{1}{\Pi_f^\pm}=q_b + r_\pm,
    \quad |q_b|\gg|r_\pm|.
    \label{eq:sm_error_inverse_hierarchy}
\end{equation}
Thus \(\Pi_{\rm phys}^\pm\simeq\Pi_b\), and the desired spinon information
is a small correction to a large common background in \(R_\pm\). Let the measured response contains a small error \(\delta\Pi_{\rm phys}^\pm\), so that the measured response is
\(\widetilde\Pi_{\rm phys}^\pm=\Pi_{\rm phys}^\pm+
\delta\Pi_{\rm phys}^\pm=(1+\alpha_\pm)\Pi_{\rm phys}^\pm\), where
\(|\alpha_\pm|\ll1\). The real and imaginary parts of \(\alpha_\pm\)
describe the amplitude and phase errors, respectively. Expanding
the reciprocal to first order gives
\begin{align}
    \delta R_\pm
    &\simeq-\frac{\delta\Pi_{\rm phys}^\pm}
    {(\Pi_{\rm phys}^\pm)^2}
    =-\alpha_\pm(q_b+r_\pm)
\end{align}
Obviously, when $\alpha_\pm$ vanishes, the error $\delta R_\pm$ also vanishes. The relative error in the full inverse response remains of order \(|\alpha_\pm|\), however, its size relative to the much smaller spinon term $r_\pm$ is enhanced by \(|\Pi_f^\pm/\Pi_b|\). Specifically, we have
\begin{align}
    \frac{|\delta R_\pm|}{|r_\pm|}
    \simeq |\alpha_\pm|\left|\frac{\Pi_f^\pm}{\Pi_b}\right|.
    \label{eq:sm_error_inverse_amplification}
\end{align}
Whether this large background error survives the extraction depends on the correlation between the two channels. 

For the Hall channel, define
\(\alpha_c=(\alpha_++\alpha_-)/2\) and
\(\alpha_d=(\alpha_+-\alpha_-)/2\). Using
\(r_\pm=r_f^L\pm i r_f^H\), the error in their measured difference is
\begin{align}
    \delta r_f^H
    &=\frac{\delta R_+-\delta R_-}{2i}
    \simeq-\alpha_c r_f^H+i\alpha_d(q_b+r_f^L),\notag\\
    \frac{|\delta r_f^H|}{|r_f^H|}
    &\lesssim |\alpha_c|+
    |\alpha_d|\left|\frac{q_b}{r_f^H}\right|.
    \label{eq:sm_error_Hall_calibration}
\end{align}
The last estimate uses \(|q_b|\gg|r_f^L|\) and assumes
\(r_f^H\ne0\). An identical calibration error $\alpha_c$ in both $\pm$ channels cancels the large background and primarily rescales the Hall response. A small relative error $\alpha_d$, however, leaves a residual proportional to \(q_b\), which can exceed the spinon signal. Accurate Hall extraction therefore requires \(|\alpha_d|\ll|r_f^H/q_b|\).

The longitudinal absorptive channel is similarly sensitive to phase
calibration. For example, there could be a small phase error in the measured response \(\widetilde\Pi_{\rm phys}^\pm=e^{i\phi_\pm}
\delta\Pi_{\rm phys}^\pm\simeq(1+\alpha_\pm)\Pi_{\rm phys}^\pm\), or equivalently, $\alpha_{\pm} = i\phi_\pm$. In this case, the relative phase error $\alpha_+=-\alpha_-$ does not affect the longitudinal response, but an identical phase error \(\alpha_+=+\alpha_-=i\phi\), with real \(\phi\), gives
\begin{equation}
    \delta[\operatorname{Im}r_f^L]
    =\operatorname{Im}\frac{\delta R_++\delta R_-}{2}
    \simeq-\phi\operatorname{Re}(q_b+r_f^L)
    \simeq-\phi q_b.
    \label{eq:sm_error_longitudinal_phase}
\end{equation}
Here \(q_b\) is real below the chargon absorption threshold. For an accurate reconstruction of the spinon longitudinal response, one may require the common phase error to satisfy \(|\phi|\ll|\operatorname{Im}r_f^L|/|q_b|\). These estimates explain why a small chargon kernel may impose stringent error contraol in the amplitude and phase calibration measurements.

Another possible error source is the calibration of the spinon interband threshold \(\Delta_s\), which fixes the KK subtraction constant through
\(r_{f,\infty}^L\simeq2.49788/\Delta_s\). If the inferred threshold is deviated from the true value by a small relative error as \(\widetilde\Delta_s=(1+\epsilon)\Delta_s\), then
\begin{equation}
    \widetilde r_{f,\infty}^L
    =\frac{r_{f,\infty}^L}{1+\epsilon},\qquad
    \frac{\delta r_{f,\infty}^L}{r_{f,\infty}^L}
    =-\frac{\epsilon}{1+\epsilon}\simeq-\epsilon.
\end{equation}
With the optical input held fixed, this produces a frequency-independent
shift in the reconstructed \(\operatorname{Re}r_f^L\). However, the final
inversion \(\Pi_f^\pm=1/(r_f^L\pm i r_f^H)\) converts this offset into a
frequency-dependent error, especially where the inverse spinon response
is small. The reconstructed geometric spectra can therefore undergo
changes in line shape rather than a simple overall rescaling.

\begin{figure}[!t]
    \includegraphics[width=1\linewidth]{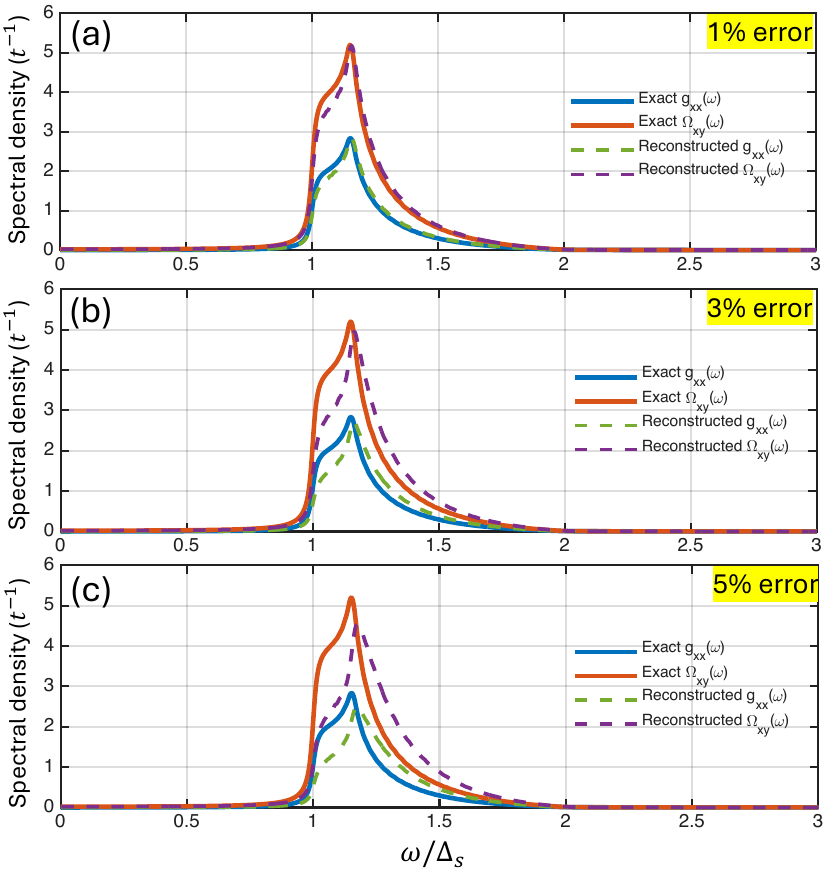}
    \caption{Spectral density of Berry curvature and quantum metric when the spinon gap $\Delta_s$ is measured with $1\%$ error (a), $3\%$ error (b) and $5\%$ error (c). These errors are simulated by manually deviating the spinon gap $\Delta_s =2 \sqrt{3}t_f$ in the reconstruction procedure. The dashed lines corresponds to reconstructed results while the invariant solid lines corresponds to the exact results. The calculation is performed at $U/t=15$ with positive flux $\pi/2$ and $\eta_s/t_f=0.05$.}
    \label{fig:Delta_s_error}
\end{figure}

Figure~\ref{fig:Delta_s_error} illustrates this sensitivity for threshold
errors of \(1\%\), \(3\%\), and \(5\%\), with the underlying response
otherwise unchanged. The discrepancy grows with the calibration error;
for the parameters shown, a \(5\%\) error produces a large deviation from the exact result for both the quantum-metric and Berry-curvature spectra, including reduced peak heights and enhanced high-frequency shoulders. We admit that our reconstruction protocal of spinon QGT spectral is feasible in principle, but quantitatively accurate experimental reconstruction demands precise measurements of the complex optical response and \(\Delta_s\), especially in the deep-mott regime. We expect that a smaller Hubbard interaction \(U\) (but still in the CSL phase) would reduce the precision requirement for the reconstruction when the virtual doublon-holon excitations become less forbidden, yielding a larger chargon response.

% \bibliography{references}

\end{document}